\documentclass[12pt,preprint]{aastex}

\newcommand{\hst}{{\it HST}}
\newcommand{\halpha}{H$\alpha$}

\slugcomment{To appear in AJ.}

\shorttitle{Image-Processing Techniques}
\shortauthors{Rector, Levay, Frattare, English \& Pummill}

\begin{document}


\title{Image-Processing Techniques for the Creation of Presentation-Quality Astronomical Images}


\author{Travis A. Rector}
\affil{University of Alaska Anchorage, 3211 Providence Drive, BMB 212, Anchorage, AK 99508}
\email{rector@uaa.alaska.edu}

\author{Zoltan G. Levay and Lisa M. Frattare}
\affil{Space Telescope Science Institute, 3700 San Martin Drive,
Baltimore, MD 21218}

\author{Jayanne English}
\affil{Department of Physics and Astronomy, University of Manitoba, Winnipeg, MB R3T 2MB, Canada}

\and

\author{Kirk Pu'uohau-Pummill}
\affil{Gemini Observatory, 670 N. A'ohoku Place, Hilo, HI 96720}




\begin{abstract}

The quality of modern astronomical data, the power of modern computers and the agility of current image-processing software enable the creation of high-quality images in a purely digital form.   The combination of these technological advancements has created a new ability to make color astronomical images.  And in many ways it has led to a new philosophy towards how to create them. 
A practical guide is presented on how to generate astronomical images from research data with powerful image-processing programs.  These programs use a layering metaphor that allows for an unlimited number of astronomical datasets to be combined in any desired color scheme, creating an immense parameter space to be explored using an iterative approach.  Several examples of image creation are presented.

A philosophy is also presented on how to use color and composition to create images that simultaneously highlight scientific detail and are aesthetically appealing.   This philosophy is necessary because most datasets do not correspond to the wavelength range of sensitivity of the human eye.
The use of visual grammar, defined as the elements which affect the interpretation of an image, can maximize the richness and detail in an image while maintaining scientific accuracy.  By properly using visual grammar, one can imply qualities that a two-dimensional image intrinsically cannot show, such as depth, motion and energy.  In addition, composition can be used to engage viewers and keep them interested for a longer period of time.  The use of these techniques can result in a striking image that will effectively convey the science within the image, to scientists and to the public.

\end{abstract}




\section{Introduction}

For many decades astronomical color images have been generated using large-format photographic plates
and traditional darkroom techniques, e.g., \citet{mal92}.  In the early 1980s, charge-coupled device (CCD)
detectors began to replace photographic plates as the instrument of choice for astronomical research.  However,
until recently CCD arrays lacked the number of pixels necessary to compete with the fine grain of photographic plate emulsion.  In recent years CCD detectors have grown in size and the physical size of pixels has decreased.  And
mosaics of CCD arrays have been implemented in many instruments.   The large number of pixels in these
cameras now allows for high-quality optical images to be generated in a purely digital form.  

Furthermore, the
continuous improvement of imaging capabilities in non-optical windows of the electromagnetic spectrum have 
enabled the creation of high-quality images at other wavelengths as well.  Historically, astronomical images have been made by combining grayscale images taken through red, green and blue optical filters.  But often images
are now made from datasets that are either outside the optical window or do not match the characteristics of the color-detecting cones in the human eye.

The development of advanced astronomical instrumentation has been contemporaneous with the advancement of computing power and, in particular, digital image-processing (IP) software for commercial applications.  These IP programs, e.g., Adobe\textregistered\ Photoshop\textregistered\footnote{Adobe and Photoshop are either registered trademarks or trademarks of Adobe Systems Incorporated in the United States and/or other countries.}, Photoshop\textregistered\ Elements and The GIMP\footnote{The GIMP is written by Peter Mattis
and Spencer Kimball, and released under the GNU General Public License.}, offer unprecedented power, flexibility and agility in digital image generation and manipulation.  The combination of
these technological advancements has led to a new ability to make color astronomical images.  And in many ways, it
has created a new philosophy towards how to create them.  No illustration of this is more apparent than the famous \citet{hes95} ``Pillars of Creation" image of the central region of the Eagle Nebula (M16), with the {\it Hubble
Space Telescope} (\hst).   This image demonstrated the tremendous resolution of the \hst\ Wide-Field Planetary Camera~2 (WFPC2) camera.  It also showed how narrow-band imaging can change our view of an object.  And it
demonstrated how color schemes can be used to imply depth, motion and texture in an
astronomical image.  Just as importantly, it illustrated how effectively such images can inspire the public and
generate enthusiasm for astronomy in general.  

The success of the \hst\ image of M16, and images like it, inspired the creation of the {\it Hubble Heritage
Project} \citep{nol01}.  Since its inception in 1998, the project has released a new color image from \hst\ every month.  And the
success of the {\it Hubble Heritage} project has inspired many other observatories to invest resources into
the creation of  images from astronomical research data that are primarily intended for the lay person. 
Many scientists have also become interested in ways to create such images from their data, not only for
public consumption but also as a visualization tool for colleagues.  The goal of this paper is to demonstrate many of the
techniques used by image creators of the {\it Hubble Heritage} team, the National Optical Astronomy Observatory, the Gemini
Observatory, the Canadian Galactic Plane Survey and others to generate astronomical images from research-quality data.  

This paper is intended for a range of audiences, from professional astronomers to outreach specialists.  The procedural steps, from projecting the data into images to the final electronic and print output, are discussed in \S 2.  This section serves primarily as an instructional guide.  It will be of most interest to those with little image-processing experience who wish to learn how to generate a color image from their data.  As many of the steps described herein are subjective, different philosophies regarding image creation are discussed in \S 3.  This includes details on how to select colors and use composition to engage the viewer.  This section will be of greater value to those who have prior experience in assembling images but wish to improve their skills, although in many ways the steps described in \S 2 will be more clear after reading \S 3.  Step-by-step examples are provided in Appendices~A~\&~B.  These are of particular value as tutorials on how to implement the steps that are described in mostly general terms in \S 2.  Novice image makers are encouraged to work through the examples in Appendices~A~\&~B after reading \S 2 and before reading \S 3.  
Appendix~C provides examples of the cosmetic cleaning steps often necessary on images generated from optical and near-IR data.  Appendix~D presents an explanation on how to create photometrically calibrated images.  
Appendix~E gives an example of image preparation for color process printing in Photoshop.  

\section{Procedure}

Generating a color image can be distilled into the following steps: (1) intensity scaling and projection of 
each dataset into grayscale images; (2) importing the datasets as layers into an IP software package; (3) assigning a color and intensity rescaling to each layer; (4) fine tuning
the image, which includes overall color balance, the cleaning of cosmetic
defects, and orientation and framing; and (5) preparing the image for electronic distribution and print production.  Prior to image generation, it is very important that your monitor is color calibrated and a color management workflow is established, steps that are described in \S 2.6.1 and \S 2.6.2.

The techniques described herein assume that more than one dataset,
preferably three or more, will be combined to produce a color image.  
A dataset is defined as a two-dimensional image of a particular waveband, polarization or other
distinct characteristic; e.g., an optical image through a single filter, or a radio image at or in a particular
waveband and/or polarization.  
These techniques are designed to take advantage of the distinct structural information in each dataset
for which they were obtained.  
For comparison, a popular technique among amateur astronomers is known as the LRGB method, e.g., \citet{gen03}, wherein an image is first generated in the traditional ``natural color" scheme (see \S 3.2.1) from datasets obtained through red, green and blue filters.  To improve the image quality, an unfiltered, ``luminosity" image is added that lacks color information but has a higher signal to noise ratio.  This
technique is well suited for small-aperture telescopes because most objects are limited by relatively poor signal to noise.  The LRGB method is effective for decreasing the noise in an image, but
results in a loss of wavelength-specific structural information.
This method is not well suited for use with scientific data, which is rarely obtained unfiltered.  Indeed, narrow-band observations
are obtained specifically to increase the contrast between emission-line and continuum-emission regions of an object.   Thus the
quality of an image can actually be improved by the exclusion of particular wavebands.

The data must first be
fully reduced, e.g., optical data are bias and flat-field corrected, with a standard data
reduction package, such as IRAF\footnote{IRAF is offered by the National Optical Astronomy Observatory,
which is operated by the Association of Universities for Research in Astronomy (AURA), Inc. under cooperative
agreement with the National Science Foundation.}, IDL\footnote{IDL is offered by Research Systems, Inc., a
wholly-owned subsidiary of Eastman Kodak Company.} or AIPS\footnote{AIPS is offered by
the National Radio Astronomy Observatory, which is a facility of the National Science Foundation operated
by Associated Universities,Inc. under a cooperative agreement.}.  The data must also be stored in a data
file, e.g., FITS format, that can be projected into an image.  The image will then be imported into an IP
program which can handle layers, e.g., Photoshop, Photoshop Elements or The GIMP.  The steps in this section will be
discussed in general terms, without reference to a particular software program, where possible.  A straightforward example
using IRAF, Karma \citep{goo95} and The GIMP is given in Appendix~A.  And a more complex example using IDL and  Photoshop is given in Appendix~B.  
The reader is encouraged to read both examples, as different techniques are illustrated in each.  
It is important to note that the {\it ESA/ESO/NASA Photoshop FITS Liberator} \citep{chr04} is a very useful plug-in that can scale and project FITS datafiles directly into Photoshop, versions 7.0 and higher, and into Photoshop Elements, version 2.0 and higher.  It provides a graphical interface for many of the steps described in \S 2.1 and is a viable alternative to IRAF, KARMA and IDL for most of the steps described.  
The tutorials were written for IRAF version 2.12, Karma v1.7, The GIMP v2.0, IDL v.6.0 and Photoshop~CS v8.0.
Of course, as new versions of these software packages are released, features and options will be added, moved and removed.  Note that the power of
these programs is far beyond what can be explained here.  The reader is therefore expected to have a
basic familiarity with the data reduction and IP programs to be used.

While not explicitly discussed in the rest of \S 2 or in Appendix~B, Photoshop Elements is a less-expensive alternative to Photoshop which includes all of the features essential to implement the procedures described herein.  As of this writing, Photoshop Elements is in version~3.0.  It is capable of layers, adjustment layers, 16-bit channels and many of the other valuable features of Photoshop.  Important features that Photoshop Elements lacks include the patch tool, channel editing, some masking features and CMYK separations.  The last feature is particularly useful for preparing images for press in the professional journals as well as for public consumption, as discussed in \S 2.6.

While presented as a linear step-by-step process, it is important to emphasize the iterative nature of
image creation; e.g., while assigning colors to a dataset it may be noted that a different intensity scaling may
improve the image, thereby prompting a return to the original dataset.  Similarly, the flexibility of IP
software such as  Photoshop allows the exploration of many different realizations of an image
simultaneously.  With the variety of intensity scalings and colorings possible for each dataset, the parameter space for combining all of the datasets into a single image is immense.  It is recommended that the user explore many different realizations of the image, e.g., by using thumbnail sketches as described in \S 2.4, before settling on a final composite image.  It is also recommended that you show multiple versions of the image to unbiased viewers to see how the image engages them before making a decision on the final image.  

\subsection{Scaling and Projecting Data}

One of the most critical steps in the image generation is also the first step after the data are reduced. 
Each dataset must first be projected into a grayscale image.  It is important to distinguish between a ``dataset"
and an ``image."  A dataset consists of the measurements made by an instrument on a telescope, e.g., the number of digital units in each pixel of a CCD camera.  A dataset is ``projected" into an image by using
a scale function\footnote{The scale function is also often referred to as the ``transfer" or ``stretch" function.} that converts the data values in each pixel into one of a finite number of pixel values.  In a grayscale image, these pixel values are shown as shades of gray.  In an index-color (color-mapped) image, these pixel values are shown as pre-determined color values.  Only 256 pixel values ($2^8$) are available in an 8-bit image; whereas 16-bit images have  65,536 pixel values ($2^{16}$).

Most optical CCDs have 15-bit or 16-bit converters which can create datasets that have dynamic ranges of $>$10,000; and radio interferometric observations can result in datasets with dynamic ranges of $>$100,000, e.g., \citet{per99}.  An 8-bit image is therefore incapable of displaying the structure and detail in a dataset over the  entire dynamic range without some compromise.  Sixteen-bit images are therefore preferable because they can retain the fine detail within a dataset over a much larger dynamic range.
Unfortunately, this does not necessarily result in a more detailed image because computer monitors are currently capable of displaying only 256 shades per channel.  Thus, while 16-bit grayscale images better retain the detail within a dataset, ultimately they cannot display it any better than an 8-bit grayscale image.  

While current display technologies cannot show the detail in a 16-bit image, the advantage of the format is that it better retains the fine structure contained within the original datasets.  Note that, after the initial scaling is done in IRAF, IDL or {\it FITS Liberator}, every rescaling of the dataset, e.g., with the levels or curves tool in Photoshop, results in a loss of pixel values.  Within an 8-bit dataset, the loss of even a few of the 256 pixel values can have a significant impact on the quality of an image.  Every time a 16-bit image is rescaled, there is also a loss of some of the 65,536 pixel values.  But the effect of this loss on the image itself is minimal because  the remaining 16-bit pixel values
are redistributed among the 256 values available for image display, resulting in no loss of image quality. 
Thus, high-contrast scalings are not likely to introduce posterization (see below) into the final image, allowing more flexibility in choosing the initial scaling for projecting the data into an image.

As of this writing, 16-bit grayscale and 16-bit-per-channel color image formats are an emerging technology.  Currently only an handful of programs, including IDL and the {\it  FITS Liberator}, are capable of projecting 16-bit grayscale images.  And only Photoshop~CS~v8.0 and Photoshop Elements~v3.0 are capable of manipulating them in the means described in this paper.  As mentioned, the advantage of the 16-bit format is that it allows a great deal more flexibility in choosing the initial scaling.  The value of this, however subtle, is nonetheless tremendous.
The disadvantage of the 16-bit format is that file sizes are effectively doubled and more CPU processing power is needed, a considerable concern when working with large datasets.  The 16-bit format is therefore strongly recommended unless the sizes of the datasets are prohibitive.

Ultimately, the 16-bit image format currently offers only a modest advantage over 8-bit, as
24-bit color image formats, the dominant standard, allocate only 8~bits each to the three color channels.  While this may sound like a limitation, 24-bit color images still can generate over 16~million distinct colors, which is more than current monitors can display; and, as discussed in \S 2.6, it is  far more than can be printed.  Thus, to some degree, scaling each dataset into an 8-bit
grayscale image is an undesirable but unavoidable step.  Fortunately, with the careful selection of a scaling
scheme, the structure and detail of the original dataset can be retained with one or more scaling solutions.

\begin{figure}
\plottwo{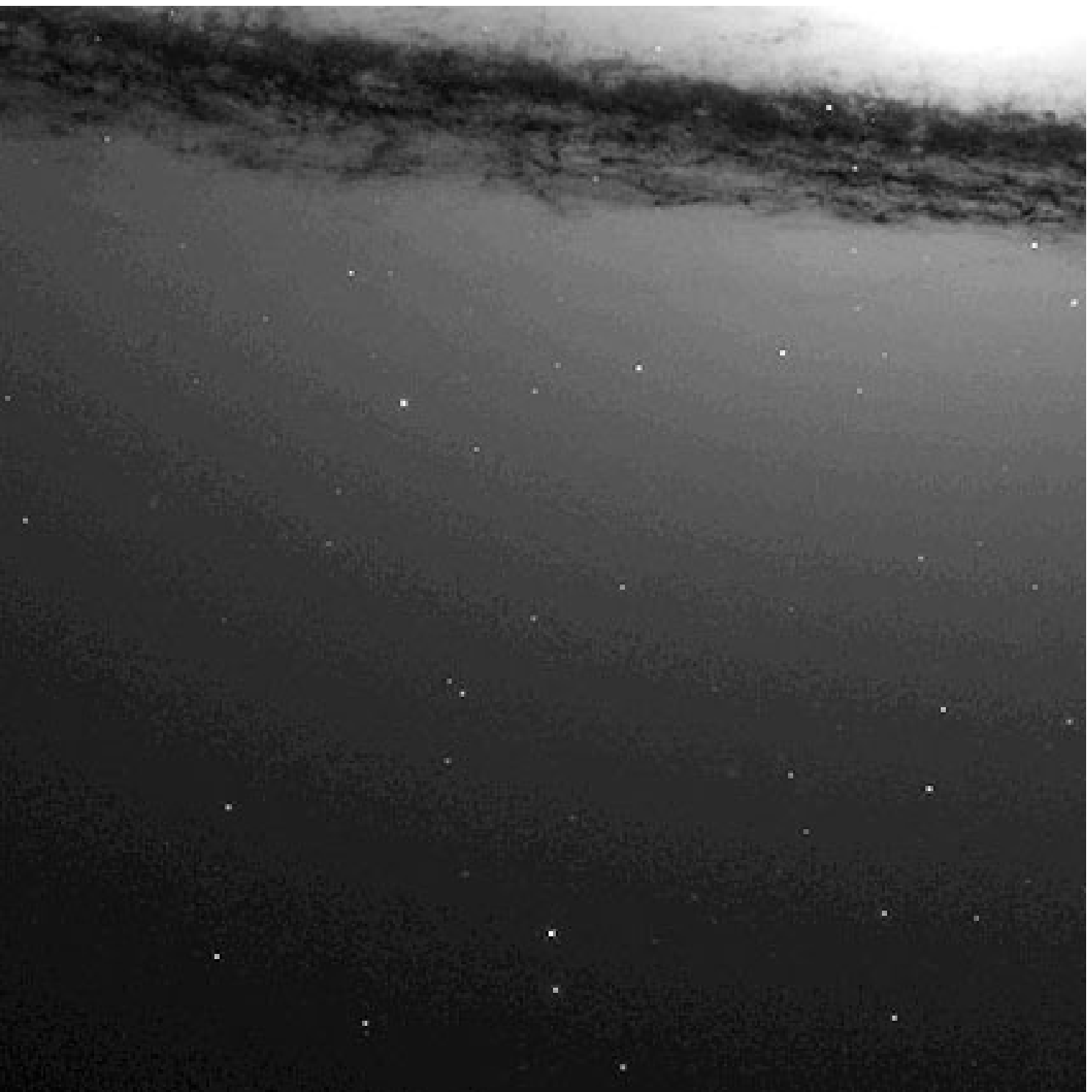}{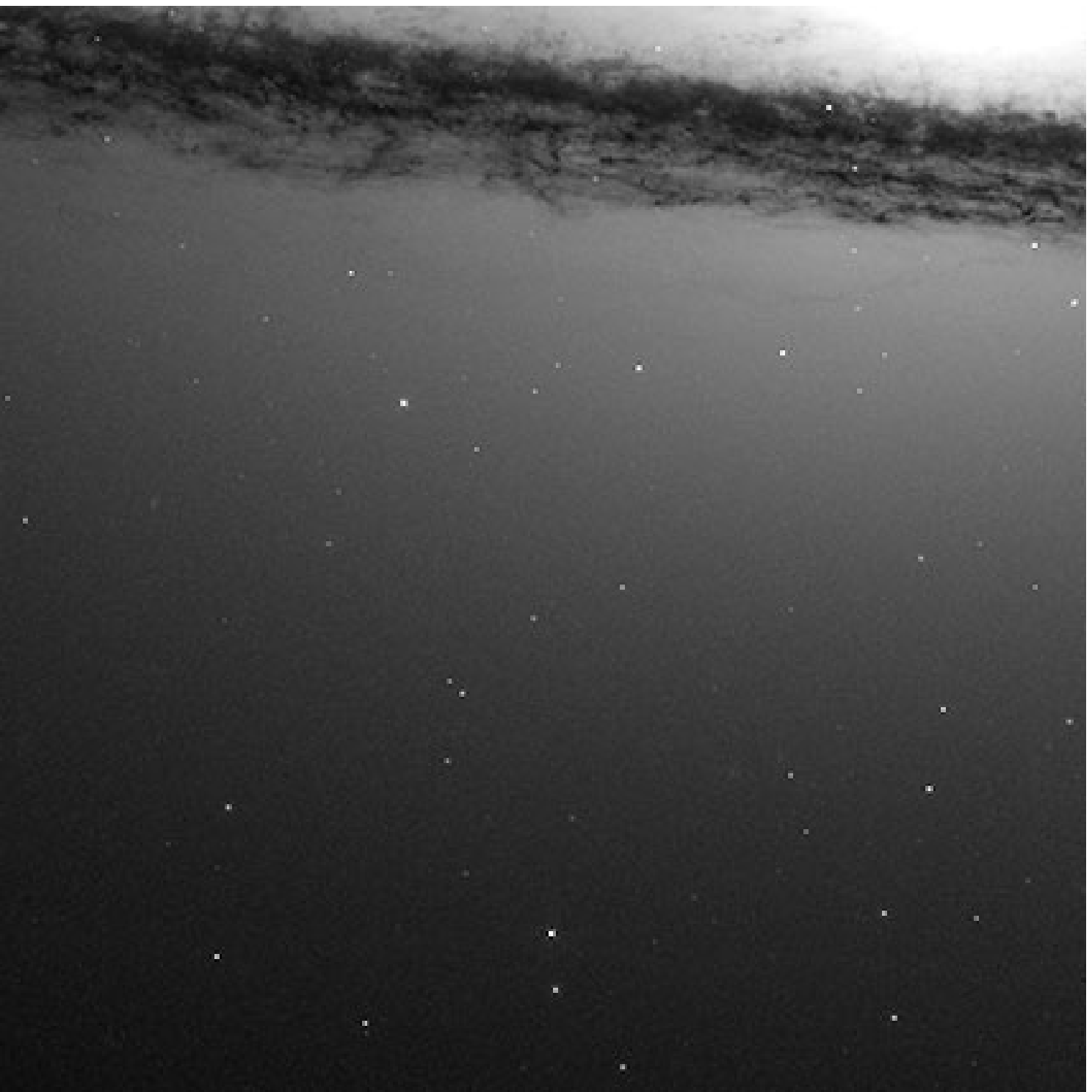}
\caption{This is a portion of M104 that was observed with the \hst\ ACS camera.   Both images were projected from the same dataset
with a square-root function and identical minimum and maximum scale values, and then identically rescaled with a high-contrast curves adjustment in Photoshop.  The image on the left was projected into an 8-bit grayscale image and suffers from ``posterization" due to the high-contrast rescaling.  The
image on the right was projected into a 16-bit grayscale image and does not suffer from posterization.  This is an extreme example that
can be avoided if the dataset is instead projected with a logarithmic function. \label{fig-1}}
\end{figure}

\begin{figure}
\plottwo{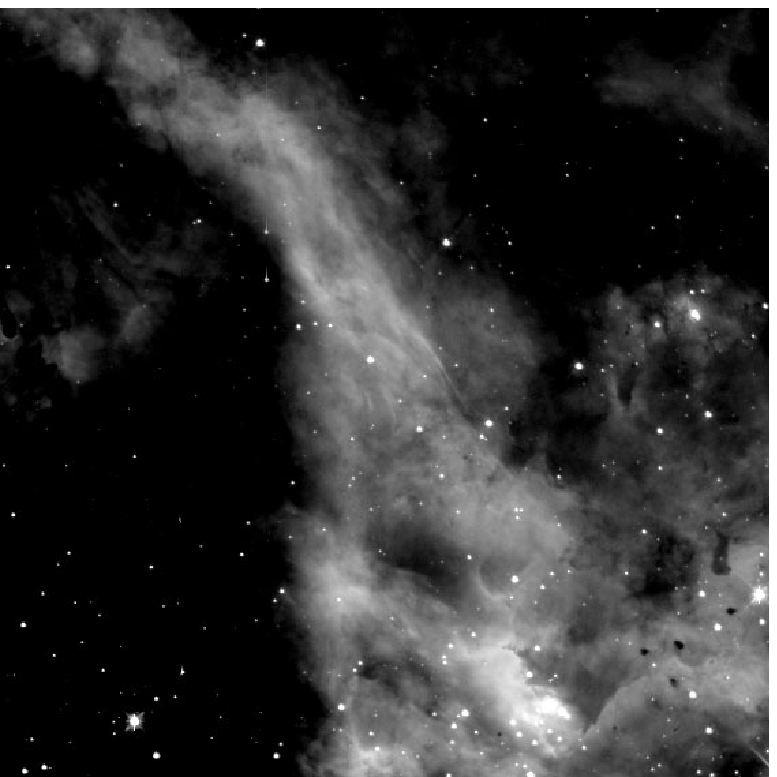}{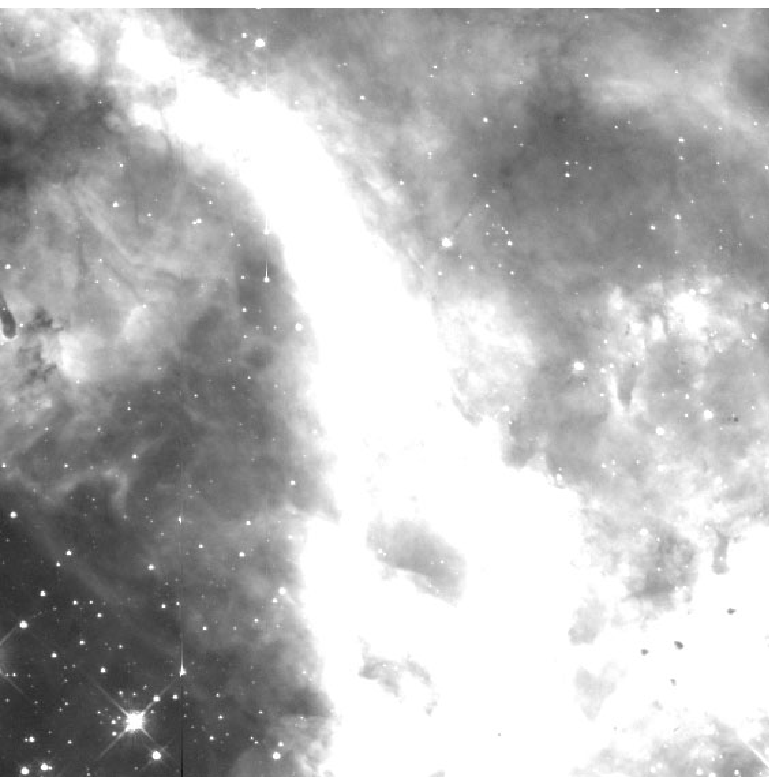}
\caption{Two images projected from the same dataset but with different scaling functions.  The image on the left contains large undersaturated (black) regions.  And the image on the right contains large oversaturated (white) regions.  In both cases detail in the saturated regions is lost.  To recover the detail in these regions the minimum and maximum intensity scale values must be changed.  \label{fig-2}}
\end{figure}

For any scaling system, minimum and maximum data values for the scale must be specified.  All data values
below the minimum scale value are ``undersaturated" and will be displayed as a pure black; and all data
values above the maximum scale value are considered ``oversaturated" and will be displayed as a pure white. 
Each intermediate data value will be displayed as one of 254 shades of gray.  When choosing a scaling system
consider the following compromise:  If the difference between minimum and maximum scale values is large, the
scaling will be coarse and subtle differences in intensity will be lost; e.g., in a linear scale from 0 to
32,768, each shade of gray will represent a range of 128 in data values (since $32,768 / 256 = 128$).  Thus, if the
values of two pixels both fall into the same bin of 128, they will be shown as the same shade of gray and
therefore will be indistinguishable.  In extreme cases, when large sections of a
dataset are assigned the same shade of gray, the resulting image can look posterized
(Figure~\ref{fig-1}).   If the difference between the minimum and maximum scale values is small,
minute differences in data values will be better resolved.  In such a case, all data values outside of the scaling range
will be undersaturated or oversaturated; and detail in these regions will be lost (Figure~\ref{fig-2}).   If a scaling
undersaturates or oversaturates a particular region of the image, that region will be shown as pure black or
pure white respectively; and detail within the saturated regions cannot be recovered once the dataset is
projected into an image.  

\begin{figure}
\includegraphics{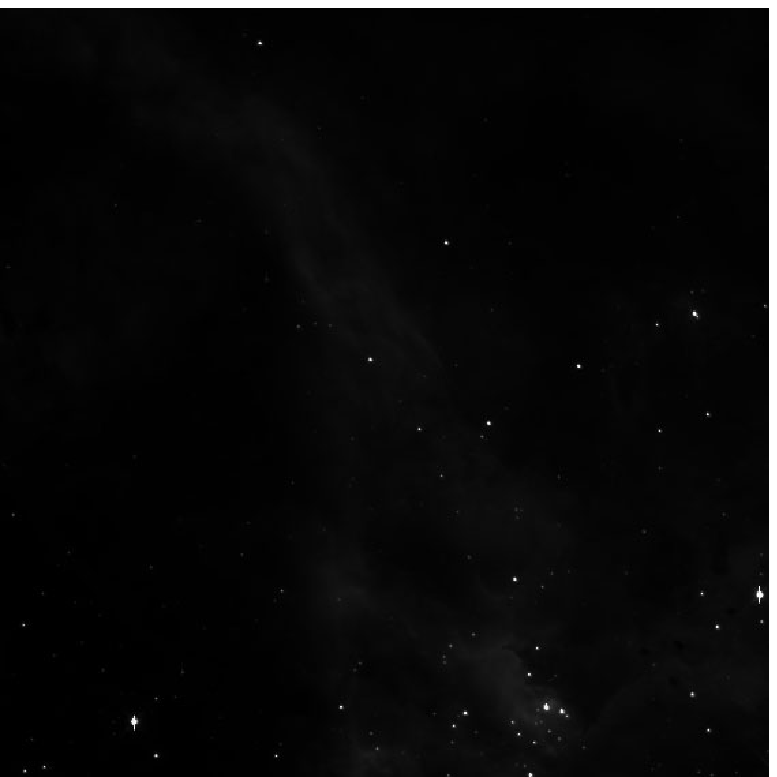}
\includegraphics{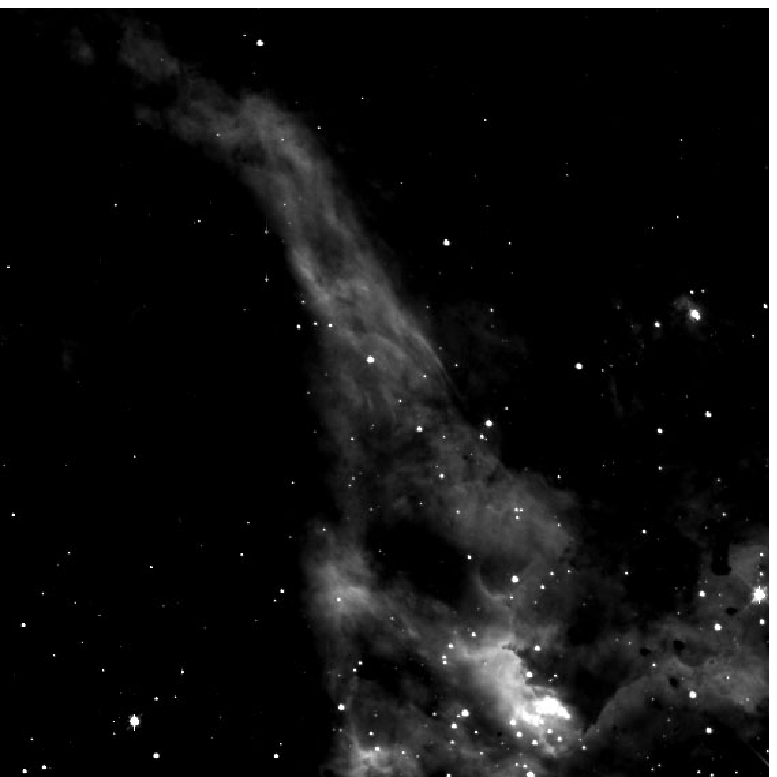}
\includegraphics{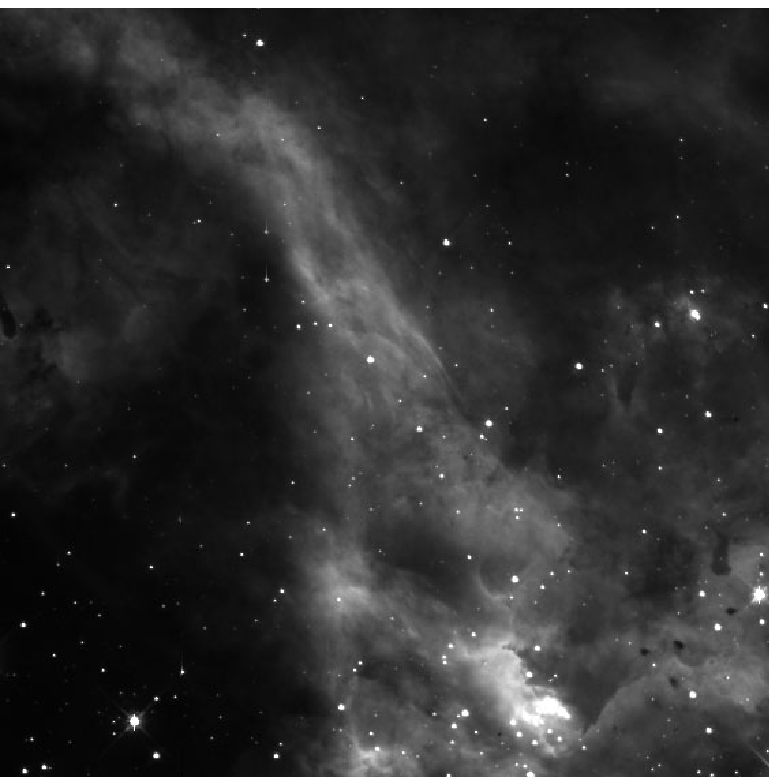}
\includegraphics{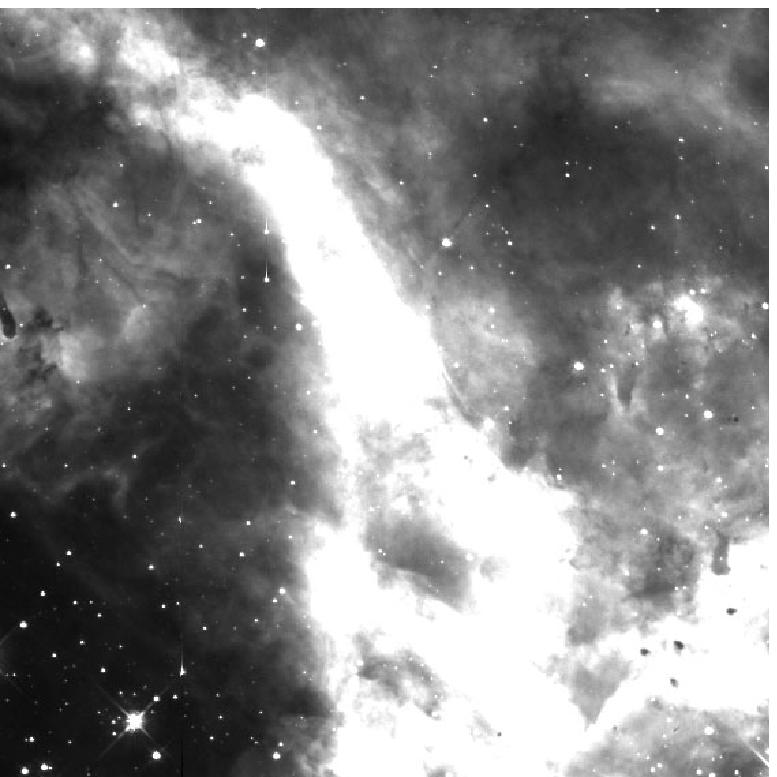}
\caption{Four examples of images projected from the same dataset of 30~Doradus, observed with the \hst\ WFPC2 camera and F656N filter.  Each projection is done with a linear scaling but with different minimum and maximum scale values.  The upper-left image shows a scaling of $(0:4096)$ that encloses nearly all of the data values but as a result does not show the faint structure.  The upper-right image shows a scaling of $(100:500)$ that better shows the midtones but undersaturates the shadows and oversaturates a bright portion of the nebula in the lower right.  The lower-left image shows a scaling of $(0:500)$ that does not saturate shadows.  The lower-right image shows a scaling of $(0:150)$ that best shows the shadows but oversaturates the highlights.  None of these scalings is sufficient to show the detail in the dataset, suggesting that an alternative to the linear scaling is necessary.  \label{fig-3}}
\end{figure}

In general, a good scaling will maximize the tonal range without saturating the detail of interest.  A good starting point is to choose a
scaling with a minimum data value at or slightly above the noise level of the image.  
When choosing the minimum scaling value it is important to consider the target audience.
Scaling the image so that the noise is visible is valuable for scientists to ascertain the sensitivity of the data
but it tends to be distracting to the public.  Because they don't understand the nature of astronomical data, the public
tend to interpret noise as a defect in the image.  The  maximum scaling value should be set
to just above the highest datapoint value in the source structure.  In the case of optical images it is
generally not a concern if the cores of stars exceed the maximum scaling value because brighter stars 
usually are scaled to appear white in the final images; and 
the unsaturated halos around these stars will indicate the color of the star.  A histogram of data values may
prove useful in identifying an appropriate scaling range; e.g., the IRAF {\it histogram} routine, part of the
stsdas.graphics.stplot package, can plot a histogram.  However, the histogram will contain stars and bad
pixels which in general should be ignored when choosing a scale.  
Figure~\ref{fig-3} shows examples of the same dataset with different scalings to illustrate the difficulty in choosing a good scaling.
It is best to search iteratively for good
scaling values by displaying an image, e.g., with ximtool or SAOImage~DS9\footnote{SAOImage~DS9 is provided by the
Smithsonian Astrophysical Observatory / High Energy Astrophysics Division Research \& Development group.},
and examining the pixel values in the regions of interest.  Karma's {\it kvis} tool is well suited for this task because it has an interactive histogram that immediately shows the effect on the image of selecting different values in the histogram.  
The {\it FITS Liberator} does this as well.  It also shows which pixels will be saturated.

\begin{figure}
\includegraphics{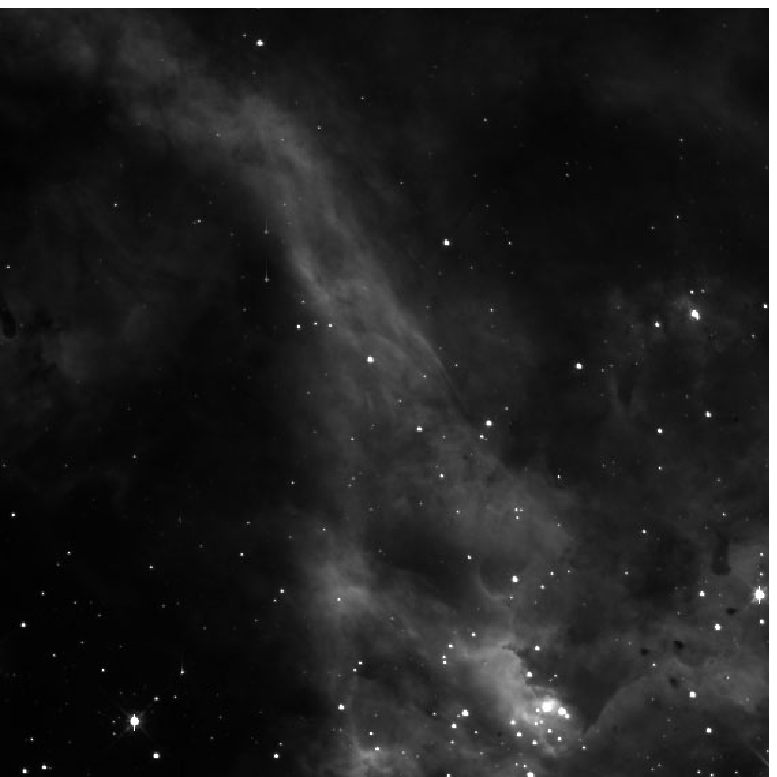}
\includegraphics{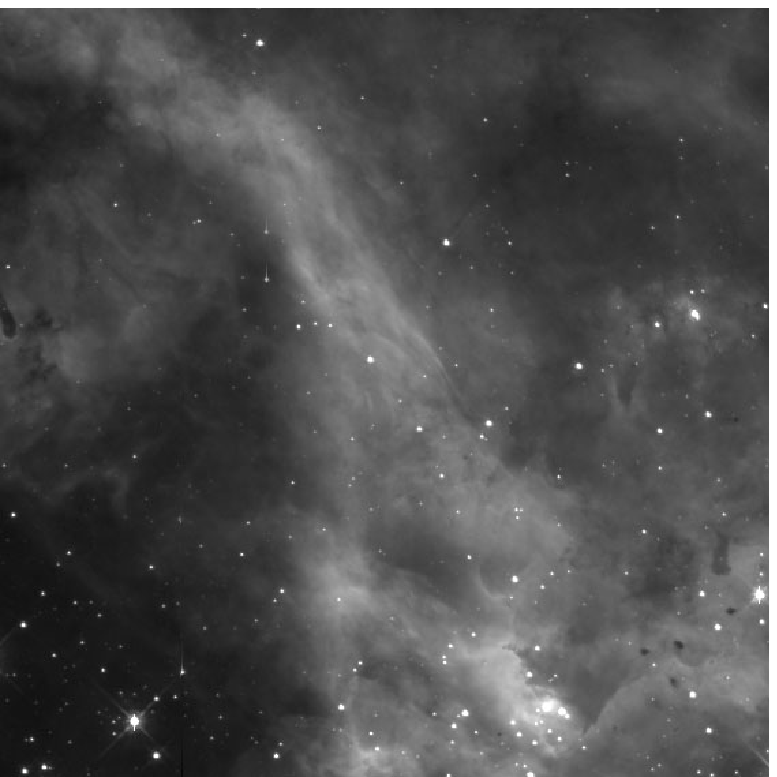}
\includegraphics{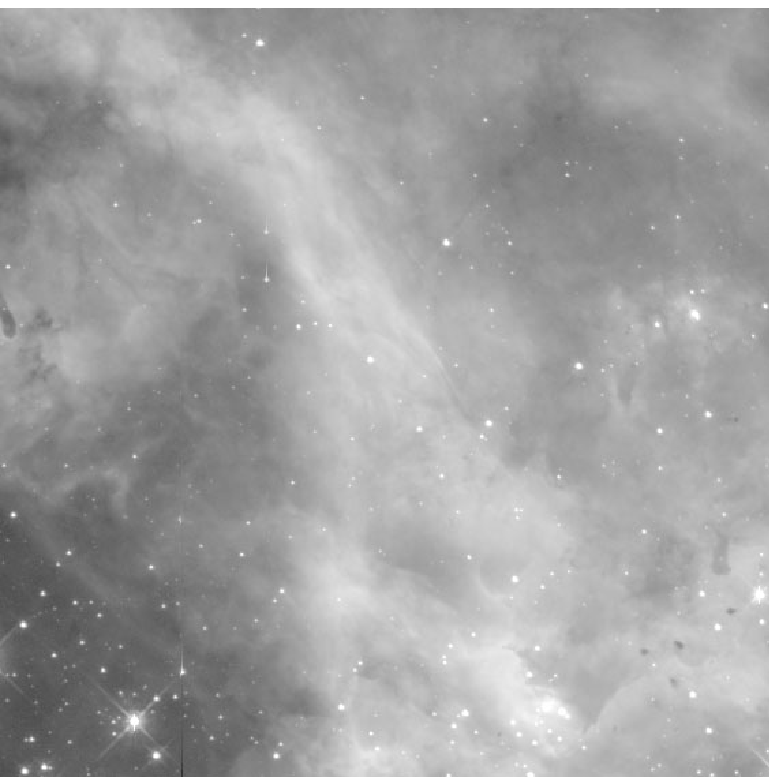}
\caption{Three examples of images projected from the same dataset of 30~Doradus, observed with the \hst\ WFPC2 camera and F656N filter.   Each image is projected with the same maximum and minimum scaling values of $(0:750)$ but with different scaling functions: linear (top left), square-root (top right) and logarithmic (bottom left).  The square-root scaling provides the best usage of the 256 shades of gray available.  The linear image is too dark and the logarithmic image has too low of contrast.  \label{fig-4}}
\end{figure}

A scaling function needs to be chosen that will optimize the contrast and detail throughout the tonal range.  In general, a linear scale function works well when most of the structure of interest in a dataset has a modest dynamic range and structural detail is very subtle, e.g., a
diffuse emission nebula.  The most common nonlinear scaling transformation functions are square-root and logarithmic scales, e.g., as shown in Figure~\ref{fig-4}.  A logarithmic scale often works well when there is a large dynamic range in the structure of
interest but most of the structure is faint, e.g., a spiral galaxy that has a bright, compact bulge and large, faint spiral
arms.  Another scaling system, commonly used in digital cameras, is histogram equalization.
In histogram equalization, the pixel values of an image are sorted and divided into 256 bins so that each bin contains an equal number of pixel values. The bin containing the lowest pixel values is assigned a grayscale value of $0$ and the bin of highest pixel values is assigned to $255,$ with all other bins receiving intermediate values.  The original pixel values are then replaced by the grayscale value of their corresponding bin.  Histogram equalization works well when the number of bad pixels is very small.
Other nonlinear functions, such as log-log and those given in \citet{lup03} should also be considered.  IRAF, IDL, Karma and other programs allow user-defined scaling functions. 
Unless scaling each dataset to a photometric scale, as described in Appendix~D, 
it is important that a separate scaling system be determined for each dataset
because the data values for each dataset are different.  This may seem counter-intuitive at first.  
However, using the same scaling system for each dataset
can lead to undesirable results.  For example, a chosen minimum and maximum scaling range may be appropriate
for one dataset but too wide for another, resulting in poorer resolution of the intensity scale; i.e.,
only a fraction of the 256 shades of gray will be available to show the detail in the dataset.  Worse,
if the chosen scaling range is too narrow for a particular dataset, regions of the dataset will be
undersaturated or oversaturated.  Therefore, be certain to apply a different scaling range for each dataset.  Also consider using a
different type of scaling function for different datasets; e.g., a logarithmic scaling for broadband filters and a
linear scaling for narrow-band filters.  

In some cases the dynamic range of a dataset is too large; and there is interesting structure in both the
bright and faint regions of the dataset which cannot be shown well with a single scaling solution.  
Several solutions are possible.
The ``unsharp mask" is an image manipulation technique that compresses dynamic range
by dividing the original version of an image by a low-resolution, i.e., blurred, version.  The end result is that
large-scale structure is cancelled and fine detail is retained, even over a large dynamic range.  The
technique has been used with great success in several astronomical images, e.g., \citet{mal79}.  
Note that the {\it Unsharp Mask...} filter in Photoshop functions somewhat differently.
In the image to be filtered, it locates pixels that differ in value from surrounding pixels by a specified threshold and 
then increases these pixels' contrast by the specified amount; i.e., for the neighbors of a selected pixel
the lighter pixels become lighter and the darker pixels become darker based on the specified amount.
Another option to compress dynamic range is to subtract or divide a function from the dataset before
projection; e.g., to generate an \hst\ image of M51 \citep{her01} the original datasets were divided by a galaxy disk
model.
In some
cases it is useful to use more than one scaling of the data, with the intention of showing the detail in different
regions of the image.
The two scalings are then exported
as separate grayscale images, imported into the IP software and combined.   One approach is to blend
the two images as layers with decreased opacity; e.g., in the \hst\ image of M51 described above, two 
scalings of each dataset were blended.  
One scaling was a logarithmic stretch; and the other scaling was linear but divided by the
galaxy disk model.  When blended, the overall dynamic range was compressed and
the contrast in the arms was enhanced. 
Alternatively, the two scalings can be combined via a mask, a technique that is
described in \S 2.3.3.

The {\it FITS Liberator} plug-in is an efficient and effective way to project each dataset directly into Photoshop, thereby eliminating several steps.  However, 
if IRAF, IDL or another program are used to project the datasets into images, it is best to export each dataset into a 24-bit graphical file format that does not use ``lossy" file
compression schemes that sacrifice the quality of the image for a smaller
file size.  The JPEG file format uses a lossy compression scheme and therefore should be
avoided.  The TIFF and PPM file formats use either no compression or a lossless compression
scheme and are therefore ideal choices.  The CompuServe GIF file format does not use
compression; however it is an 8-bit file format and should also be avoided unless the image is stored as a
grayscale image.  The IRAF {\it export} command has a limited selection of file formats available,
none of which can be directly imported into Photoshop.  One option is to export a dataset as a SUN rasterfile, which is
a 24-bit format, that can be opened by The GIMP but not by  Photoshop.  Karma can export PPM files,
which can also be opened by The GIMP but not by Photoshop.
However, it is simple to use a separate utility, e.g., Graphic Converter\footnote{Graphic Converter is shareware from
Lemke Software.} for the Macintosh or the {\it convert} command and {\it pbm} packages for Unix, to convert SUN rasterfiles and PPM-format files into a TIFF-format file that can be opened by  Photoshop.  In addition to the SUN rasterfile format, Karma's {\it kvis} task can export directly to EPS files that can be opened by Photoshop, thereby saving a step.

It is important to note that the exported graphical file is no longer a dataset in the sense that it is no longer photometrically accurate.  The intensity scaling is chosen to show the detail within the object, and cannot be used to make scientific measurements.  In this paper the terminology of ``dataset" is used when referring to the data and ``image" is used when referring to a dataset that has been projected into an image via one of the scaling functions described above.

\subsection{Importing Data into Layers in an Image-Processing Program}

Once each dataset has been exported as a separate grayscale image file, the next step is to import each
image into an IP program that is capable of combining multiple ``layers" of images into a single image. 
Photoshop, The GIMP, and some other IP programs, use a layering metaphor, wherein multiple images can be stacked
onto each other and combined via different algorithms.  To create a color image, each grayscale image is imported
into the IP program as a layer.  Each layer is then assigned a color.  Each layer will probably also need to
be rescaled within the IP program to enhance the detail within the layer.  The layers then combine to form the resultant color image.

It is important to note that layers are different than ``channels," which are the orthogonal colors which
define the color space; e.g., in ``RGB color" the channels are the primary colors red, green and blue.  In RGB
color, all other colors are reproduced by combining these three colors in varying amounts.   Primary colors and the RGB colorspace are discussed in \S 3.1.

A traditional method
for creating a color image is to combine three datasets by assigning each dataset to a channel, such that
each dataset is, in effect, assigned to one of the primary colors.  The layering metaphor allows three major
improvements over this method.  First, an image can be created with any number of datasets; it is not limited
to the number of channels.  Second, any color can be assigned to a layer, not just the channel colors, e.g., red, green and blue in RGB color.   Third, it is possible to 
temporarily hide a layer such that it is not visible.  This is useful when modifying one layer at a time,
as in the following steps.  
The power and flexibility of layers are discussed in more detail in \S 2.3.

The first step is to create a master image file in the IP program that will contain all of the grayscale
images, each as a separate layer.  To do this, open each grayscale image and copy it into the master file as
a new layer.  In Photoshop and The GIMP, the default combination algorithm for the layers is ``normal," which is analogous 
to laying a piece of opaque paper with an image onto another piece of paper.  If the opacity is set to 100\% the 
image below it will be completely obscured.  
It is important that the combination algorithm for each layer is set to ``screen" and 100\% opacity.  

The screen algorithm acts to increase the pixel values of a layer, effectively brightening the image.  Assuming an 8-bit channel depth, it combines the selected, ``foreground" layer with the ``background" layer underneath it by the following equation \citep{bun00}:

$$r = 255 - \frac{1}{255}(255-r_f)(255-r_b)$$

Where $r_f$, is the foreground layer red-channel pixel value, $r_b$ is the background red-channel pixel value, and $r$ is the resultant red-channel pixel value.  The same equation is applied to the green and blue-channel pixel values as well.  From the equation it should be apparent that the resultant channel pixel value will be higher unless either the foreground or background pixel value is zero, corresponding to a pure black.  Thus, if an image layer contains nonzero values in only one channel, combining it with other layers will only affect the values in that channel.  Using the screen algorithm to combine three ``R, G and B" layers, each of which contains nonzero values in only the R, G or B channel respectively, is functionally identical to copying these layers into the channel values directly.  The advantage of using layers with the screen algorithm is that more than three layers can be combined.

Be certain that
all layers are set to the screen algorithm; otherwise some layers may not be visible.  In The GIMP, the combination algorithms can be found in the mode menu on the {\it Layers, Channels, Paths and Undo} window.  The other combination algorithms have been used with success, but the screen algorithm is recommended first.  The master file should be saved in the native Photoshop (.psd) or GIMP (.xcf) format to retain the layers.

In the layering metaphor, it is possible to modify each layer without affecting the other layers.  In
addition, the appearance of a layer can be changed without modifying the layer itself. 
In recent versions of Photoshop (v5.0 and higher), it is possible to use ``adjustment layers," which are
layers that do not contain image data, but rather modify the appearance of the layer(s) below with an algorithm.  The primary
benefit of an adjustment layer is that it does not actually change the image layers below it; thus the effects
of an adjustment layer can be easily undone by deleting the adjustment layer itself.  The effect an adjustment layer has on the overall image can be quickly seen by hiding the adjustment layer.   A good
way to visualize how an adjustment layer works is the example of a ``colorize" adjustment layer.  It is
analogous to laying a color gel over an image in that both the gel and the adjustment layer will cause the image below to take on the
monochromatic color of the filter.  The effect of the adjustment layer can be removed by deleting or hiding
the adjustment layer, which is analogous to removing the filter gel.  By using adjustment layers, it is simple
to modify or undo adjustments and try new ones, thereby making it much easier to try many different color schemes and
scalings.  The agility of adjustment layers allows this parameter space to be explored more efficiently.
Adjustment layers are therefore especially useful for the two important adjustments made to the layers in an image, colorizing and intensity scaling. 
Another advantage of adjustment layers is the ability to develop the ideal color and intensity scaling scheme with a scaled down ``thumbnail" file (\S 2.4) and then simply copy the adjustment layers from the thumbnail file and apply them to the full-resolution image.  

If adjustment layers are unavailable or not desired,
an alternative approach is to duplicate a layer before modifying it.  If unhappy with
the result, simply delete the modified layer and restart.  At each step duplicate the layer to be modified.  And then after the modification blink between the original and modified layers to see the effect.

This is a good point to confirm that the scalings used to export the data worked properly.  If there are unexpected
regions of saturation, return to the original data, rescale and project it again.  At first the master
file will contain only grayscale image layers, thus it needs to be converted into and saved as an
RGB color image before the layers can be colorized.  Note that the image should be assembled with RGB color.  The differences between the RGB and CMYK color spaces are discussed in \S 2.6.

\subsection{Alignment, Rescaling and Masking of Layers}

Once all of the images are copied as layers into the master file, the layers need to be prepared before colorization.  The preparation includes the alignment of the layers to a common astrometric frame, a fine-tune intensity rescaling of each layer to better show detail, and the masking of layers which combine more than one dataset, or more than one scaling of a single dataset.

\subsubsection{Alignment of Layers}

In general the datasets, and therefore the layers, will not be {\it a priori} astrometrically aligned.  If all of the data were
obtained from the same telescope and instrument during the same epoch, astrometric registration is usually as simple as a
linear shifting of the layers.  Rotation is also usually necessary for data from the \hst\ and for alt-azimuth telescopes.  Most polar-mounted telescopes maintain a fixed orientation for the instrument relative to the sky, however if the data are from the same telescope and instrument but from a different epoch it may also be necessary to slightly rotate the image if the instrument was removed and reinstalled between epochs.  Large-format optical instruments such as the NOAO Mosaic cameras have significant geometric distortion across the field of view.  The varying thickness of filters can also introduce differential geometric distortions.  If geometric distortions are not removed, a process commonly done during data reduction, it can result in problems with layer registration as well as with mosaicing of images with significant offsets.

The layers can be aligned by selecting each layer and shifting it to a common reference layer.  
Both  Photoshop and The GIMP have a move tool that can be used to linearly shift images.  
Layers can be more effectively aligned by temporarily inverting the grayscale of a layer and reducing its opacity, making its offset relative to other layers more apparent.  Once the layer is properly aligned, invert it again to return it to normal, and restore the layer's opacity to 100\%.
Note that neither package can shift images
by fractional pixel values.  Such shifts should probably be done in a data analysis package prior to exporting the image, although a simple workaround is to rescale the entire image to a higher resolution, shift the image, and then scale down the image back to its original resolution.

Manipulations such as rotation, pixel rescaling and higher-order geometric distortions should probably be done in a data analysis package so that the characteristics of the datasets match as closely as possible before registration is attempted.  However,  Photoshop and The GIMP have several tools that are useful, and usually more intuitive.  
Both packages are capable of rotation, pixel scaling and shear.   Photoshop and The GIMP use a graphical interface for most manipulations.   Photoshop and The GIMP can also rotate a layer via numeric entry; however, as of version 8.0, Photoshop can rotate no more precisely than a tenth of a degree, a limitation which is significant
in large format images.

It is possible to combine data from different telescopes and/or instruments such that one dataset partially replaces data from another, e.g., insetting a high-resolution \hst\ WFPC2 image into a wide-field ground-based optical image, by using
a mask in a manner similar to that described in \S 2.3.3.  After the layers are aligned, the region in the larger image that is overlapped by the
smaller image can simply be masked out by copying the smaller image into the larger image's pixel layer mask.  
Using a Gaussian blur and/or painting in the mask along the boundary of the datasets should be done so that the transition between the datasets is less conspicuous.  \hst\ images of M51 \citep{her01} and the Helix Nebula \citep{her03} are examples of this technique in practice.
Such combination techniques should be done only with data of a similar nature, e.g., optical data through filters with similar bandpass characteristics.  To do otherwise can be misleading as to the nature of
the source.

\subsubsection{Rescaling each Layer}

As discussed in \S 2.1, it is important to properly scale each dataset before exporting it as a grayscale
image.  However, additional fine tuning of the intensity scaling for each layer is usually desirable; and most IP programs have a
wide range of image scaling tools that usually exceed what is possible with most data reduction
packages.  These scaling tools map the 256 shades of gray in each layer into the 256 shades shown
on the computer.  Thus it is possible to remap a layer so that detail within the layer is more visible and so
that there is a better balance between the regions of shadow, midrange and highlight.  When rescaling a layer, all of the other
layers should be hidden so that the effect on the layer of interest can be seen more clearly.

In Photoshop, there are two tools which are particularly useful, the {\it Levels...} and {\it Curves...} tools.  These can be
applied directly to a layer via the {\it Adjust} submenu under the {\it Image} menu, or preferably with a levels or
curves adjustment layer.  When creating an adjustment layer, it is important to select the ``Group with
Previous Layer" option so that the adjustment layer only modifies the data layer below it.  Otherwise the
adjustment layer will modify all of the layers below it.  

The {\it Levels...} tool allows one to set the minimum and
maximum scale values in the same way as done before in \S 2.1.  If the scale values were chosen appropriately
there, little modification of these levels will be necessary here.  However, if they do need to be changed significantly it is a good idea to project the original datasets again but with better scale values.  This is especially necessary if the image is being assembled in 8-bit mode.  

The {\it Levels...} tool also allows
a ``gamma correction" to be applied to the layer, which is a simple, nonlinear way to remap the 
pixel values to the displayed intensity values on the monitor. It is a power-law function, i.e. $intensity = pixel^\gamma$. Most monitors have a gamma value between $1.7$ and $2.7$.  A gamma correction consists of applying the inverse of this relationship to the image before displaying, i.e. $newpixel  = oldpixel^{(1/\gamma)}$. 
A gamma correction greater than one brightens the fainter regions of the image but will tend to saturate the brighter
regions.  Conversely, a gamma correction of less than one will better show detail in the bright regions but will
darken the faint regions.  

The {\it Curves...} tool enables a more complex mapping of the data values to the image values.  By adjusting the slope of the line within the {\it Curves...} tool,
the contrast can be changed for different levels of intensity.  Note
that each layer can be rescaled as many times as necessary.  An advantage of using adjustment layers
is that multiple scalings of each layer can be tried before one is chosen.  After colorizing each
layer, as described in \S 2.4, adjust the intensity scaling of each layer again to achieve
a better color balance.  As mentioned before, achieving the best intensity scaling and colorization for each layer is an interactive, iterative, process.

\subsubsection{Using a Mask to Combine Multiple Scalings}

In some cases, a single scaling range and transformation function, e.g., linear, log, or square root, may not adequately render detail with optimal tonal range and contrast throughout an image that has a large dynamic range. One technique that can be useful to improve the contrast as well as show the structure in faint regions (shadows) and bright regions (highlights) is a combination of images generated with different intensity scalings. That is, more than one image may be projected from each original dataset using a different scaling, i.e., by using different minimum and maximum data values, and/or a different transform function. The image should be optimized to best show the contrast in each particular intensity region: shadow, midrange, and highlight. If multiple images are needed to achieve this goal, these images can then be combined into a single image in Photoshop or The GIMP with a mask that defines which spatial region of each layer will be visible. This technique can be used on individual grayscale filter exposures, later composited into a full-color image as described previously; or it can be applied to separate color images that have already been composited from sets of individual grayscale images.


In general, two separately scaled images will suffice to achieve good tonal range and contrast throughout the image. In principle, any number of intensity scalings and different transforms may be combined, but with correspondingly increased complication in reassembly. Each of these images then represents a narrower tonal range from the full range of intensities in the original image, and each step in gray value then translates into a smaller change in original data value, preserving more tonal detail. Using a transform function appropriate for each image section permits optimizing the contrast and tonal range with less chance of quantization or posterization, as described in \S 2.1.

For example, compress the large tonal range of a galaxy nucleus with a log scale that completely clips the faint outer arms and background. In addition, use a separate stretch with a linear transform to increase the contrast of spiral arms, dust lanes, etc. in the disk. This would not unduly enhance the noise near the sky background, but would saturate the bright nucleus. One image will be scaled to optimize the tonal range and contrast in the shadows at the expense of overly saturated highlights. The other image will be scaled to optimize the tonal range and contrast in the highlights at the expense of undersaturated shadows.  The example in Appendix~B illustrates this technique.


To recombine the images, place each in a separate pixel layer. On the visible, upper layer attach a pixel layer mask.  In Photoshop, this is done by selecting {\it Add Layer Mask} from the Layer menu.  In The GIMP, right-click on the upper layer's name in the  {\it Layers, Channels, Paths and Undo} window and select {\it Add Layer Mask} from the menu.
A layer mask is a special gray pixel channel, also known as an ``alpha channel," whose pixel values indicate the transparency/opacity of each matching pixel in the accompanying image layer. A black (zero value) mask pixel hides, or ``masks," the image pixel, allowing the underlying layer to show through. A white (255 value) mask pixel completely hides the underlying layer and allows the upper layer to be visible. Intermediate mask values indicate intermediate transparency. Note that the opacity of an entire layer may be specified independently of the mask as well.  Note that adjustment layers may also have masks, thereby limiting the effects of the adjustment layer to the masked regions.


The mask can be generated automatically, manually, or some combination of these.  One can manually edit the mask by painting the mask pixels.  To activate the mask, rather than the image, control-click on the mask icon in The GIMP and simply click in Photoshop.  Both packages offer a large range of options for editing the mask manually. Alternately, the magic wand tool or {\it Select / Color Range} can be used to create a selection region and fill the pixel layer mask with black to apply as the mask. Specify a wide value range, e.g., in Photoshop either with the tolerance option with the magic wand tool or fuzziness with {\it Color Range}, to produce a mask that is not hard-edged in the overlap region between the images. The intent is to blend the two layers seamlessly; a hard-edged mask will be readily apparent.

It is possible to produce a mask semi-automatically by using a gray pixel layer as the mask. Specifically, the brighter, saturated layer, i.e., the one optimized for the fainter features in the image, can be copied to the other layerÕs pixel layer mask. It helps to blur this mask to blend the two stretches. Applying a Gaussian Blur filter works well. A pixel radius of $\sim$1-2\% of the linear size of the image is a good starting point for trying the blur, but watch how the overlap area changes appearance as the radius is changed to find an optimal value.
In The GIMP, the blurred image can be produced as a separate file, copied and pasted into the layered file.  The blurred image can then be copied into the mask by clicking on the anchor icon.

The power of these IP packages is evident in being able to immediately see the results of hand-editing a layer mask, either by operating on a single layer independently or by watching an underlying layer appear and disappear. In this way, how the two layers merge can be fine tuned. It can also be instructive to view the mask as it is edited by using the ``Quick Mask Mode," particularly for making very small, precise edits.

This multi-mask layering technique presents ample opportunities and challenges for subjective choices, which can lead to criticism of undue manipulation. If applied sparingly, judiciously, and honestly, it is appropriate and acceptable because it can render a greater amount of detail in a single image. It is analogous to the techniques of dodging and burning long used in traditional darkroom photography to compensate for the limited tonal range available in a photographic print relative to the broader range inherent in the negative. It also can be likened to making two exposures of a given scene: one exposed for the shadows, and the other exposed for the highlight.  The two are then combined with a mask, extending the tonal scale.  Such techniques actually better match the performance of the human eye, which uses contrast as a means of discerning detail within regions that contain large variations in brightness.

\subsection{Color and the Image}

\subsubsection{The Parameters of Color}



Once each layer is appropriately aligned and rescaled, the image is ready to be colorized.  
Color is parameterized in most IP packages in an ``HSV" colorspace wherein a specific color is defined by its hue, saturation and value.  
``Hue" describes the actual color, e.g., yellow or yellow-green.  ``Saturation," also often referred to as intensity, is a measure of a color's purity.  A purely desaturated color is mixed with an equal amount of its ``complementary" color.  
Drawing a line across the color wheel from a color, through the center of the wheel, to the color on the direct opposite
side of the wheel defines a color's complement. For example,
purple is the complement of yellow.  A purely saturated color contains none of its complement.
``Value" is a measure of the lightness of a color, i.e., the amount of white or black added.  
Other terms used to modify color are ``tint" and ``tone."  A ``tint" is a color with added white; e.g., pink is a tint of red.  A ``tone" is a color with added gray and a ``shade" is a color with added black.  

\begin{figure}
\plottwo{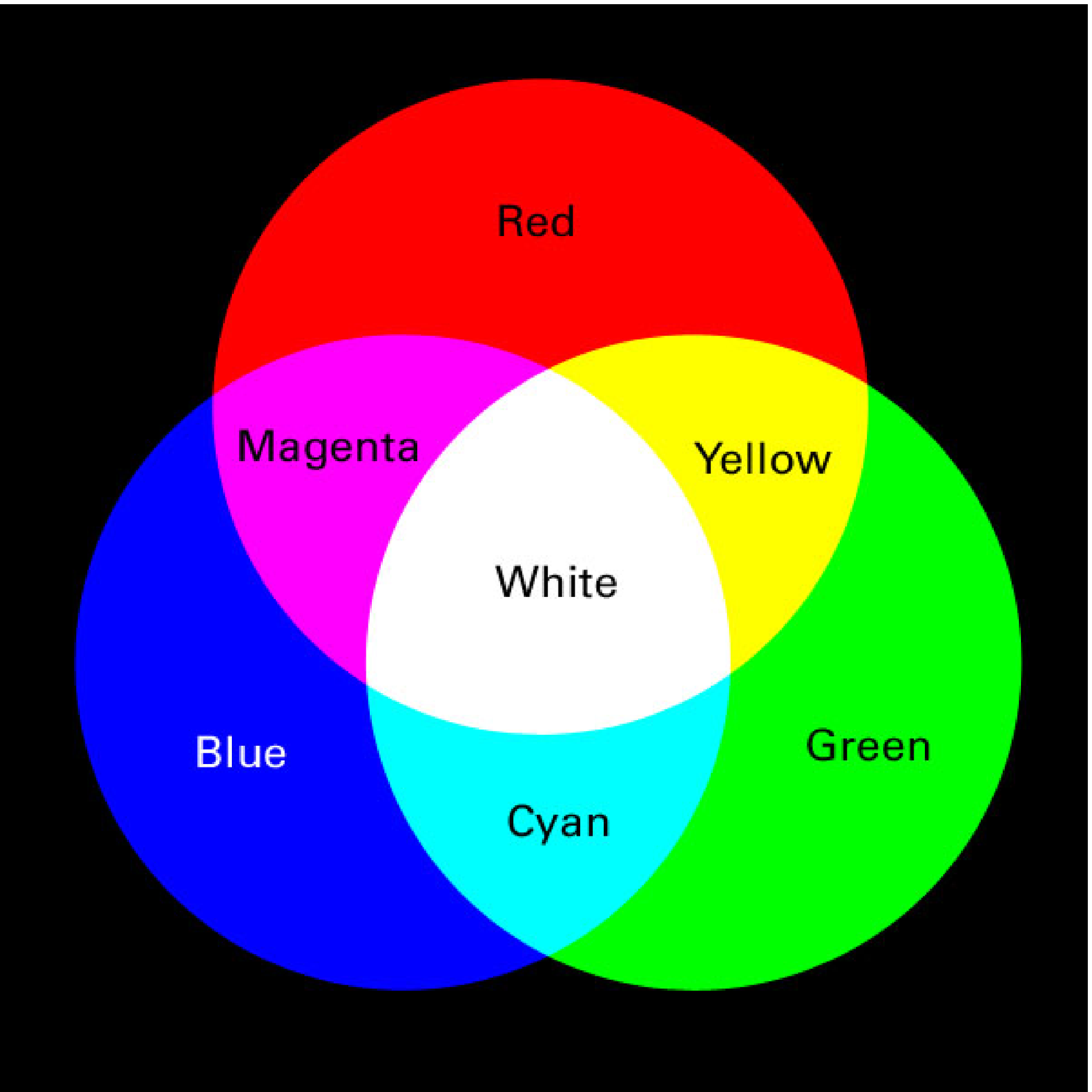}{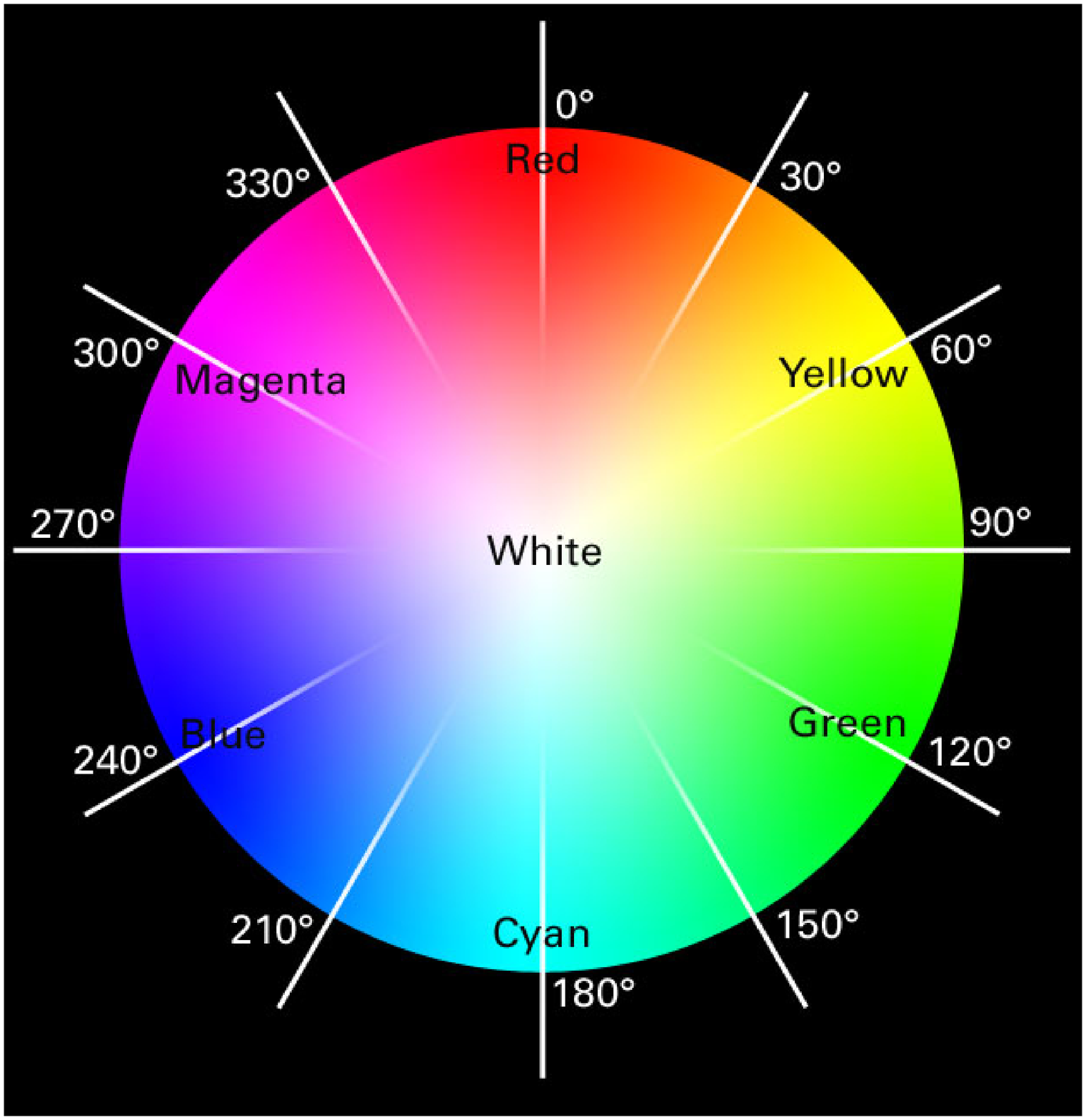}
\caption{The color wheel model based on Newton's studies.  
The figure on the left shows the primary colors in two systems of color mixing.  The additive primaries of
red, green, and blue combine to produce white at the center of the image.  The secondary colors in this system, cyan,
magenta, and yellow are primaries in the subtractive system that mix to produce dark
neutral gray. In the color  wheel on the right the angular positions of hue are given on the perimeter.  Colors directly opposite of each other on the color
wheel are complementary colors.  For example, yellow (60\arcdeg) is the complement of blue (240\arcdeg) because they are 180\arcdeg\ apart. 
\label{fig-5}}
\end{figure}

The HSV colorspace has its origins in Newton's color circle or wheel.  The simple
color wheel used in most IP packages
is shown in Figure~\ref{fig-5}.  It is reminiscent of many color systems including those of Johann
Wolfgang von Goethe and Albert H. Munsell.  It combines the color
systems of physicists studying light with the system required for print
reproduction.  For a history of color wheels see \citet{bir69}.  The subject is also elaborated upon below and in \S 3.1.

Hues are created by combining
primary colors that, by definition, are the fundamental colors that
cannot be created by any combination of other colors. This simple color wheel
model uses two different systems for creating color, each system
consisting of three primary colors. One system is the additive color system in which
the primaries, labeled Red, Green, and Blue (RGB), add to produce
white.  This system is used when combining light, and is the standard for computer displays and
projection devices.  The RGB primaries
are based on the spectral sensitivities of the three classes of
photoreceptors in the retina, which peak at wavelengths of about 5800\AA,
5400\AA, and 4400\AA\ respectively, e.g., \citet{mac99}.  One can
use samples of filter gels for stage lighting, for which
transmission curves are available, or a rendition of the CIE\footnote{CIE is the Commisision Internationale de I'Eclairage, the international clearing house for color research.}
chromaticity diagram, e.g., \citet{wys82}, to acquire a sense of the color associated with
each peak; R is reddish-orange, G is yellowish-green, and B is
purplish-blue.

A display device's RGB peaks fall into the additive system's
passbands, but the peak for R is 7000\AA\ \citep{wil83}. 
Since all monitors vary, and users can set various skews
to the color system, it can be difficult to get an accurate
sense of the additive primaries when using a computer monitor.  Also, the
range of colors that a monitor can produce are quite limited compared
to the range available to human vision \citep{mac99}.

The other color system widely used is the subtractive system, in which the primary colors are
Cyan, Magenta and Yellow (CMY).  They add to produce dark neutral
gray; and, when combined with a pure black channel (K), they serve as the standard system for printing devices. This latter system is used when combining pigments; and inspection
of color printer cartridges or of the calibration squares on commercial
packaging will give an accurate notion of the color of these
subtractive primaries.  Note that the addition of two of the
subtractive primaries in the CMY system produces a primary in the additive RGB system.
Adding magenta and yellow produces a reddish-orange
which is the primary red (R) in the additive system.  Similarly, cyan
and magenta add to produce the purplish-blue (B) while magenta and
yellow produce yellowish-green (G), Analogously, adding two primaries together in the additive
system produces a primary in the subtractive system.

In the HSV system, the three parameters of hue, saturation and value define a three-dimensional color cylinder wherein all colors are contained.  Hue is measured as an angular position on the wheel.  In the RGB color system, 0\arcdeg\ is red, 120\arcdeg\ is green and
240\arcdeg\ is blue.  The radial axis is saturation, wherein fully (100\%) saturated colors lie upon the outer edge of the color wheel and fully-desaturated (0\%) neutral gray sits at the center.   The simple color wheel bisects the color cylinder, and value is the axis orthogonal to the color wheel.

RGB primaries are used for color assignments in the IP software. However
as discussed in \S 2.2, any color can be selected  
for each layer; and \S 3.1 describes several approaches to consider.
Again, when using adjustment layers it is simple to modify the
coloring at any time, so be certain to explore many different
color schemes.

\subsubsection{Color Assignment to Each Layer}

The multi-layered image file is now ready for the creation of the color image.  This is the most difficult, time-consuming and enjoyable step in the
process.  As mentioned before, this is a highly iterative process.  If working with large images ``thumbnail sketches" are very useful.  That is, reduce the dimensions of the master file and make multiple copies to explore different color schemes; e.g., if the image is 1500x1500 pixels, then make copies that are 500x500 or
300x300 pixels, and then make multiple copies.  Try many color and scaling combinations and develop different strategies with these thumbnails.  Once settled on the best combination,
apply this combination to the full-resolution master file.   The primary advantage of this approach is that the manipulations are faster because the thumbnail file sizes are smaller.   Another advantage is that one is less likely to have a vested interest in a single color combination simply because it took a long time to generate.  

In Photoshop a colorization can be applied
directly to a layer via the {\it Hue/Saturation...} tool, on the {\it Adjust} submenu under the {\it Image} menu,
or by creating a hue/saturation adjustment layer that is grouped to the layer to be
colorized.  This tool will allow the adjustment of the hue, saturation and lightness of the layer.  To colorize
the image, select the colorize option.  Once the colorize option is selected,
the hue determines the angular position on the color wheel.  
As each layer is adding color and intensity to an image, it is easy to
saturate areas of the overall image (which will be shown as pure white) if each layer is too bright.  Thus it is
necessary that each layer be darkened when choosing a color assignment, particularly if more than three layers
are to be combined.  In Photoshop, a layer
can be darkened with the {\it Hue/Saturation...} tool, or in an adjustment layer.  Setting the lightness parameter to $-50$ will ensure that the brightest values within that layer do not saturate.
Details for colorization of a layer with The GIMP are given in Appendix~A.
To generate the color image, apply a different colorization to each grayscale image layer.  If using adjustment layers be certain
to group each adjustment to its associated image layer.  Strategies on choosing colors for each image layer is discussed in \S 3.1.

\subsection{Fine Tuning the Image}

A successful image keeps the viewer's mind focused on the content of the image, not on how it was created.  Thus, it is important to remove undesirable instrumental effects, e.g., excessive noise, cosmic rays, CCD charge bleeds from bright stars, and seams from mosaics, that will distract the viewer.   It is often the case that some datasets will suffer from poorer signal to noise or from more defects than other datasets.  If the effects are particular to an individual layer, e.g., noise and cosmic rays, it is usually better to fix each layer individually before flattening the image, a step described in \S 2.5.2.  Once the image is cleaned and flattened, it is ready for the final color and intensity scaling.  

\subsubsection{Cleaning the Image}

Table~\ref{tbl-3} summarizes several of the most common artifacts that are found in telescope CCD data; and \citet{von99} list several more. If these defects are not treated in earlier data-processing steps, e.g., dithering and flat-fielding, they can be removed with tools in the IP program.
 It is important to note that these tools should only be used to fix cosmetic defects in the image due to instrumental effects. 
Care must be used not to introduce artifacts or change intrinsic structure from the image.  Sharpening filters should be used sparingly, if at all, as they can introduce dimples and exaggerate the source structure.
It is recommended that cosmetic cleaning be done while working on a duplicate of the layer that is to be edited. The duplicated layer can then be blinked on and off to compare the cosmetically corrected image to the original.  Another good practice while cleaning an image is to apply a temporary ``Levels" adjustment layer that brightens the image. Periodic inspection of the image in a brighter mode may more easily show unwanted artifacts introduced by the cleaning process that should be removed.

\subsubsubsection{Reducing Noise and Fixing Bad Pixels}

In Photoshop, there are several filters available which are effective at reducing noise and removing cosmic rays.  The {\it Gaussian blur...} and {\it Median...} filters are useful for reducing noise, but with some reduction in the resolution of the image.  The {\it Dust and Scratches...} filter is effective at removing cosmic rays as well as reducing noise.
Images generated from radio interferometric data often suffer from dimples on the scale of the longest baseline, which can be ameliorated with a Gaussian blur.

Three tools in Photoshop, as of version 7.0, are particularly useful for fixing undesirable pixels in an image: the clone stamp tool, the healing brush tool and the patch tool.  The clone stamp tool copies pixels from a preselected sample region onto another region.  The healing brush tool is similar to the clone stamp tool in that it copies pixels from one region to another.  However, when doing so it also matches the texture, lighting, and shading of the sampled pixels to the pixels in the region to be repaired. As a result, the repaired pixels are often better blended into the image.  The patch tool blends pixels like the healing brush tool.  It is different in that it repairs an entire  selected area with pixels from another area.  In some cases it is better to set the opacity on these tools to less than 100\% as it will help to blend good regions with the cloned regions more seamlessly.  At the time of this writing, only the clone stamp tool is available in The GIMP.  And Photoshop Elements has only the clone stamp and healing brush tools.  Naturally, the user manuals for these programs should be consulted to better understand these tools.

\subsubsubsection{Fixing CCD Charge Bleeds}

A common problem for optical images is CCD charge bleeds, i.e., ``blooming," from bright stars in the field.  
The bleeds can often be fixed with the clone stamp and healing brush tools.  However, this can be difficult if the bleed overlaps diffraction spikes from the secondary-mirror support.  One can take advantage of the symmetry of the diffraction spikes to fix the CCD charge bleeds by copying the region of the image that contains the untainted diffraction spike, rotating the copy 90\arcdeg, and pasting it over the bleed.  The clone stamp and healing brush tools can then be used to blend the copy into the image, which can be
difficult in areas with complex structure.  An example is given in Appendix~C.  A logical question is whether or not the diffraction spikes themselves should be removed, as they too are an instrumental effect.  However, they should be retained because they serve as an important mental cue to the viewer that it is an astronomical image.
Again, the purpose of cleaning the image is to remove flaws which will distract the viewer.  Because of their familiarity, the diffraction spikes serve as an aid to the viewer's interpretation of the image.

\subsubsubsection{Fixing Mosaic Chip Seams}

Another common characteristic of optical and near-IR telescope data is the presence of mosaic chip seams and gaps. Many large format cameras are a composite of individual CCD chips that are mosaiced or stitched together to form a larger field of view. Some cameras  have a physical gap between the detectors, e.g., the \hst\ Advanced Camera for Surveys (ACS) and the NOAO Mosaic cameras. A common artifact is a seam or gap at the boundaries of the CCD chips when the data from the individual chips are projected into a single image. A chip seam will still have some signal within it; but the area is usually noisier, and often brighter or darker, than the image areas that have pixel positions well away from the seam. A chip gap will simply be a lack of data within a given area.  This area is usually primarily along a single dimension but may be more complex depending on the method of data projection.  

On some telescopes, it is customary to dither observations over a chip gap to improve the signal to noise ratio within the gap and to fill in missing pixels. If the signal-to-noise ratio is low, a dithered chip gap may still be obvious and may require further cleaning.  To repair chip seams and gaps, the clone stamp, healing brush and patch tools can be used in the same manner as for bad pixels. For a chip seam, where data exists on both sides of the seam, use of the stamp pad in Normal, Darken or Lighten mode with varying levels of opacity is effective. One may wish to experiment with blending a 50\% opacity clone from one side of a seam with a similar clone from the other side for a closer match.  

Data within a dithered chip gap may also appear to be systematically brighter or darker than data in surrounding regions due to the combination process.  This may be the result of inadequate flat fielding or an inaccurate estimate of the relative scaling levels.  Data within the gaps can be rescaled to match the surrounding data by first selecting the data within the gaps, e.g., with the marquee, lasso and color range selection tools in Photoshop.  The selected regions can then be feathered and rescaled with the same methods as described in \S 2.3.1.  If a discontinuity is present between the gap and surrounding data regions they can be repaired in the same manner as seams as discussed above.
 
\subsubsection{Overall Color and Intensity Balance}

After each layer has been rescaled, colorized and cleaned hopefully the image is close the desired result.  
The final, overall color and intensity scaling should be done after the image is cleaned because the cosmetic defects can be distracting.  If it has not yet been done, flatten the image into a single layer.  The flattened image should be saved as a different file so that the original, unflattened image can be returned to if necessary.  The flattened image will have a reduced file size and will speed up future editing of the image.  In Photoshop, select {\it Flatten Image} from the menu in the layers window.  At this point it is a good idea to apply overall color balance, brightness and contrast adjustments to the image, either directly or with adjustment layers.  In Photoshop, the {\it Color Balance...} tool or adjustment layer can be used to add or remove color from the image; e.g., if the image appears to be too red, add cyan, the complementary color to red, to balance the image.  Also fine tune the brightness and contrast of the image directly or with adjustment layers.  Keep in mind that the image will be viewed on many computer systems with monitors of varying brightness and quality.  Since astronomical images usually contain mostly black, they often appear to be too dark on many monitors.

As discussed in \S 3.1, the eye perceives quantities such as color and brightness relatively, and uses seven contrasts to 
perceive information from an image.  One of these is the ``light-dark" contrast, or simply a relative measure of intensity.  To
take advantage of the full range of the light-dark contrast, each image should contain regions of pure black and pure white.
Regions of pure white are usually always present in optical images, from the cores of bright stars.  However, some scaling may be necessary to ensure a pure black is present.  Once regions of pure black and white are identified, some color balance is
usually necessary to ensure that the black and white regions are a neutral color.  In Photoshop and The GIMP, the eyedropper tool and
the info panel can be used to measure the RGB value of any pixel.  To get better statistics, it is recommended to set the sample size for the
eyedropper to ``5 x 5 Average" to sample the pixels near the cursor.  Move the cursor around the chosen region
for pure black and make sure that the values of the RGB {\it channels}, not layers, are less than 10 each and
that the value of each channel is on average roughly equal to the others.  If one channel is systematically
higher or lower than the other two, either adjust the color balance directly, or use a color balance adjustment layer,
to add or remove intensity in the shadows from that particular channel.  For example, if the selected region of pure black
consistently shows the R channel to be $\sim$5 counts higher than the B and V channels, then balance the colors by
overlaying a color balance layer, selecting the shadows, and then subtract red (which is equivalent to adding cyan) until
the three channels are more in balance.  The same process should be repeated for the regions of pure white, 
adjusting the color balance in the highlights until the channels are well balanced, each with a value greater than 245 or so.

\subsubsection{Cropping and Orientation}

Traditionally astronomers have used an orientation in which north is
up and east is to the left, as one would conceivably see the object if
he or she were laying on the ground, looking upward, with the top of his or her head
pointed northward. However, for the sake of composition, other
croppings and orientations should be considered.  Clearly, cropping should
be used to remove instrumental defects often found on the edges of an image.
However, cropping and orientation should also be used to emphasize the regions
of interest and to keep the viewer engaged with the overall image.
The use of composition for emphasis and
to keep a viewer engaged with an image is discussed in \S 3.3.
Selecting the orientation and cropping is also an iterative
process. It
is best to produce many thumbnail sketches of crops and orientations
for comparison and to visually inspect them to see which are the most
successful.  

Photoshop and The GIMP offer cropping and rotation tools.  The rotation tools of both programs are discussed in \S 2.3.1.  Both programs allow a cropped region to be drawn graphically, or be entered numerically.  As of version 7.0, the Photoshop cropping tool contains a shield which can be used to block out the region of the image that will be cropped.  Once a region is selected with the cropping tool, select the shield, set the shield color to black and change the opacity to 100\%.  This will completely obscure the region to be cropped.  The effect of cropping can now clearly be seen by changing the selected region, looking away from the image for a moment, and then viewing it again.  Notice how your eyes engage the image.  
Avoid the temptation to include every portion of the image that is interesting.  The result may be a cluttered image wherein no particular element is sufficient to keep the viewer interested.  Images from \hst\ are often successful because of their tight croppings to the subject; although this is sometimes an unintended consequence of the relatively small field of view of WFPC2.  In particular, the \citet{hes95} \hst\ image of M16 benefits from a very tight cropping, giving the pillars the illusion of immense size.

A factor to consider when cropping the image is how the image will ultimately be used.  Most images will appear in
the press no larger than 10~cm on a side.  Thus, the detail in a large image may become too compressed to see.
Another factor to consider is the ratio of an image's linear field of view to the resolution.  Since images will be seen at roughly the same physical size, either in press or on a computer screen, point sources will appear to be more compact as this ratio increases; e.g., ground-based optical images can appear to be as sharp or sharper than \hst\ images if their field of view is sufficiently large.  For this reason, point sources in images generated with interferometry and adaptive optic systems tend to appear as ``blobs" in images despite their excellent intrinsic resolution because this ratio tends to be small for interferometers and non-multi-conjugate adaptive optic systems.  Thus, too tight of a cropping can cause the image to look poor.

\subsection{Color Display and Reproduction of the Image}

Image generation by the above steps will hopefully generate a color image that looks good on your computer's monitor.  But the ultimate goal is to deliver the image to others and for it to look as intended.  Unfortunately, a major difficulty with the electronic distribution and print production of digital images is that the appearance of an image can vary significantly on different output devices.  For example, an image which looks properly balanced on one computer monitor may be too dark on another; or the image may look too red when printed on a particular printer.  

It is anticipated that the image will be used in one or more of the following three ways:  The image will be distributed electronically to be viewed on another display device, e.g., via the web or an electronic journal.  Or the image will be printed directly to an in-house printer, e.g., on a personal or widebed printer.  Or the image will be sent as a file to be printed on an out-of-house printer, e.g., in a paper journal or on a professional press.  Each of these scenarios requires that the image be prepared so that it will look as expected.  Fortunately, the effective use of ``color management," before and after image generation, can give reliable image reproduction on other output devices.  Precise color management and calibration can be difficult; and it has led to the development of an entire sub-industry.  Appendix~E gives an example of image preparation for color process printing in Photoshop.  Unfortunately color management is not yet available in The GIMP.  For more information, \citet{fra03} offers a thorough review of the subject.

\subsubsection{Color Calibration of Your Monitor}

Prior to creating images on your computer, it is important to color calibrate the monitor so that it will display the fullest possible range of tones and the most accurate colors, including a balanced neutral gray.  An uncalibrated monitor may not properly display all of the colors that are possible in RGB color; e.g., a monitor that is too dark will not resolve dark grays from pure black.  

Relatively simple software tools are available, e.g., ColorSync and Adobe~Gamma, which can be used to calibrate your monitor systematically and accurately.  
It is important to note that color calibration will be affected by external light sources incident upon the monitor.  More robust monitor calibration is possible using additional software along with a hardware device that measures the brightness and color of the monitor directly, thereby removing the effect of external light sources as well as removing the uncertainty caused by subjective calibration by eye.   This is strongly recommended if you intend to generate images regularly.

\subsubsection{Establishing a Color Management Workflow}

To help ameliorate the problem of unreliable color reproduction, most operating systems and IP software, including Photoshop, perform color management with the ICC\footnote{Founded in 1993, the International Color Consortium (ICC) was established to create a color management solution that would deliver consistent color between applications, across computing platforms and across imaging and printing devices.} system, an established standard that uses ICC profiles.  An ICC profile is a file that characterizes how a device displays, inputs or prints color.  ICC profiles exist for display devices, e.g., computer monitors, for input devices, e.g.,  scanners and digital cameras, and for output devices, e.g., inkjet printers and printing presses.  
An ICC profile establishes the relationship between a device's color space and the profile connection space (PCS).  Device color space is defined as the coordinate system used to input, display or print color on a particular device; e.g., the device coordinates for a color monitor are the three 8-bit RGB values.  The PCS is the absolute color space defined by the range of colors that the average human eye can see.  In the ICC system, the PCS is parameterized with either the CIELAB or CIEXYZ color spaces \citep{for97}.  A profile contains a look-up table or function that is used by a color-matching method (CMM) to translate the device color space to and from the absolute PCS.  

In a color-managed workflow, ICC profiles are used to ensure that an image will appear as similar as possible on any output device, e.g. computer monitor, inkjet printer, high-end proofing device or printing press.  The workflow operates as follows.  When an image is opened, the color values in the file are converted into the working color space (see below) with the ICC profile for the working space.  The ICC profile for the working space and the ICC profile for the monitor are then used to convert the colors from the working color space first into the PCS and then into the device color space particular to your monitor.
When the image is prepared for another output device, the ICC profile specific to the output device is then used to convert pixel values from the working color space to the PCS and then to the device color space specific to that particular output device.  Thus, accurate ICC profiles are necessary for both the display device used to generate the image and for the intended output device.  
In addition to improving the performance of your monitor, most monitor calibration software can generate a custom ICC profile that characterizes the monitor's performance.  Most operating systems contain ICC profiles for different types of displays, however a custom profile, generated for your specific display and work environment, is preferred.


Prior to generating an image in Photoshop, a working color space should be selected so that your image will display as desired.  A working color space establishes the ``gamut,"  defined as the set of colors from the PCS that can be represented.  Colors excluded from the gamut cannot be reproduced in an image.  Establishing the gamut is necessary because only a finite number of data values are available to display different colors.  It is analogous to an artistÕs palette with a set number of colors that may be used to create a painting.  When choosing a gamut the following must be considered:  A wide gamut spreads the available values over a larger portion of the color volume\footnote{The three coordinates in each of the CIELAB and CIEXYZ systems define a color volume of all the colors visible to the average human eye.}, allowing a wider range of colors to be included in your image.  A narrow gamut compresses these values into a smaller color volume such that a smaller range of colors is available, but subtle variations between different colors are better resolved.  The problem of choosing a working color space is similar to that of choosing a scaling function when projecting a dataset into an image, as described in \S 2.1.  In practice the chosen color space does not severely restict image generation because in a 24-bit color image there are $2^{24}$, or over~16 million, color values available.  Choosing the working color space is more important for reliable reproduction of your image on other output devices.

Unfortunately, numerous working color spaces are defined and used.   Photoshop currently defaults to an RGB color profile called ``Adobe~RGB~(1998)" that provides a relatively large gamut of colors.  It is also well suited for documents that will be converted to CMYK for press production.  The ``Apple~RGB" color profile also has a large gamut and is well matched to the display characteristics of Macintosh monitors, which tend to be brighter than most.  
The ``sRGB" color profile is a popular standard among PC computers but it has a relatively small color gamut.  Because of its popularity the sRGB color profile is good for images to be distributed electronically, but is a poor choice for images intended for press production.  Because of its wider gamut, popularity for press work and its standard usage in Adobe products, the Adobe~RGB~(1998) is a good overall choice of color profile to use when generating an image.

\subsubsection{Preparing an Image for Other Display Devices}

Once your image is done the final ``master" image should be saved at full resolution and with no, or non-lossy, 
compression, e.g., in TIFF format.  From this image, smaller and/or compressed images can be generated as necessary, e.g., JPEG images for web distribution.  The working color space should be embedded in the saved file.  The embedded profile will indicate how the image was intended to be displayed.  When an image is opened, the color values in the file will be remapped into the gamut of the working color space, if the embedded color space does not match the working color space.  
The ICC profile for the computer monitor, if one is available, will then be used to properly display the image.  Color profiles can be embedded within many image formats, including TIFF, JPEG, EPS  and PDF.  

Naturally, color management works best if the destination computer is properly calibrated, which is almost always the case for a professional print lab but seldom the case for the typical computer user.  Unfortunately, many low-cost computers and their monitors display too dark.  To assist in color calibration, a standard 10 or 19-step grayscale ramp, e.g., on the Kodak Q-60 and IT8 targets, can be included with the image.  If the monitor is properly calibrated, as described in \S 2.6.1, each shade in the ramp should appear as distinct.  Placing the ramp in the image file can help but it may be distracting, especially if its use is unclear.  Alternatively, it can be included as a separate file, with calibration instructions. 

\subsubsection{Preparing of an Image for In-House Printing, Publication and Press Production}

The issue of color management and calibration is further complicated when preparing an image for print, whether it be for an in-house printer 
or for out-of-house printing such as for a research journal, magazine or professional press.  In most cases the image must be converted from the RGB to the four-color CMYK color space.  
Images are created in RGB color, which is an additive color scheme in the sense that, as color is added to the image, the
brightness of the image increases; i.e., it becomes whiter.  This is in contrast to the CMYK color space, which is a
subtractive color scheme in that, as color is added, the image becomes darker.  
The RGB system is used for electronic display devices, such as computer monitors, because increasing the pixel color values whitens the color.  The CMYK system is used for most print media because adding ink to an image darkens it.
Thus in most cases it is necessary to transform the image from the RGB color space to the CMYK color space, which is a nontrivial process.  

The problem of converting from RGB to CMYK is twofold.  First, most color print output devices use gamuts that contain fewer colors than 24-bit RGB images.  The more colors on the palette the more accurate the reproduction of the original RGB image will be. 
Currently most color print devices use four-color (CMYK) process printing, which can theoretically reproduce only about 3~million of the more than 16~million colors that a computer monitor can display.  Furthermore, four-color lithography can in practice usually only reproduce about 100,000 distinct colors.  Second, the RGB gamut is widely different than the CMYK gamut used by most printers.  Some areas of the color spectrum, including intense greens, purples and oranges, are virtually impossible to accurately recreate with the CMYK gamut; e.g., much of the pink in the Horsehead Nebula image of \citet{rec01a} and the greens and violets in the Trifid Nebula image of \citet{mic02} are outside the gamut of the CMYK color space.

For in-house printing, most printers do not require the image to be converted to CMYK prior to printing.  It is done
automatically by the printer's driver.  For most out-of-house printing it is requested that the image be converted.  In either case it is a good idea to manually convert the file to see the effect to the image.  Naturally, the journal should be consulted for specific instructions.

\subsubsubsection{Manually Converting from the RGB to CMYK Color Spaces} 

The GIMP, as of version 2.0, offers only preliminary support for converting an RGB image into CMYK space.  However,  Photoshop has several tools which assist in the preparation of an image for press.  The image can be converted into
the CMYK color space by changing it under the {\it Image~/~Mode} submenu.  Doing so will result in noticeable changes to
the image, as CMYK has an intrinsically smaller gamut than RGB.  The results of these changes can be previewed with
the {\it Proof Colors} and {\it Gamut Warning} options under the {\it View} menu.  First be certain that the proof setup is
set to ``working CMYK."  The {\it Proof Colors} option will then show how the image will look when converted to CMYK.  Toggle this option quickly to see the effects.  In addition, one can toggle the gamut  warning option to see which
colors, as shown in gray, will be unprintable in CMYK.  These colors will be assigned nearby color values in CMYK
space, but the results may be undesirable.  

The color balance for the image may have to be fine tuned before printing.  The image's color balance should be adjusted to bring all of the colors in the image, as much as possible, into the CMYK color gamut.  Another problem is that off-whites and subtle gradients are difficult to print.  Since pure white in the CMYK system is a complete absence of ink, there is a significant step between pure white and slightly off-white colors.  Such steps can appear as banding in the image.  Similarly, the CMYK system cannot produce subtle gradients as well as RGB.  Thus, such gradients can appear as posterization.  To mitigate these problems the entire image may need to be darkened so that pure whites do not appear in the press-ready version of the image.

\subsubsubsection{Preparing an Image for Publication}

Manual conversion into CMYK colorspace is a necessary step for all figures that are to be printed in color in scientific journals.  Most journals require, for the hardcopy version, that each color figure be provided as an encapsulated Postscript-format (EPS) file in CMYK format for the paper version of the article, and a JPEG or TIFF image in RGB format for the electronic version.  Most journals do not expect their authors to use color management.  Thus, an author-approved hardcopy of each figure is usually requested as a comparison to ensure proper color balance.  
Unfortunately many people do not have access to a printer that can print CMYK files, as most use their own driver to convert to print RGB files directly. 

The optimal resolution usually requested to be 300~dpi or better.  Scientific journals are typically printed on No.~5 groundwood paper stock, which is the lowest-quality press paper available.  As a result, the reproduction of color figures in most journals is not optimal.  Magazines are usually printed on No.~2 paper stock, which is of much higher quality.  Image reproduction is therefore usually much better.  Most magazines prefer to convert the image themselves.  Thus, an RGB image with an embedded color profile may be all that is necessary.
Magazines and journals will want the master file, i.e., the image at maximum resolution and with no compression; although most journals want the figure to be in Embedded PostScript (EPS) format instead of TIFF.  Thus, when generating the EPS file ascii or binary encoding should be used.  Note that some on-line journals, e.g., the  arXiv.org e-Print archive, have image size limits.  In these cases images should be saved in EPS format with JPEG encoding to reduce the file size.




\subsubsubsection{Preparing an Image for In-House Printing}

Most color inkjet and laser printers do not require that the image be converted into CMYK.  The conversion is done automatically by the printer's driver, which uses the ICC profiles for the working color space and for the printer itself.
To alleviate any potential problems, the latest version of the printer driver should be installed.  The printer's ICC profile must also be installed, if it was not already installed with the driver.  Most manufacturers provide profiles for their printers, however in some cases third-party ICC profiles are available and may work better, especially for older color inkjet and other non-Epson printers.

Prior to opening an image and preparing it for printing, it is important to establish what ICC profiles are to be used.  The RGB working space should be set to that used to generate the image and the CMYK working space should be set to the ICC profile for the printer.  In Photoshop, this is done in the {\it Color Settings...} window.  If it does not appear as an option in the CMYK working space menu, the printer's ICC profile can be loaded by clicking on the {\it Load...} button.  In this window, also select the advanced mode to reveal the conversion options.  Select ColorSync as the CMM engine.  Do not select black-point compensation when printing to an ink jet printer.  In the print options be sure that ColorSync is selected so that Photoshop can communicate with the printer driver.  This will allow the two profiles to be used to properly convert the image to the device color space for printing.

\subsubsubsection{Preparing an Image for Press Production}

In general, the preparation of an image for process color printing will depend on the capabilities of the printing facility to be used.  Before preparing for commercial printing, contact the printing facility or publisher to determine how the file should be prepared.  In most cases the printer or publication will provide an ICC profile specific to their press.  In some cases the printing facility will suggest using one of the pre-installed ICC profiles available in Photoshop.  If a publication or printing company provides the ICC profiles they will need to be installed in a specific directory for access in Photoshop. Printing companies that do not use ICC profiles should be avoided.  

If a custom ICC profile for the press is to be used, it needs to be installed into Photoshop, as described above, before
the image is opened.  It can then be converted to CMYK with the {\it Convert to Profile...} option under the $Image > Mode$ menu.  Select the vendor-provided target ICC profile in the {\it Destination Space} pull-down menu. Then, under {\it Conversion Options}, select the Apple ColorSync engine.  Selecting the perceptual intent will use the printerÕs specific algorithms as part of the installed ICC profile. Black-point compression should be selected if the image will be used for process color printing.  Note that this method of CMYK conversion will not affect the behavior of the {\it CMYK Color}, under the Image menu, nor affect the color settings.  

Check that the resolution of the image matches the specification provided by the publisher or printing company, which is usually $133$~lpi for magazines or journals and $150-175$~lpi for sheet fed printing on high quality glossy coated paper stock.  Finally, save the image file with a new name.  Be certain to not overwrite the original RGB image. For separations the file should be saved as a TIFF with the provided ICC profile embedded.

Prior to going to press it is important to get a commercial ``contract" proof from the company or publisher that will reproduce the image. Printing companies and publishers produce contact proofs from negatives. The negatives are then used to produce plates for the printing press.  Contract proofs are the most effective way to communicate color intentions for the print job, and it helps protect both parties from problems once the job is on the printing press.

The problem of proper color balance is also likely to arise once the image is at press.  Very minute changes in the inks and levels, as well as moisture content in the paper, can result in significant color shifts.  This problem is 
compounded for astronomical images in general because they
lack cues that press monitors use to determine the accuracy of the color balance in images of terrestrial scenes, such as skin tones.  Proper press production is an iterative, and very time consuming, process.  It also requires persistence and
patience, as the desires of the printer, i.e., quickly completing a press run, can conflict with the goal of producing the best print possible.  It is essential to insist on proofing the final product.

Like an artist with a limited palette, the four-color process can create breathtakingly beautiful reproductions.  Still, to have more colors on oneÕs palette will create an even more striking work. The terms ``expanded gamut" and ``Hi-Fidelity" (Hi-Fi) printing describe techniques that expand the number of colors printers can accurately reproduce on the final printed piece. 
Among them are ``bumpÓ plates, which augment the four-color process technique with additional Pantone colors. Other expanded gamut efforts involve various systems using five, six, or even seven colors to better recreate the full spectrum of visible colors.
In particular, ``Hexachrome" is currently perhaps the most reliable method of process printing available for accurately reproducing the full range of colors in the RGB color space. 
Hexachrome is a six-color process that dramatically expands the CMYK gamut. Hexachrome process printing can actually exceed the color spectrum displayed on computer monitors in all but a tiny part of the color range. 
Examples of Hexachrome printing suggest its greatest strength lies in reproducing demanding images, where it yields colors that were once missing but seldom noticed, such as florescent blue, turquoise, acid green, a wide range of soft pastels and other subtle tones. The result is sometimes subtle but always noticeable and memorable.  Because of the dramatic colors that Hi-Fidelity, and in particular Hexachrome, can produce, it is worthwhile to generate images that take advantage of their expanded gamut.  

\section{Philosophy of Image Creation}

Historically, a small percentage of telescope time has been used to create 
public outreach photographs to appear in the news, textbooks and
magazines.  However photography is now rarely used; and
currently two other sources for outreach images dominate.  One source is
images from research data that are directly provided by individual astronomers or by image-making teams.  These
images are created with the specific intention of being released to the public.  
In the other case, institutional news and public relations
teams appropriate images and figures that were created by scientists with the sole intention of communicating to their colleagues.  These images are often released to the public stripped of their captions and therefore lacking a guide to understand them.  It is
particularly this latter case that makes the issue of how
images are constructed, and subsequently interpreted by the non-expert,
of relevance to astronomers.

In general, the lay person has little understanding of how astronomical images are made.  
And inevitably astronomers and image processors are asked about the authenticity of the representation, with such questions as, ``is
that what it really looks like?" and, ``is this what I would see if I were standing right next to it?"  It
is beyond the scope of this paper to discuss the physiological and psychological response of
the human eye to the intensity and color of light, however it is worth noting that
the answer to the above questions is always ``no," regardless of how the image was generated.   Images can be generated that do render the ``intrinsic" colors of an object as closely as possible by using a set of filters that cover a wavelength range similar to the human color-vision range and by photometrically calibrating the filters relative to each other in the manner described in Appendix~D.  And yet these images also differ from human perception for several reasons, including the eye's poor sensitivity to red light and the inability to detect color from faint light.  
Additionally, contemporary astronomical images tend to display observations in selected, limited-wavelength ranges; and these ranges are distributed throughout the entire electromagnetic spectrum, not just the optical.  Thus, an image serves as an illustration of the physical properties of interest rather than as a direct portrayal of reality as defined by human vision.   After all, the reason for using a telescope is to show what the human eye cannot see.

This is not to argue that one is free to perform any
image manipulation, as the intention is to
retain scientific accuracy. To move, modify or delete
structure within the data for aesthetic
reasons is not allowable if the goal is to portray the true nature of the object.  
What manipulations are then allowable, and even desirable?
Color and intensity scaling schemes should be used that
maximize the richness and detail within the object to convey
the most information about the source while retaining the visual
qualities that make it naturally intriguing.
In particular, ``visual grammar," defined as the
collection of elements that affect the interpretation of an image, can be used as a guide for finding a composition and color scheme
 that highlights aspects of the
astronomical object  while retaining scientific accuracy.

The employment of visual grammar when choosing the intensity scalings, color scheme and composition is crucial for producing an image that is clear without
a legend, thereby making it legible to the public, and sufficiently engaging.  The process is similar to that used by 19th-century painters who portrayed the newly-explored American West for a public 
that was as unfamiliar with it then as they are now of the far reaches of space \citep{kes04}.  As was also the case
back then and now, the challenge is to create an image that accurately conveys the nature of an unknown world in a way that is also exotic and inviting.
While this may appear to be a conflict between art and science, it is 
the contention of the authors that it is possible, and indeed worthwhile, to address the aesthetics of the image while
simultaneously articulating the scientific content of the data. The use of color and composition to achieve these goals is discussed below.

As mentioned in \S 2.4.1, it is important to generate an image which focuses on the astronomical content and not
on the method of data acquisition or image construction.  In this regard, it is important to note that the public does have expectations based upon past
experience; e.g.,  spiral galaxies are expected to have bluish disks because most images of spiral galaxies are that
color.  From prior experience the public has developed a mental link between the color and morphology of spiral
galaxies.  Thus, an image of a green spiral galaxy can be distracting.  In contrast, the public does not necessarily
expect
nebulae to have a particular color because each nebula has a unique morphology.  And few nebulae are well known to the public, although notable exceptions do exist, e.g., the Horsehead Nebula.  Therefore, if the public is not likely to be familiar with previous images of a nebula, one is allowed greater freedom in
choosing the color scheme, scaling and framing.  Datasets outside of the visual regime, e.g., radio, IR and X-ray, are also cases where color schemes are less constrained.

To iterate, it is best for the image processor to create many versions of the image, with different color schemes and compositions, before deciding on the final version of the image.  An iterative process of trying different scalings and colorizations will yield the best results.



\subsection{Choosing a Color Scheme}


When working with
optical data, a traditional approach to color assignment, often referred to as natural color, is to assign to each dataset the ``visible"
color of its filter.  The visible color is essentially the
passband of the filter as perceived by the human eye; i.e., what one
would see if one looked through the filter against a bright, white
light.  For this approach to work well several filters
that cover most of the visible range of light must be used \citep{wai97}. However, datasets in enough filters may not be available.  And
the overall color of the image can become skewed if the colors of the
filters aren't balanced; e.g., an
image that is created from $V$, $R$, $I$ and \halpha\ datasets
would contain too much red because the three of the four datasets would be assigned a reddish color.  
Furthermore, modern images often include data from wavelengths outside the visible regime.  Thus, there is a need to define a more generalized methodology for color selection.


Choosing a color assignment will depend on the datasets to be
combined, the science to be illustrated, and personal
aesthetic.  When choosing colors it is also important to consider the
fact that our eyes, and indeed our senses in general, function by
detecting relative differences; e.g., a line appears to be long only
when a shorter line is present for comparison.  Color in an image is
similarly intensified or weakened by the contrast of a color's
properties with other colors in the image \citep{itt70}.  Thus,
contrasts between colors can be used to highlight or de-emphasize
elements of the image.  
 Black and white images have only one contrast scheme, that of light
to dark.  However, there are seven contrasts in color images
\citep{itt70,itt90}. These contrasts are presented in
Table~\ref{tbl-1}.  An example of an astronomical image
that strongly shows a particular contrast is also given for each contrast, although an image usually consists of more than one
contrast.  Other examples are also given in \S 3.2.  To produce a striking image, attention to these contrasts
is strongly encouraged.  These contrasts have greatly influenced the work of artists such as
Delacroix, Turner, the Impressionists, van Gogh, and C\'ezanne, who were
influenced by studies of color by the chemist M.E. \citet{che67}.  The color contrasts are described in much
more detail in \citet{itt70,itt90}.

In this section, the different color contrasts are noted as recommended color assignment schemes are described.  
When assembling an image, notice which contrasts can result after the initial color assignment.  And then adjust the color scheme, either by choosing different colors or by swapping the color assignments amongst datasets, to achieve these contrasts.  The next section describes examples of color schemes and their resultant contrasts in astronomy images.  


In general a good starting strategy when assigning color is to space the color assignments evenly around the color wheel; e.g., if creating an image with three datasets, 
assign undiluted colors that are 120\arcdeg\ apart on the color wheel.  This is known as a triadic color scheme and is the simplest example of hue contrast.  The 
traditional RGB color scheme falls into this category because the primary colors red, green and blue are separated by 120\arcdeg.  This selection represents the extreme instance of the hue contrast in the same way that black-white represents the extreme of the light-dark contrast.  
Similarly, the secondary colors cyan, yellow and magenta are also separated by 120\arcdeg\ and may be used.  However, one is free to use any colors that are separated by
120\arcdeg.  
Similarly, if there are four datasets, space them at 90\arcdeg\ intervals around the colorwheel, which is known as a tetradic color scheme. ÊNote that white and black can be included as colors
in the contrast of hue scheme, so any pure whites and blacks that appear in the final
image will strengthen this contrast. ÊThree or more adjacent colors, i.e., within 30\arcdeg\ of each other on the color wheel, also create a contrast of hue, although the effect is subtle.


By choosing colors that are evenly spaced on a color wheel, it improves the sampling of the color space; i.e., there will be a wider range of colors within the image.  Also, for this reason it is important to assign fully-saturated colors to each layer, because
only fully-saturated colors can combine to produce all of the colors available in the color wheel.  Unfortunately it is impossible to produce an image that fully samples the color space with only two datasets.  However,  this limitation can be overcome by creating a third dataset that is the intensity average of the existing two.  This technique has been successfully used in several images, including an HST/NOAO image of the Helix Nebula \citep{her03}.




Alternatively, if there are only two datasets one can assign
colors that are separated by 180\arcdeg, which is known as a
complementary color scheme. The complements are used in two contrasts:
complementary and simultaneous.  It is worth noting that
selecting exact complementary colors and mixing them together, creates
what are called compensating tones.  When combined, they will produce
neutral gray in the CMY subtractive system or pure white in the RGB additive
system.  Two or more colors are defined as mutually harmonious if their mixture
yields a neutral gray \citep{itt70}.  However if the colors are not exact complements then a brown is produced; this defines the case which is considered non-harmonious.  
A harmonious combination of three colors can be made by splitting the complement.  For example, if yellow is chosen then
its complement, blue, can be split by instead selecting two blues, one that is more purple and one that is more cyan; 
i.e., in Figure~\ref{fig-5} the colors at 270\arcdeg\ and 210\arcdeg\ are the split complement of 60\arcdeg.

The simultaneous contrast is particularly important since colors
are effected by their surrounding hues.  When a color is viewed,
our physiology produces its complement as an after image. One can
experience this phenomena by staring directly at a color for several
seconds.  After looking away one will see an after image of the
complementary color.  Hence if there are two non-complementary colors,
physiologically four colors are seen, two of which are after images. If there is a
sharp common border between the two actual colors, it becomes
ill-defined and appears to ``vibrate\footnote{Adjacent red and blue
colors can be disorienting because the eye cannot focus on both
colors simultaneously.  This is because of the relative locations of the red and blue
detectors in the eye.  The result is known as vibration.};"  The same
physiological color mixing also occurs if daubs of color are beside 
each other, rather than mixed.  Although the
resultant hue will be the same, it will appear more luminous.  This
method was used in art by the Pointilists in particular.

Therefore, based upon the number of available datasets, one should
attempt to distribute the hues assigned to the images in a manner that maximizes the usage of
the color space.  And, of course, the effect should be pleasing.  If, after the
initial assignments, the overall color of the image is unattractive, try swapping
the color assignments given to the layers.  Also try changing the initially assigned hues
overall; e.g., instead of starting with red, green and blue (RGB) assignments, try using cyan, magenta and yellow (CMY).
This is easily accomplished in Photoshop by using an overall {\it Hue/Saturation}
adjustment layer that is placed above all of the other layers.  In this case, do not
turn on the colorize option, and adjust the hue only.  This shifts the hue assignments, as
shown in the two color rainbows at the bottom of the dialog box.  It is analogous to
``spinning the color wheel" for the entire image; e.g., changing the hue 180\arcdeg\ will change an
RGB scheme into a CMY scheme.  Slowly adjust the hue to see the
effects on the overall image.  Once approaching a scheme which appears to be
close to attractive, adjust the hues for the individual layers separately.  Even changes
of only a few degrees in hue can have a significant impact on the appearance of the
image.  Choosing the color assignments for each layer is
perhaps the most important step of the process, so attempt multiple schemes and
weigh decisions carefully.





\begin{figure}[hbtp]
\begin{center}
\includegraphics{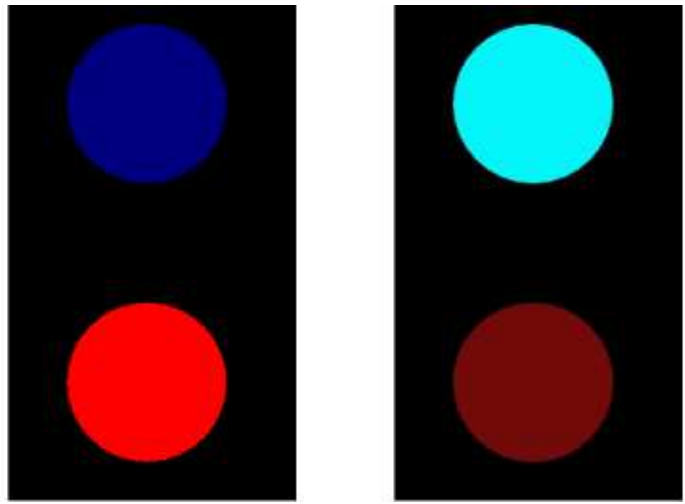}
\caption{Warm and Cool colors.  When viewing the figure on the left, cover the figure on the right with your hand.
Although the two circles in the figure on the left are of identical size, the red circle
appears to be closer than the blue one. This demonstrates the
differences between warm and cool colors.  To use blue to
represent motion towards the viewer, this can be achieved by
warming up the blue by adding yellow and using a contrast of
saturation, i.e. using a dull red.  An example is given in the figure on the right.}
\label{fig-6}            
\end{center}
\end{figure}

Aside from its aesthetic aspect, color can be used to imply physical 
characteristics, such as depth, temperature and motion.  For example, 
our minds perceive
hues that contain blue, collectively known as cool 
colors, to fall into the background while hues that contain red, 
known as warm colors, jump towards the viewer 
(Figure~\ref{fig-6}).    This is a result of our everyday 
experience, wherein distant objects, such as mountains, appear to be 
bluer because of scattered light from the foreground airmass;  
see \citet{lyn01} for examples.  Thus, the warm-cool contrast can be used
to create an image that has a three-dimensional feel by selecting colors for data layers which, when combined, produce warm colors for the objects that should come 
forward and appear closer, and cool colors for those that that should fall into the 
background and thus appear farther away.  In addition, our mind perceives cool colors to be 
literally colder than hues which contain warm colors.  This is a result of our experience with reddish flames and bluish ice.  Of 
course it is in contrast to the Planck spectrum, wherein redder objects are cooler
and bluer objects are hotter.

Note that a given contrast can be supplemented using other contrasts.
For example, if an image contains the colors of orange-red and cyan, it has the complementary contrast as well as a warm-cool contrast wherein the orange
will appear in the foreground.  Since dark colors in a light
background also jump forward, one can use the light-dark contrast
to further create depth by darkening the red-orange and surrounding it
by a light cyan. Both colors should be equidistant in value from neutral gray to maintain a harmonious relationship.
Figure~\ref{fig-6} also demonstrates how applying a second contrast can cause a reversal of how the warm-cool contrast is read.

Colors and color combinations can also evoke emotional responses \citep{whe94};
e.g., red is synonymous with the powerful emotions of love and hate.
Thus, color combinations with red are often powerful and forceful.
The color yellow, associated with the Sun, often implies life and
motion.  Blue is often associated with calmness; and violet is often
synonymous with magic, mysticism and the extraterrestrial world.  Trendy colors,
such as the pastels that were popular during the 1980's, imply
timeliness but may appear dated in the future.  Color schemes which
include conservative colors such as dark green and royal blue convey a
sense of strength and stability which may allow the
image to age better.  The contrasts of saturation and extension may used to balance or enhance emotional responses.  For
example, surrounding red by dull colors causes the red to saturate and
hence dominate.  If a large area contain greenish colors this area
may be balanced by a small area of red or pink using the contrast of
extension.

%
%

It is important to emphasize that this approach does not advocate hard and fast rules of color assignment.  Indeed, it is recommended to use a favorite color scheme as a starting point.  Then by using color contrasts and the other
factors discussed above, produce a harmonious and powerful composition.  The next section discusses some of
the different color schemes commonly used and indicates in more detail how colorizing a layer can produce a color contrast.


\subsection{Color Schemes and Examples}

\subsubsection{Definition of Color Schemes}

Unfortunately terminology is not used consistently when describing the different color schemes used in astronomical imaging.  
A traditional color scheme, often called ``natural color" or ``true color," is a specific color scheme intended to match the three color photoreceptors in the human eye.  In the natural color scheme, data are obtained in 
broadband optical red, green and blue filters (e.g., Johnson $R$, $V$ and $B$) and are assigned red, green and
blue colors respectively.  In addition, the datasets are scaled photometrically, i.e., with calibrated data, identical transform functions and maximum and minimum values that reflect the transmission and sensitivity functions of the filters and telescope, as described in Appendix~D.  Note that if the datasets in an image are subsequently rescaled as described
in \S 2.1 the resulting image is not technically a natural color image.  In the natural color scheme,
emission nebulae tend to appear as a deep red due to the strong \halpha\ emission line at $\lambda$6563\AA, a
wavelength of light which is perceived as a deep red by the human eye.  It is worth noting that the natural color scheme 
does {\it not} accurately match the eye's sensitivity to color; e.g., the Orion Nebula (M42) is a deep red in
most natural color images due to its strong \halpha\ emission.  However, M42 actually looks greenish 
when seen through a telescope due to the human eye's poor sensitivity to faint red light.

By deduction, the term ``false color" therefore applies to all images which are not made with the above color scheme.
In this paper the term ``representative color" is used when describing images assembled from multiple datasets of
different wavebands, but in a manner that does not meet the criteria of natural color.  This term is used because each dataset encodes properties of a physical phenomenon and colors are assigned to properly represent it.
Most astronomical images generated from professional
data fall into this category because the datasets are not necessarily obtained through broadband optical red, green and blue filters.  And these datasets are usually scaled and projected to maximize detail in the manner described in \S 2.1.
A representative-color scheme is said to be in ``chromatic order" if the filters are assigned color
based upon the wavelength of their passbands.  If the datasets are not assigned color in order of wavelength, it is known as ``composite order."  

Here the term pseudo-color is used to refer to a distinct technique wherein a monochromatic image is converted into a color image by mapping gray levels into colors according to a previously-defined color lookup table (LUT).
In a pseudo-color image, the color changes as a function of the value of a single physical property represented in the image, e.g., polarization, velocity or monochromatic flux density, thereby creating a multicolored image.  Note that pseudo-color image itself is not monochromatic unless all of the colors in the LUT are of the same hue; e.g., the popular ``heat" pseudocolor scheme is not monochromatic because the LUT contains  red, yellow and colors of intermediate hue.
Images generated from natural, chromatic and composite color schemes are fundamentally different than pseudo-color images. 

\subsubsection{Examples of Representative-Color Images}

An example of an image produced using the chromatic ordering scheme is the \citet{hes95} \hst\ image of M16.  The data \citep{hes95} were obtained with the
\hst\ WFPC2 camera with the F502N ([OIII] $\lambda$5012\AA), F656N (\halpha\ $\lambda$6564\AA) and F673N ([SII] $\lambda$6732\AA) narrowband filters.  To generate the image, the three datasets were colorized blue, green and red
respectively.  Filter F502N was assigned blue because it has the shortest
wavelength passband, filter F673N was assigned red because it has the longest wavelength passband, and filter F656N was
assigned green because its passband is of intermediate wavelength compared to the other two.  The chromatic
ordering for the \hst\ M16 image is not a natural color scheme because the filters were not assigned
their visible colors.  That is, when viewed against a bright, white light, the F502N filter appears to be green, not blue, to the human eye.  Similarly, the F656N filter 
looks a deep red, not green.  Only the F673N filter is assigned its perceived color, red.  
It is also not a natural color scheme because it uses narrow-band filters that pass only a very small fraction of the visible spectrum.  Furthermore, rather than photometrically calibrating the projected images they are balanced so that the \halpha\ data doesn't dominate
and turn the image completely green.  In comparison, a natural color image of M16 \citep{sho73} shows the nebula as deep
red, again because of the strong \halpha\ emission from the nebula.

The color assignment chosen is an extreme version of the hue contrast.  The image is also an example of a
split complementary color scheme. The blue (from the [OIII] filter) and the green  (from the \halpha\ filter)
combine to generate a background of greenish-cyan, whose complement
is a slightly-red orange.  Areas of the pillars are yellow-orange and
orange-red which are on either side of this orange on the color
wheel, hence the complement is split.  Also the RGB assignment to each
layer, along with the intensity stretches, ensured that the combined
datasets produced cyan, magenta and yellow, resulting in
a  strong contrast of
hue in the undiluted primaries of the subtractive system.
Where the intensity of [OIII] (blue) and \halpha\ (green) are balanced, cyan appears in the
background.  The \halpha\ (green) and [SII] (red) images combine to produce yellow along
the edge of the pillars.  And, while the center of the stars are white
due to the equal combination of RGB, their halos are magenta because the F673N filter broadens the point-spread function. \citet{itt90}, in his section on this contrast of hue states, ``the
undiluted primaries and secondaries always have a character of
aboriginal cosmic splendor as well as of concrete actuality,'' a statement that
applies to the \hst\ image of M16.  The two supporting
contrasts, (split) complementary and light-dark further boost this image's
appeal. 

Another factor to consider when choosing a color scheme is the structure within each dataset.  If each dataset
is scaled as described in \S 2.1, the intensity of the structure of interest in each layer should be about the
same.  However, sometimes this isn't possible, as some structures will appear in some datasets but not all.
For example, an image of the Flaming Star Nebula that was produced with the Mosaic~I camera on the NOAO 0.9m telescope on Kitt Peak was generated from data were obtained in three filters: [OIII],
\halpha\ and [SII].  However, no nebular signal was detected in the [OIII] filter, effectively reducing the image
to two datasets, as far as imaging the nebula is concerned.  Thus, a nearly-complementary color scheme (60\arcdeg\
yellow for
\halpha\ and 270\arcdeg\ violet for [SII]) was chosen to show the structure in the nebula.  The [OIII] dataset
was assigned a blue (240\arcdeg) color to enhance the bluish color of the hot  O--class star, AE Aurigae, at the center of
the image.  Note that this image contains two contrasts, the complementary color contrast of yellow and violet as well as the extension contrast, wherein violet occupies roughly three times the area of the image as yellow.  

\subsubsection{Examples of Images From More Than Three Datasets}

While not employing a natural color scheme, the \hst\ image of M16 is nonetheless generated with three datasets
that are assigned to the triadic RGB channel colors.  However, as discussed in \S 2.2, with
the layering metaphor it is possible to use any number of datasets, and to assign to each one any color
desired.  For example, an image of Barnard's Galaxy, NGC~6822, was generated using data from the Mosaic~II
camera on the National Optical Astronomy Observatory's\footnote{The National Optical Astronomy Observatory is operated by the Association of Universities for Research in Astronomy (AURA), Inc. under cooperative
agreement with the National Science Foundation.} {\it Blanco}
4-meter telescope at Cerro Tololo Interamerican Observatory.  The data were obtained in eight filters: 
$U$, $B$, $V$, $R$, $I$, [OIII], \halpha\ and [SII].  Since each filter provides additional information about
the object, the resulting image  \citep{mas01} contains more richness of detail; e.g., in this case a larger diversity of stellar colors.  Note that it can also be the
case that an image can be enhanced by the lack of complete wavelength coverage.  Indeed, the reason
for narrow-band filters is to increase the contrast between the emission-line and continuum regions in an object.  
Thus one needs to consider the value added by using more datasets, as it is often the ``space between the notes" that makes a composition work.

It is important to use a different color for each dataset.  Otherwise
distinct information from each of the datasets is lost.  Thus, the triadic RGB color
scheme cannot be used in an image created with more than three datasets.   In the above example, colors were
chosen for each dataset to approximately match the visible color of the filter; although this was not completely possible, as the $U$ and $I$ filters have passbands which are outside of the visible
range of the human eye.   Furthermore, each layer is scaled to best show the detail in the dataset, not to a
photometric scale; e.g., the \halpha\ layer is scaled to emphasize the HII regions within the galaxy.  Thus this image is not 
a natural color scheme.

It is important to note that there is no unique solution for the color of an individual pixel in an image generated from four or more datasets.   This is also true for an image generated with three datasets that are not assigned to a triadic color scheme.  
This ``color ambiguity" can be confusing for those who are predisposed to interpret color as the measure of a physical property, e.g., polarization or velocity, as it is used in pseudo-color images (see \S 3.1.2).  Thus it is important to clarify to the viewer the color assignments in an image to avoid confusion as to the meaning of the color.

\subsubsection{Examples of Non-Optical and Multiwavelength ``Hybrid" Color Images}

The examples above have focused on combining optical data to form a color image, however data from any
wavelength can be combined to form a color image.  An example of an image produced at non-optical wavelengths is an image of the supernova remnant Cassiopeia~A produced with radio data from the NRAO\footnote{The National Radio Astronomy Observatory is a facility of the National Science Foundation operated
by Associated Universities,Inc. under a cooperative agreement.} Very Large Array telescope \citep{rud02}.  Note that this is not a pseudo-color image because the color scales as a function of intensity at three frequency bands, 1.4~GHz, 5.0~GHz and 8.4~GHz.  

An important issue to consider when creating a non-optical image, particularly at radio and X-ray wavelengths,
is that these images often lack stars, which serve as strong mental cues that it
is indeed an astronomical image.  Multiwavelength images with optical data included can address
this issue.  Alternatively, other cues can be used; e.g., for the all-radio image of
Cassiopeia~A described above, violet was chosen to be a dominant color in the image because the color conveys a sense
of etherealness (see \S 3.1), implying to the viewer that it is not a terrestrial
object.

\begin{figure}[hbtp]
\begin{center}
\plotone{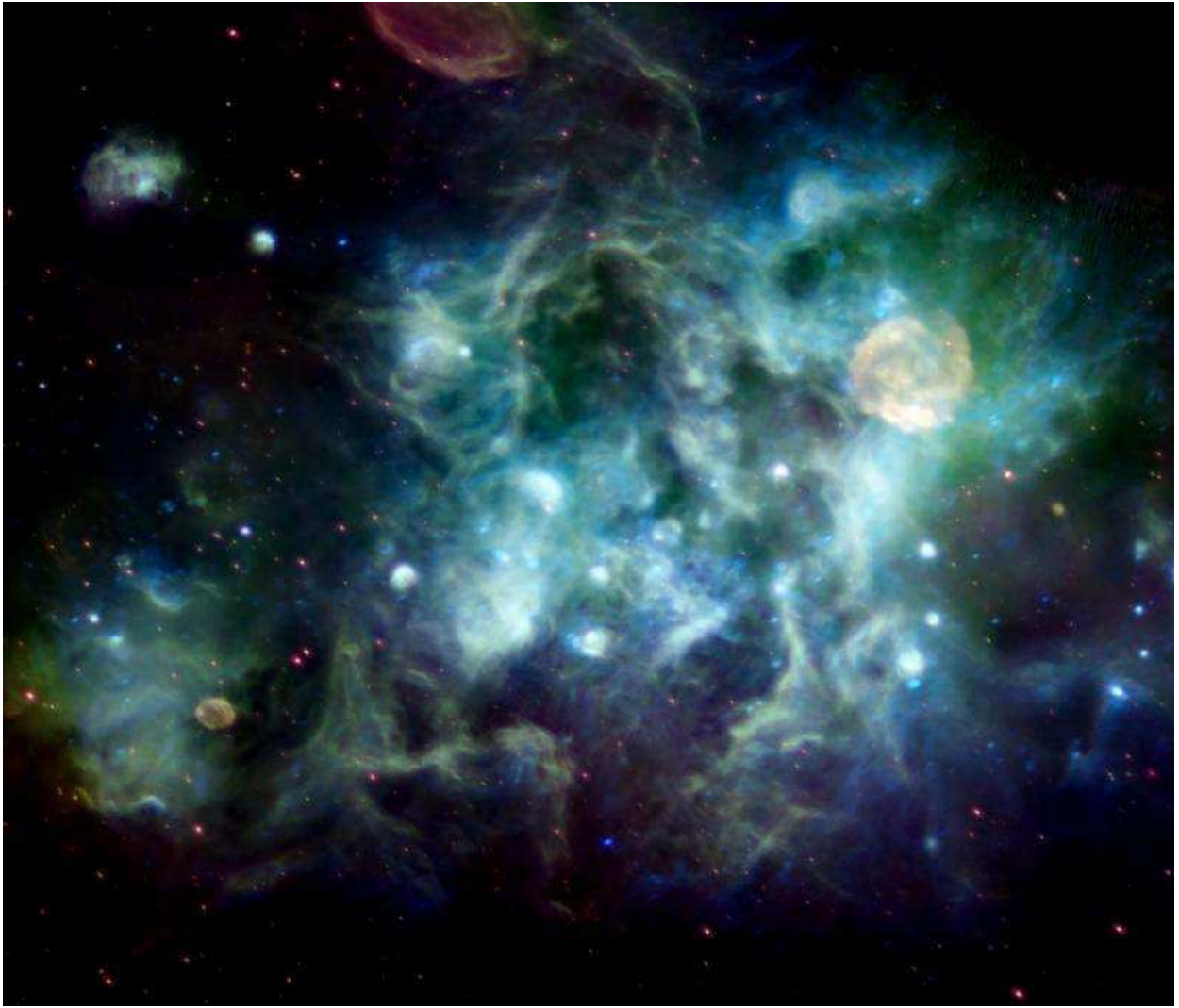}
\caption{This image of the Cygnus region by Jayanne English (U. of Manitoba) and Russ Taylor (U. of Calgary) was created by combining radio data from the Canadian Galactic Plane Survey (CGPS) with far-infrared data from the Infrared Astronomical Satellite (IRAS).  The use of color in this image reveals detail about the objects.
Bright compact, blue-white nebulae reveal the presence of a newly-formed, hot stars embedded in a dense cocoon of heated gas. The yellow and red bubbles, e.g.,  in the upper-right, top edge and lower-left of the image, are supernova remnants.  The reddish point-like sources scattered throughout the image are background radio galaxies and quasars.}
\label{fig-7}            
\end{center}
\end{figure}

A hybrid image combines data from widely separated wavebands, e.g., radio and x-ray, that have been obtained with different telescopes and instruments.  An example is that of a color image of the Cygnus region produced from radio and far-infrared data, shown in Figure~\ref{fig-7}.  In the image, 74cm and 21cm data from the International / Canadian Galactic Plane Survey\footnote{The CGPS is supported by the National Research Council of Canada and the Natural Sciences and Engineering Research Council of Canada.} were assigned rose and green colors respectively; they were then combined with 60$\mu$m and 25$\mu$m data from IRAS, which were assigned turquoise and blue respectively. 
As well as containing a light-dark color contrast,
this color scheme resulted in a warm-cool contrast causing the 
yellowish and reddish supernova remnants to feel suspended in the foreground.

Another example is a radio and optical hybrid image of the Triangulum Galaxy, M33, that was produced with five datasets.  Data through four optical filters ($B$, $V$, $I$ and \halpha), were obtained
with the NOAO 0.9-meter telescope and Mosaic~I camera on Kitt Peak.  These were combined with 21-centimeter HI spectral line radio
observations obtained with the NRAO Green Bank Telescope, NRAO Very Large Array and the ASTRON Westerbork Radio Synthesis Telescope.\footnote{The Westerbork Synthesis Radio Telescope is operated by the ASTRON (Netherlands Foundation for Research in Astronomy) with support from the Netherlands Foundation for Scientific Research (NWO).}  The optical datasets were assigned approximately visible colors, except for the I-band image.  It was assigned an orange color to differentiate it from the  \halpha\ image, which  was assigned a pure red.  The color to
assign to the 21cm dataset was not immediately clear because the data were obtained at a wavelength far longer than the
visible range of the human eye.  A chromatic color scheme would suggest assigning red to the 21cm data
because it has a longer wavelength than the optical data.  However, a violet was chosen for the
21cm dataset because violet is a mixture of red and blue.  The red in the violet implies a relationship to the HII regions that are prominent in the  \halpha\ image.  The blue in the violet implies that the gas in the HI regions is cooler than in the HII regions.  It also creates a cold-warm color contrast.  
The M33 example is therefore a composite color scheme because the 21cm dataset was assigned a bluish color
despite being the longest wavelength dataset.  A composite color scheme isn't constrained by wavelength
ordering of the datasets, nor the visible color of the dataset, if it is exclusively an optical dataset. 
Not only is the image \citep{rec02b} aesthetically pleasing, the image also provides an instructive visualization for the
public, illustrating how astronomers use multiwavelength observations for scientific purposes.

\subsection{The Composition of an Image}

One of the goals of a composition is to keep the viewer engaged
with an image.  That is, the goal is to keep the eye trained within the borders of
the image.  
Human perception of images is complex but
includes seeing ``bilaterally," that is, dividing the page into left and right
halves, and perceiving the bottom part of an image as closer than the
top part, e.g. \citet{blo90}. Simplyfing this, for 95\% of
those in western cultures, 
the eye will enter from the
left edge of the image, roughly one-quarter of the way up from the bottom.  
Travelling
horizontally for a short distance, the eye then moves along a diagonal up to
the right and exits close to the upper right corner.  If a picture
doesn't redirect this travel onto a different path, that is, onto a
trail that winds within the frame, then the viewer spends little time
apprehending the content and the picture is neither memorable nor
engaging.

A visit to an art gallery will illuminate the tactics used by
artists to redirect the eye towards the center of the image
plane. For example, they use vertical structures and
upper-left-to-lower-right diagonals to block the eye's default path.  Additionally
they will redirect the eye back towards the point of entry, close to
the left edge, by using a spot of high contrast either in terms of
brightness or of color.  Color contrasts are described
in \S~3.1.

\begin{figure}[hbtp]
\begin{center}
\plottwo{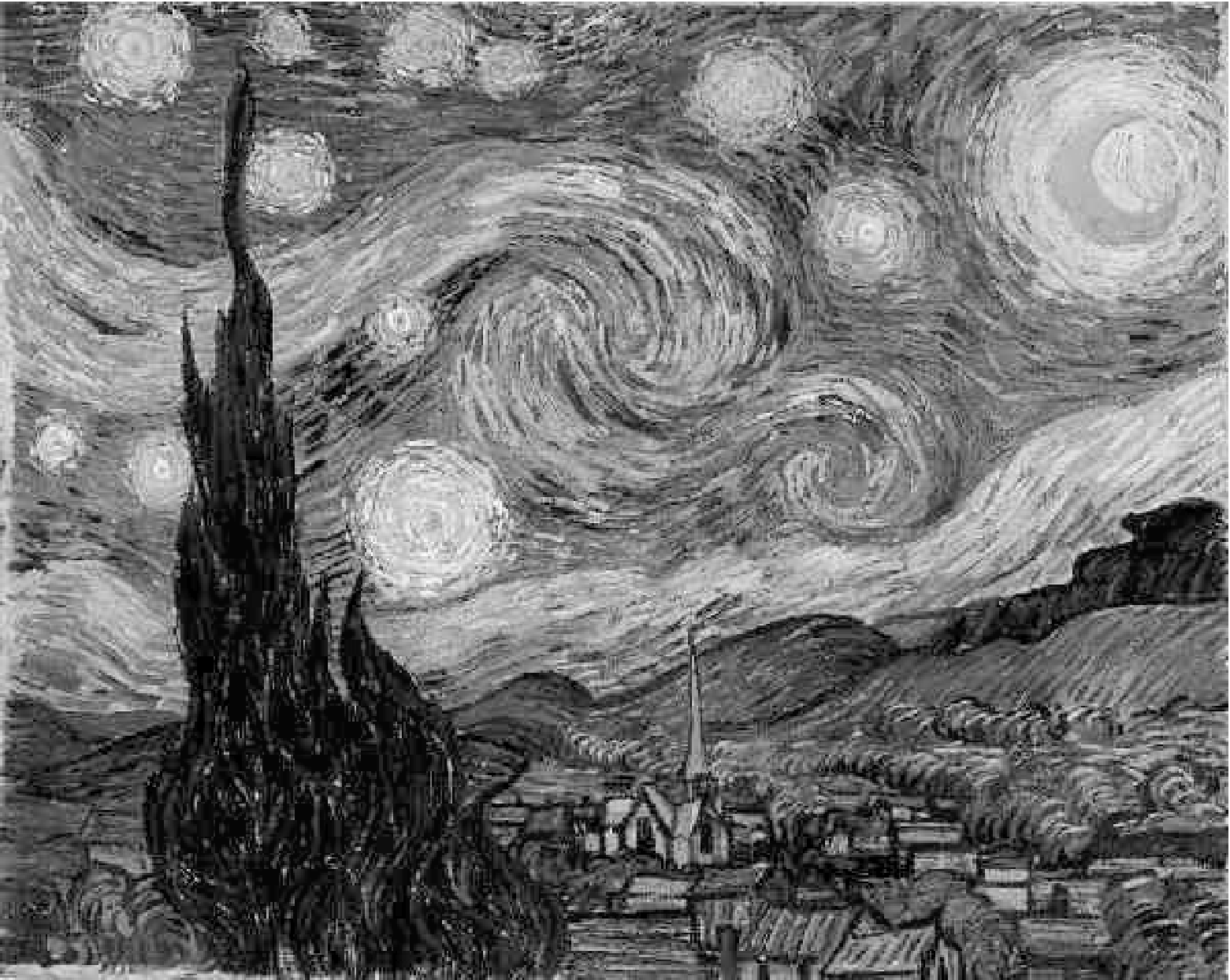}{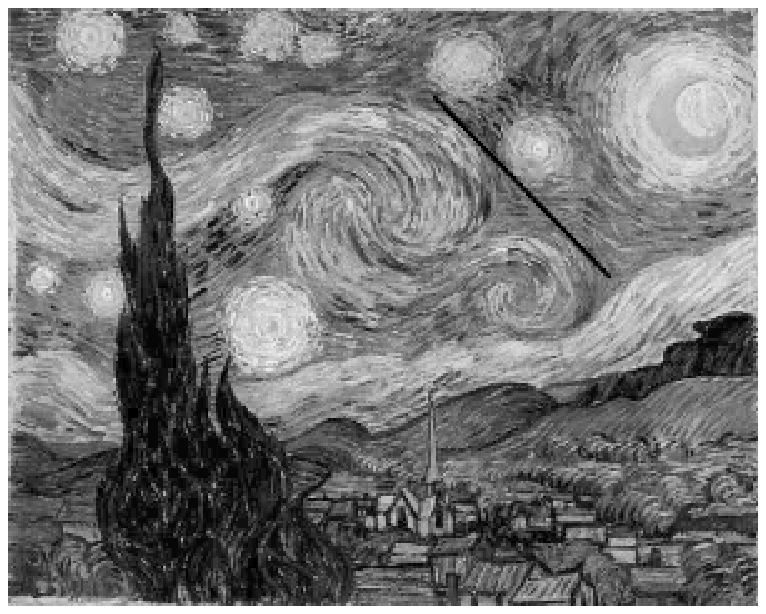}
\caption{A  black \& white rendition of V. van Gogh's ``Starry Night'' \citep{van89}.  It is shown unmodified on the left and with a diagonal black line added on the right.  The default path of the viewer's eye begins at the
lower left edge and travels on a diagonal towards the upper right.  In
order to engage the viewer's eye with the picture, this path is
blocked by structures which include the black vertical bar formed by
the trees on the left and the left-to-right diagonal formed by the
edge of the swirls in the sky.  The  diagonal line in the image on the right provides a
guide to the eye of this latter diagonal.}
\label{fig-8}            
\end{center}
\end{figure}

For example, Figure~\ref{fig-8} is a black and white rendition of
Vincent van Gogh's ``Starry Night.'' 
Note that the
cluster of trees forms a strong black vertical structure
paralleling the left edge. The word tension is often used in art to describe
a relationship between objects and the space between them, which results
in competition for the viewer's attention.  In this case, a tension is set up between the vertical structure and the
bright star to the upper right; however the dark vertical structure dominates.
The eye slides down the dark structure and starts to travel along the
rightward diagonal again, this time encountering the left-to-right
diagonal formed by the edge of the cloud swirls.  
Virtual lines like this diagonal are called ``conceptual elements" \citep{won72}.
The eye travels up the
swirls towards the upper left and again encounters the dark vertical
structure once more, slides down it and, during the next excursion,
makes an even tighter circle within the picture plane, as defined by the boundary of the picture. This is  just
one of the ways in which van Gogh creates interest in this particular
painting; he also guides the eye with color.  It provides an example of how an artist
keeps the viewer engaged with the picture.

\begin{figure}[hbtp]
\begin{center}
\includegraphics{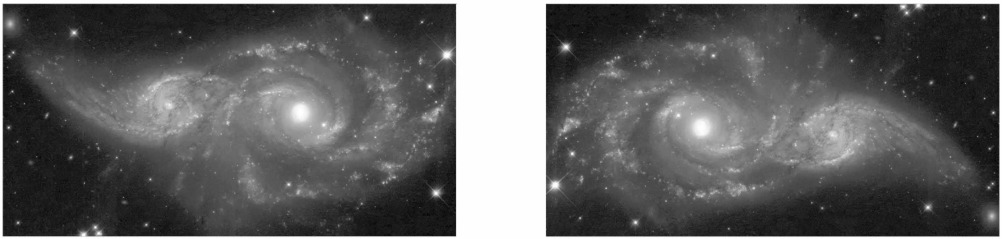}
\caption{NGC~2207 \& IC~2163: The figure on the left uses the
conventional ``North is up, East is on the left'' orientation used in
publication \citep{elm00}. The orientation on
the right produces a more 3-dimensional experience, due to the factors
described in the text, and was used for the official public release
version \citep{her99}.}
\label{fig-9}            
\end{center}
\end{figure}

Figure~\ref{fig-9} shows two orientations of the overlapping galaxies NGC~2207 and  IC~2163.  In
the conventional north up and east left display the galaxies seem to
sit flat on the picture plane.  In the ``upside down'' version the eye
first encounters the more resolved arm of NGC~2207. Additionally, the edges of the two
galaxies in the upper right corner act as an upper-left-to-lower-right diagonal, similar to the swirls in the
van Gogh image.  And the bright star at the lower left provides a point
of contrast that drags the eye back to the beginning of its path.
Not only does the eye stay longer within the image's borders than
in the conventional orientation, but the resolved arm of NGC~2207  pops forward
causing the image to feel more three-dimensional.  Hence the ``upside-down'' orientation was selected for the version released to the public \citep{her99}.

Cropping can be used to remove high contrast spots, like
stars, that would drag the viewer's eye quickly out of the picture
plane. It can also be used to place naturally occurring diagonals and
verticals in the correct position in the picture plane for slowing
down the eye's motion and to change its trajectory.  Care should also be taken in selecting the
picture's center.  If the midpoint of the target of interest is placed dead
center, then the target will appear to be sliding down when the image is viewed vertically, e.g. on a monitor or on a wall.  
Instead, the target midpoint should be placed
above the horizontal centerline of the image to keep it from appearing that it is about to fall out the
bottom of the image plane.
The target's midpoint should also be off of
the vertical centerline in order to ensure that the image appears
dynamic rather than static.  A static image risks allowing the eye to
follow the default path from lower left, quickly out the top right
corner.

\section{Summary}

The quality of modern astronomical data, and technologies now available to manipulate them, allows for the creation of high-quality images in a purely digital form.  Many groups are now exploiting this fact to create attractive images intended for scientific illustration and outreach.  The success of these images in public outreach is  responsible for much of the popularity that astronomy currently enjoys.

A practical guide has been presented to generate astronomical images from research data by using
powerful image-processing programs such as  Photoshop and The GIMP.  These programs use a layering metaphor that allows an unlimited number of astronomical datasets to be combined; and each dataset can be assigned any color.  This is an improvement over traditional methods, wherein only three datasets can be combined, and only with the primary colors of the color space, usually red, green and blue.  In the layering metaphor, each dataset can be individually scaled and colorized, creating an immense parameter space to be explored.  These IP programs offer excellent flexibility and agility in the creation of images, allowing easy exploration of this parameter space.  

A philosophy has also been presented on how to use color and composition to create images that simultaneously highlight the scientific detail within an image and that are aesthetically appealing.  Because of the limitations of the human eye, it is fundamentally impossible to create an image of an astronomical object which shows ``how an object truly appears."  This is particularly true for datasets outside of the optical window and for datasets with limited-wavelength coverage, e.g., emission-line optical filters.  Indeed, the goal of many of these images is to show what the human eye cannot see.

Thus, by properly using visual grammar, i.e., the elements which affect the interpretation of an image, it is possible to maximize the richness and detail in an image while maintaining scientific accuracy.  With visual grammar one can imply qualities that a two-dimensional image intrinsically cannot show, such as depth, motion and energy.  A basic primer has been included on some elements of visual grammar, such as the seven color contrasts, including examples of these contrasts in astronomical images.  In addition, composition can be used to engage and keep the viewer interested for a longer period of time.  The effective use of these techniques can result in an appealing image that will effectively convey the science within the image, to scientists and the public alike.






\acknowledgments



\appendix

\section{M33:  An Example with IRAF/Karma and The GIMP}

The first tutorial consists of the assembly of an image of M33, the Triangulum Galaxy, 
with data obtained with the WIYN 0.9-meter telescope and S2kb camera.  The data were 
obtained on 2002 November 12 in six filters: $UVBRI$ and \halpha.  Data reduction was
completed with IRAF in the standard manner, including overscan, bias and flatfield 
calibration.  Three 600~second exposures were obtained in $U$, $B$ and \halpha\ each.  They
were average combined by filter to reduce noise and to remove cosmic rays.  A single 600~second exposure was obtained for each of the $V$, $R$ and $I$ filters.  
In this example
an image will be assembled using either IRAF or Karma to scale the data and project it into multiple images, and The GIMP to
assemble the color image.  
This example was produced with IRAF version 2.12, Karma v1.7 and The GIMP v2.0 
in the Macintosh OS~X (v10.3.5) operating system and X11 v1.0, but the
functionality is identical with other versions and platforms.  Note that the detailed 
appearance of the figures below depends on the OS and versions of software.
It is assumed that the user has properly installed, 
and has basic familiarity with, these software packages.  The original FITS files, as well as SUN rasterfiles and The GIMP files, used in this tutorial are available for download at: {\tt http://www.stsci.edu/stsci/resources/imaging/papers/rector2005/}.

\subsection{Intensity Scaling and Image Projection}

After the data reduction, the first step is to determine proper scalings for each dataset.  In this example, this is done either
with IRAF or with Karma.  

\subsubsection{Intensity Scaling and Projection with IRAF}

\begin{figure}[hbtp]
\begin{center}
\includegraphics{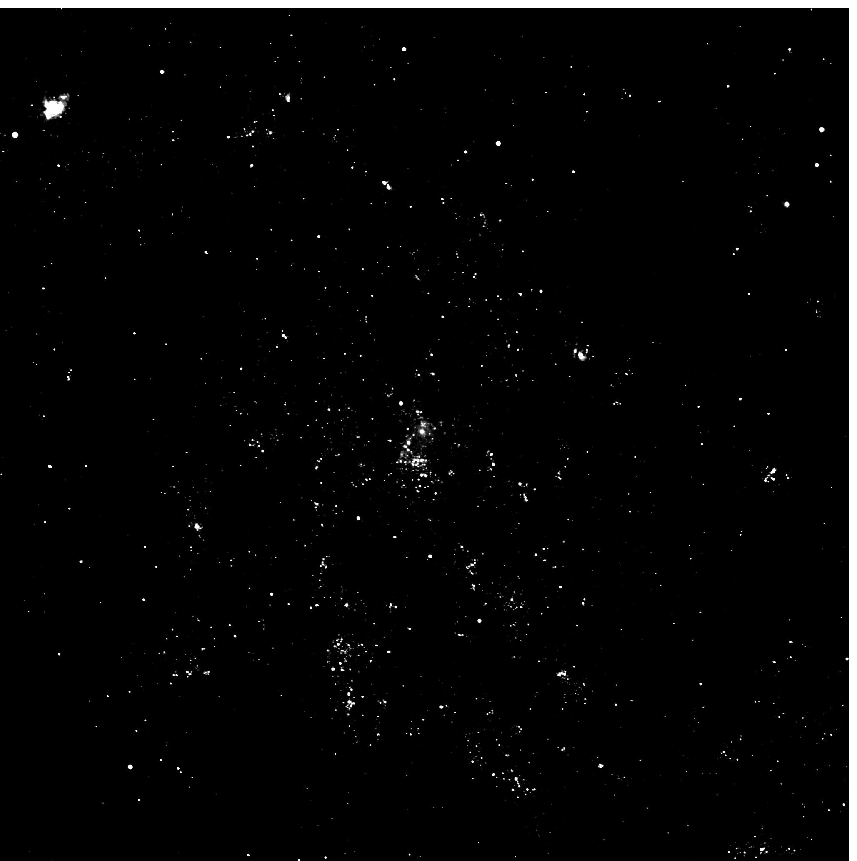}
\includegraphics{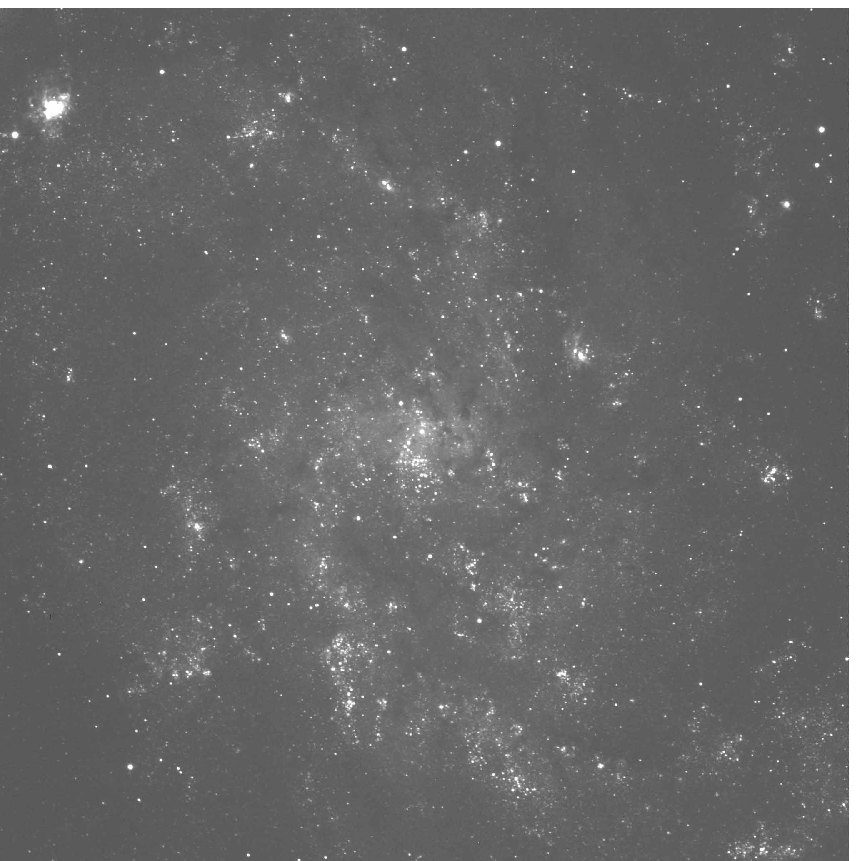}
\includegraphics{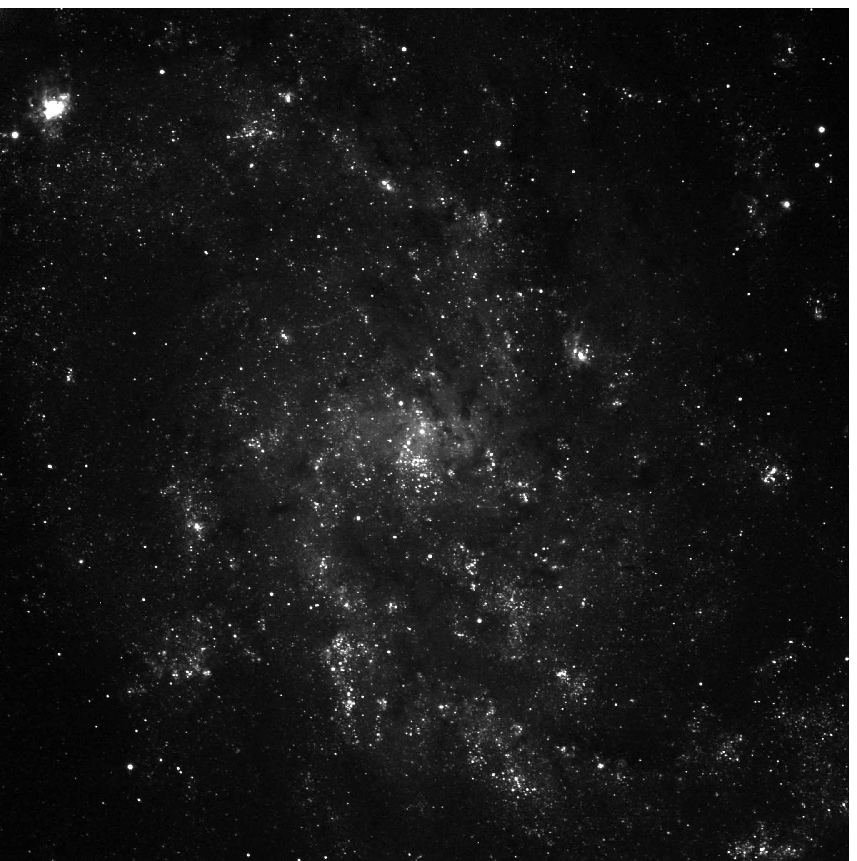}
\caption{Three examples of the minimum data value for a portion of the M33 U-band dataset.  The top figure shows large areas of  black within
the image because the minimum scale value is too high.  Note that this may not be apparent unless the contrast of the image is increased.
The middle figure shows a minimum scaling that is too low, resulting in unused shades of gray.
The bottom figure shows an ideal minimum value because the lowest data values are shown as a black with little to no posterization.}
\label{fig-10}            
\end{center}
\end{figure}

\begin{figure}[hbtp]
\begin{center}
\includegraphics{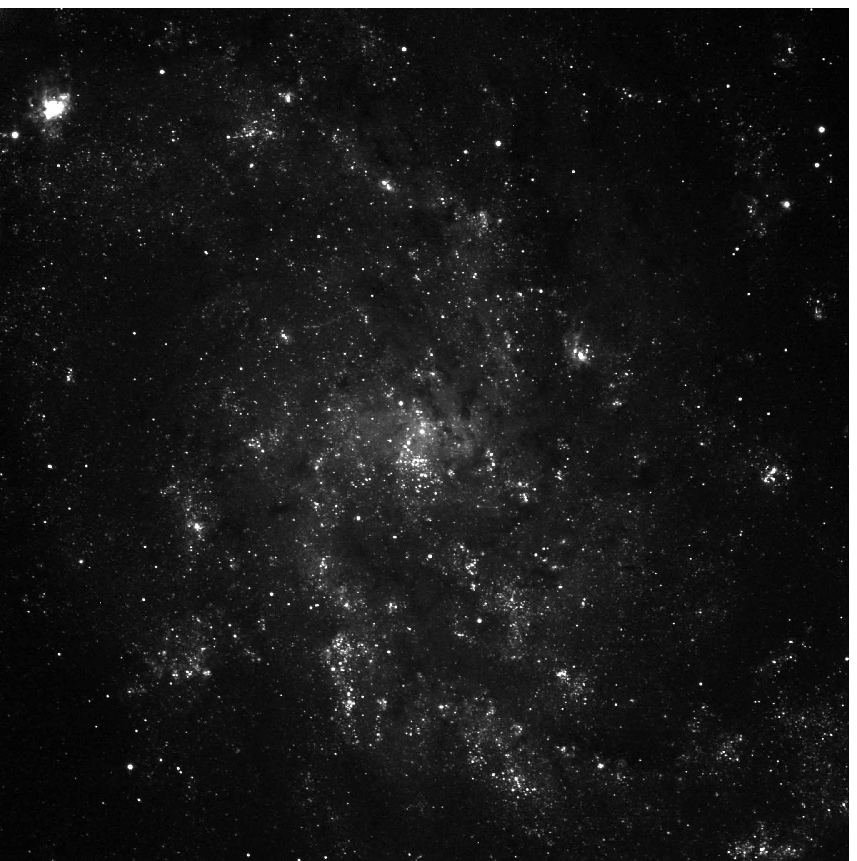}
\includegraphics{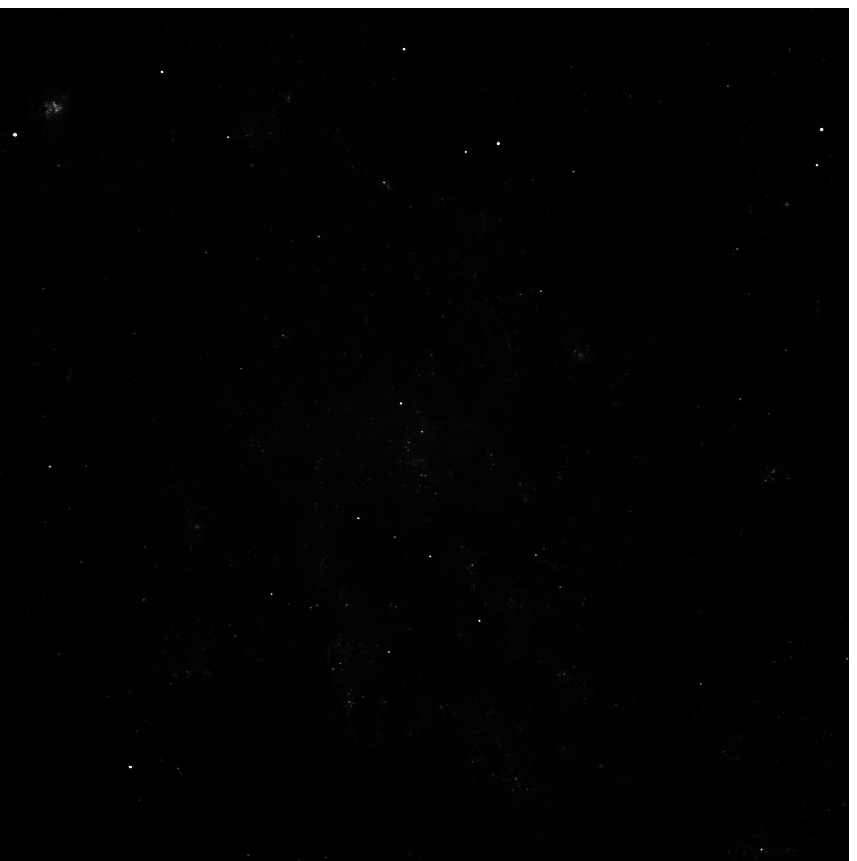}
\includegraphics{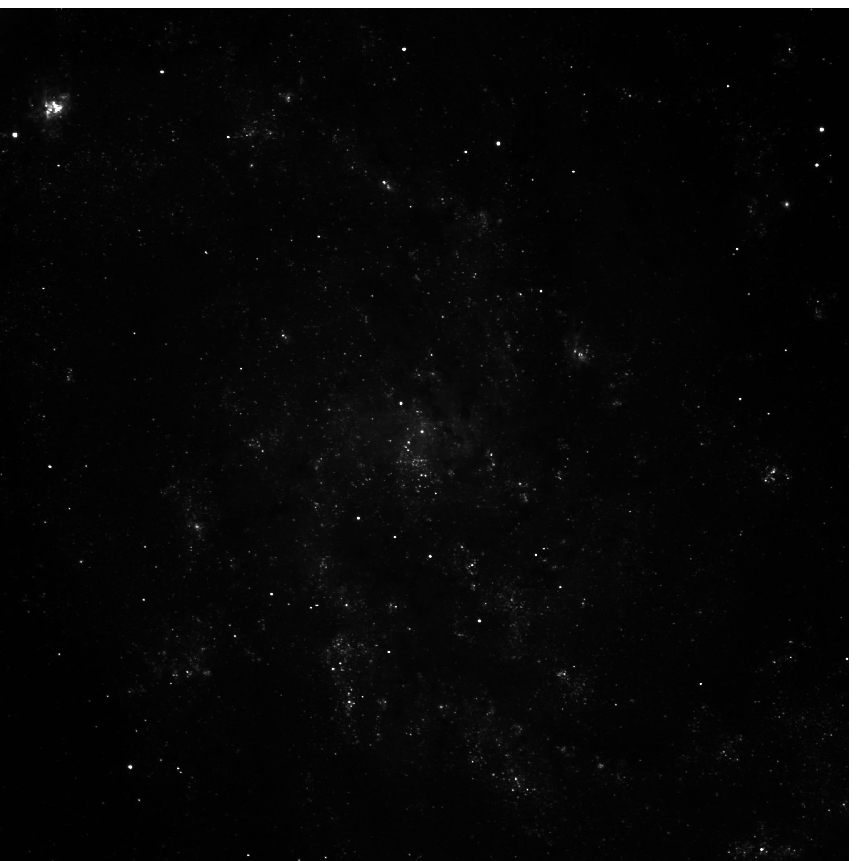}
\caption{Three examples of the maximum data value for a portion of the M33 U-band dataset.  Note the bright star-forming region, NGC~604, located in the upper-left corner of the image.  The scaling in the top image nicely shows the faint detail in M33 but the center of NGC~604 is saturated because the maximum value is set too low.
In the center figure the maximum value is set too high and shades of gray are wasted.  Despite its dark appearance,
the bottom image has an ideal maximum scale value because the nebular detail in NGC~604 is not saturated.
In The GIMP the scaling will be fine tuned to bring out the detail in the faint regions without saturating NGC~604.}
\label{fig-11}            
\end{center}
\end{figure}

In IRAF, the intensity scaling
can be done iteratively with the {\it display} command (in the images.tv package)
and SAOImage DS9.  The display program {\it ximtool} can also be used instead of DS9.  In the IRAF
command {\it display}, the minimum and maximum scale values can be set manually by
specifying values for $z1$ and $z2$ respectively.  First set $zscale=no$ and $zrange=no$
to force {\it display} to use the specified values of $z1$ and $z2$.  The minimum scale value ($z1$) should be set to
just below the noise level so that pure black will be shown as the lowest valid data value (Figure~\ref{fig-10}).  Invalid
data values such as dead pixels should be ignored.  The easiest way to determine this value is, in DS9 or ximtool,
to simply move the cursor through the darkest regions of the image and look for the lowest data values.
To determine the maximum scale value ($z2$) iteratively display the dataset until the maximum value is
just above the brightest point in the structure of interest (Figure~\ref{fig-11}).  In this example the
nebular and galactic structure will be kept unsaturated; individual stars are not important.  In the M33 datasets, the maximum
scale values are determined by inspecting the data values in the brightest region of interest, which is NGC~604, the 
bright nebula in the upper left corner of the image.  

\begin{figure}[hbtp]
\begin{center}
\plottwo{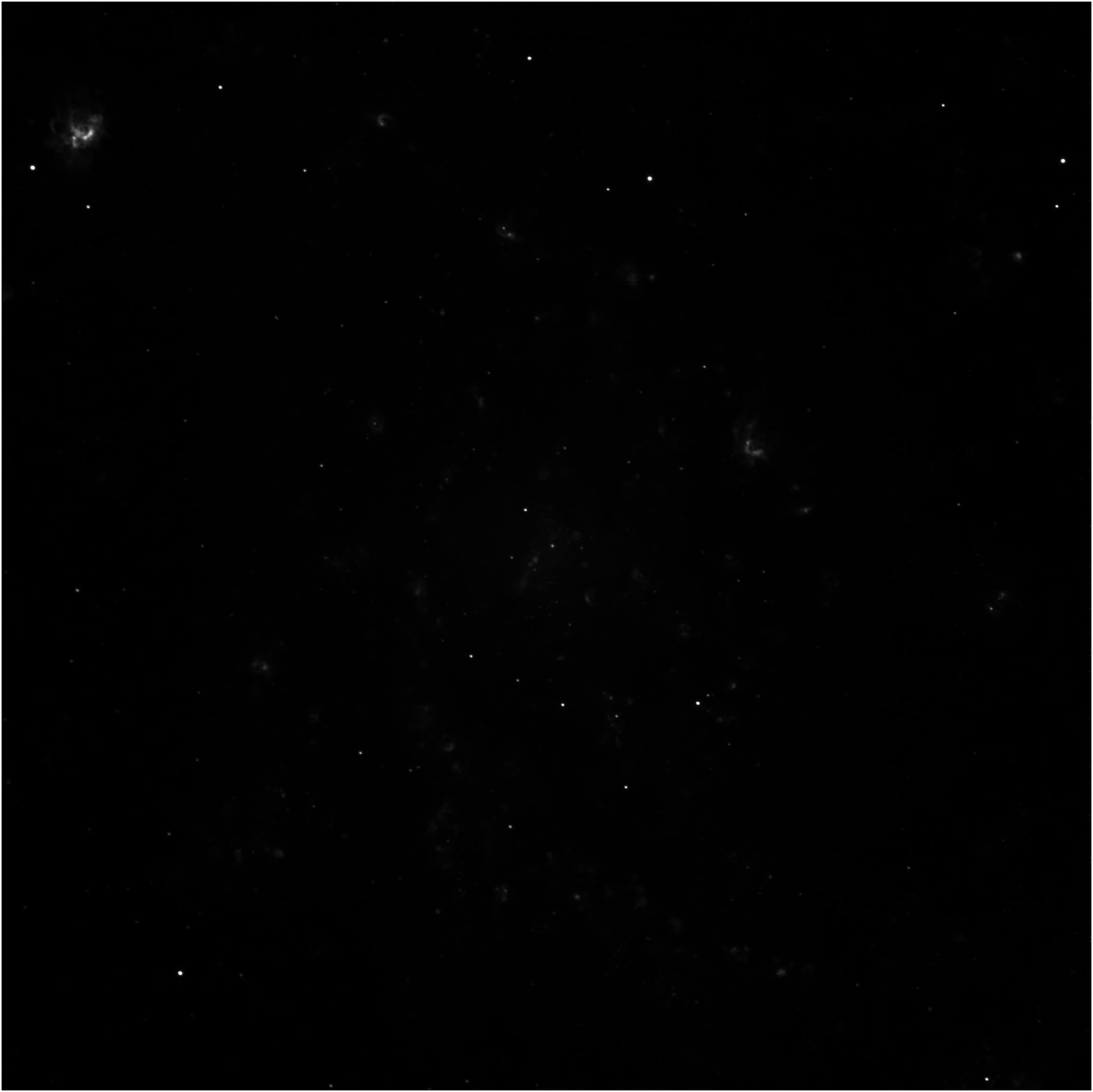}{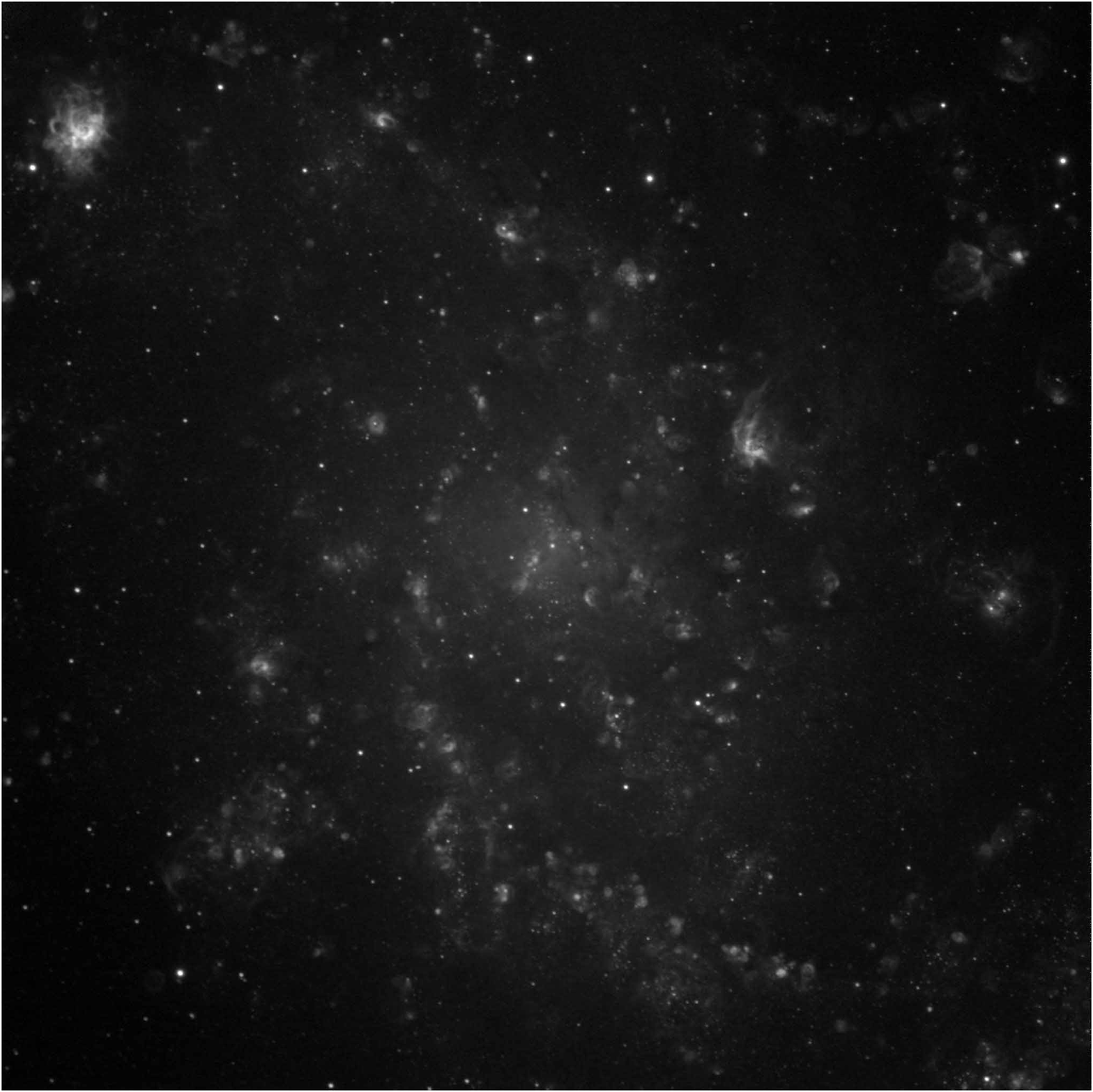}
\caption{The M33 \halpha\ dataset with the same minimum and maximum scaling values, but with linear (left) and
logarithmic (right) scale functions.  The logarithmic scale is chosen because it better shows the detail in the faint
nebulosity without saturating the bright nebulosity in NGC~604.}
\label{fig-12}            
\end{center}
\end{figure}

\begin{figure}[hbtp]
\begin{center}
\includegraphics{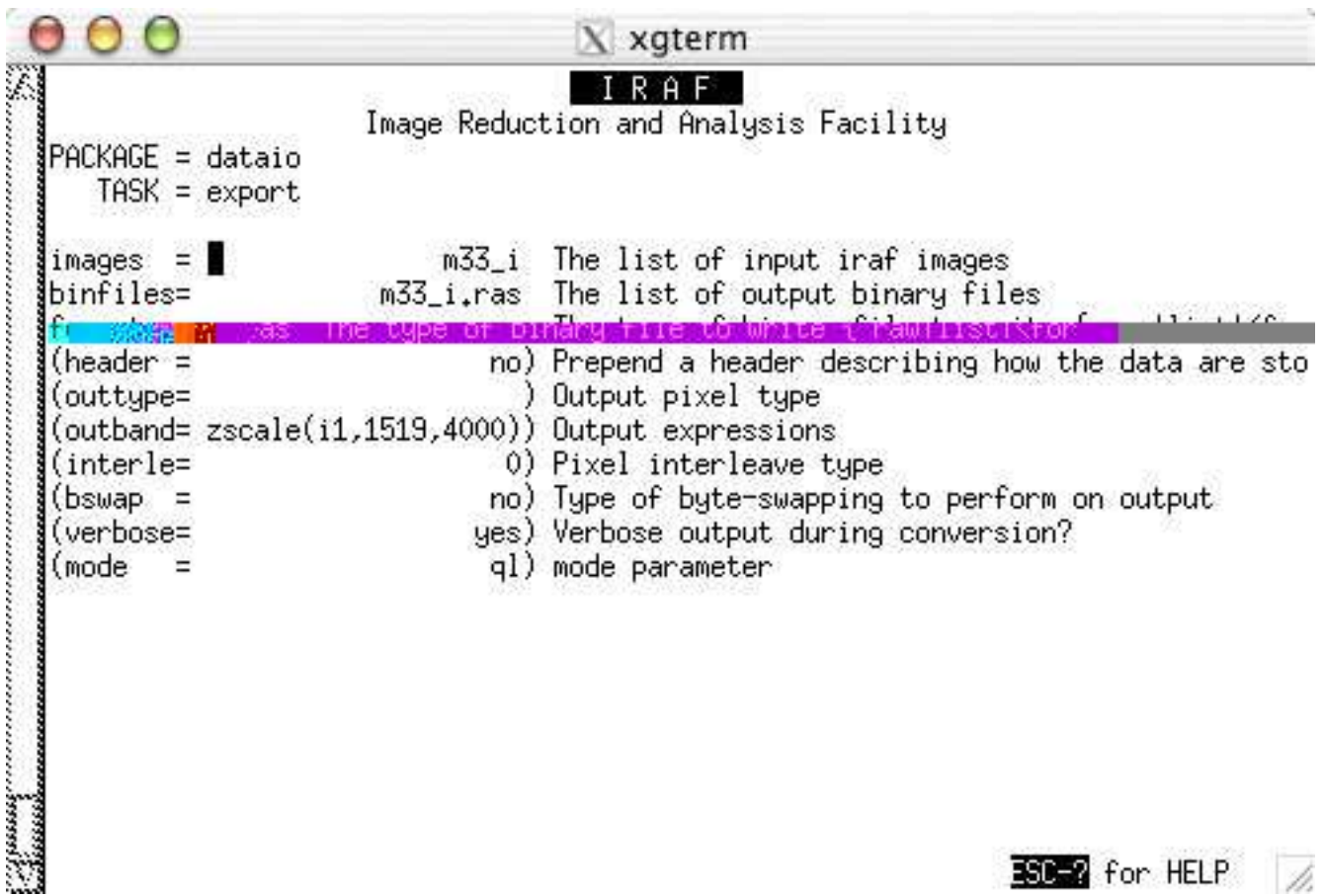}

\caption{The input parameters for the IRAF task 'export'.}

\label{fig-13}            
\end{center}
\end{figure}

The scaling values are determined separately for each dataset.  The scaling values and functions determined for each dataset of M33
are listed in Table~\ref{tbl-A}.  Column~1 is the name of the file; column~2 is the filter name; column~3 is the minimum scale value, $z1$; column~4 is the maximum scale value, $z2$; and column~5 is the scaling function.  For the broadband filters a linear scaling is chosen; and
for \halpha\ a logarithmic scale is used to better show the faint nebulosity in the galaxy (Figure~\ref{fig-12}).
The data are projected into a SUN rasterfile image using the IRAF command  {\it export} in the dataio package.  The parameters should
be set as illustrated in Figure~\ref{fig-13}.  The data scaling is achieved by using the $zscale$ function in the $outbands$
parameter.  For a linear scaling, the outbands parameter should be set to $zscale(i1,z1,z2)$.  For a logarithmic scaling, it should be set to $zscale(log(i1),log(z1),log(z2))$.
This is the last step that requires IRAF, although it is not unusual to return should it be determined that the chosen scaling
system is not ideal.

\subsubsection{Intensity Scaling and Projection with Karma}

\begin{figure}[hbtp]
\begin{center}
\includegraphics{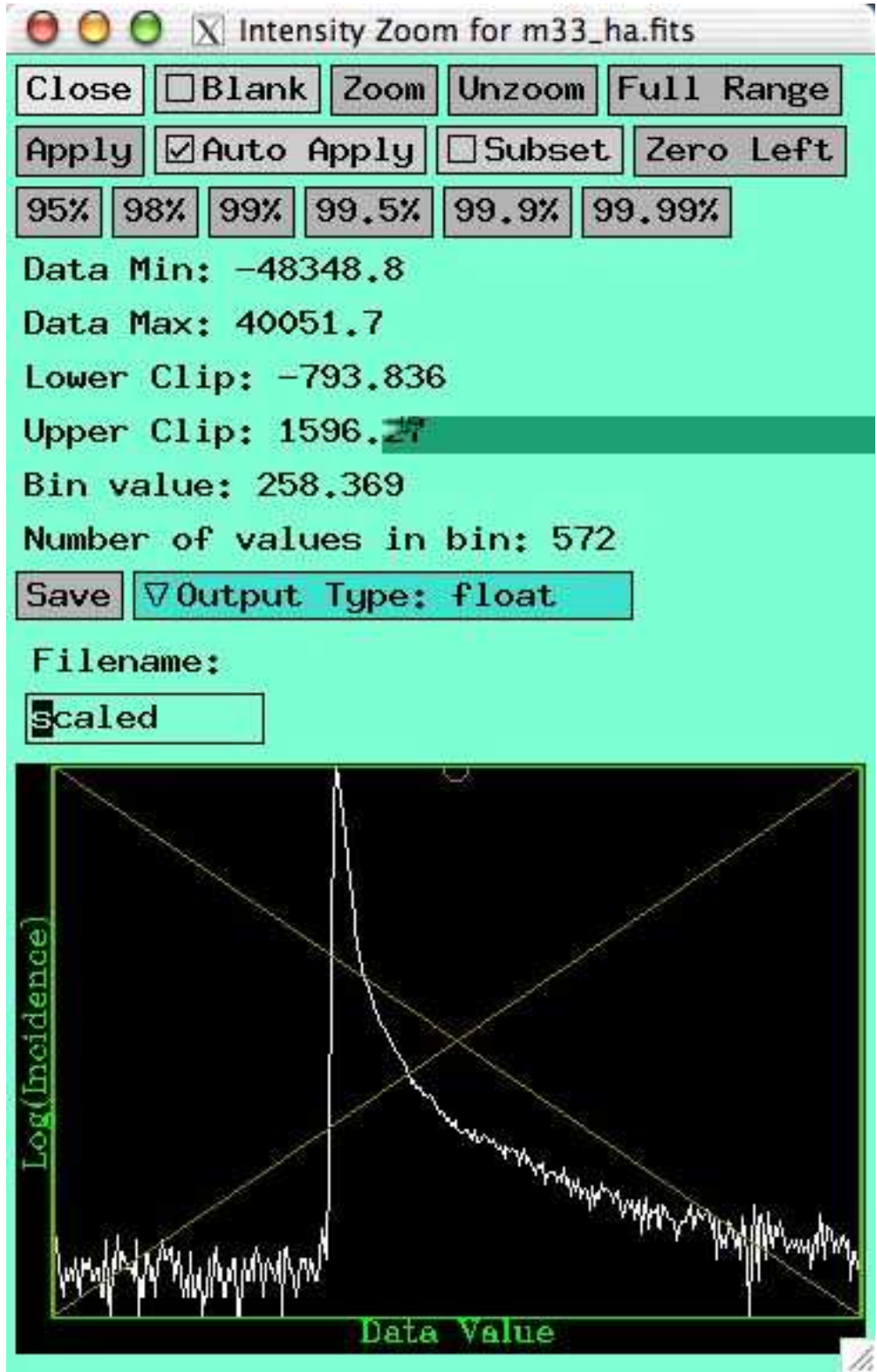}

\caption{The Karma histogram plot for the M33 \halpha\ dataset.  The minimum and maximum scale values, i.e., the lower and upper clip values, 
default to the minimum and maximum data values in the image, which are usually much too wide.  The histogram
shows a sharp peak that corresponds to the sky background level.  Values below this are due to bad pixels and
should be ignored.}

\label{fig-14}            
\end{center}
\end{figure}

The above scaling can be done more intuitively with the graphical interface in Karma's {\it kvis} program, which is included
in Karma versions 1.7 and later.  After opening the
dataset with the {\it file} button in the kvis window, the minimum and maximum scale values are automatically set to the minimum and maximum values that appear in the data.  These values can be manually set in the browser window in the low and high boxes.  The image displayed in the kvis window will automatically reflect these changes.  Alternatively, these values can be set graphically by pressing the {\it histogram} 
button in the browser window.  An intensity zoom window will open which shows a histogram plot of the logarithmic number of pixels as a function of data value (Figure~\ref{fig-14}).  Moving the cursor through the histogram will show
the bin values and the number of pixels within that bin.  The minimum scale value can be set by left-clicking
in the histogram; and the maximum scale value can be set by right-clicking in the histogram.  If the {\it Auto Apply} box
is selected the results will be seen immediately in the kvis window.  The zoom button will expand the histogram plot
to show only the data values between the minimum and maximum scale values.  The unzoom will expand the histogram to show the entire range of data values.  The percentage buttons, i.e., the 95\% to 99.99\% buttons, will set the minimum
and maximum scale values to enclose the central 95\% to 99.99\% of the data values.  They can be a useful way to start the process of searching for scale values.  The scaling function can be set
in the intensity policy window, which is accessed under the intensity menu in the kvis window.  The intensity scale can be set to a linear, logarithmic or square-root function.  If the logarithmic function is selected, the {\it log cycles} slider controls the visible dynamic range in powers of 10.  Increasing the number of cycles assigns more grayscale levels to fainter data values, in effect showing more detail in the faint regions.

\begin{figure}[hbtp]
\begin{center}
\includegraphics{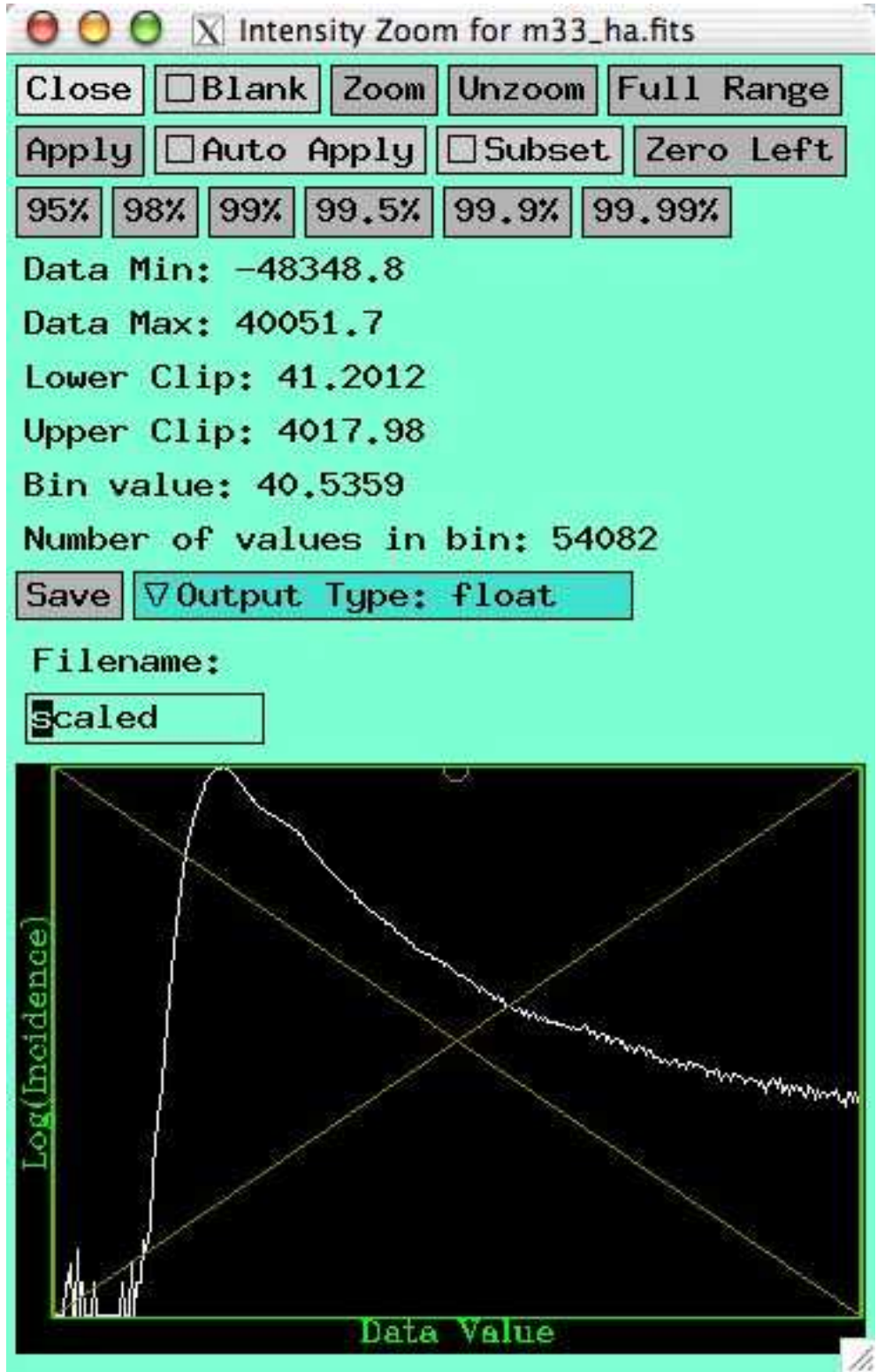}

\caption{The Karma histogram plot for the M33 \halpha\ dataset.  The histogram is zoomed in relative to Figure~\ref{fig-14}, showing only the clipped range.  The lower end should be set to just below the drop in the data values shown in the histogram.}

\label{fig-15}            
\end{center}
\end{figure}

The histogram tool is very useful for determining the appropriate scaling values.  For example, Figure~\ref{fig-15} shows the range of
data values for the M33 \halpha\ dataset.  
By moving the cursor through the histogram window it is apparent that only a handful of pixels have data values less than 30.  These
are presumably bad pixels, many of which appear on the right edge of the image.  The minimum scale value should
therefore be set to a value above this.
Note the sharp peak in data values around a data value of $\sim$48.   If
the image is mostly blank sky, i.e., devoid of source structure, this peak will correspond to the sky background level and
will be very sharp.  In this case the minimum level should be set at or slightly below this peak.  In the case of M33 most of the image consists of the galaxy itself, so the minimum level should be set to below this peak so as to not undersaturate the
sky background.  As with IRAF above, a minimum scale value of $\sim$36 is found to be about the lower end of the sky background level,
as determined by inspecting pixel values in the upper right and lower left corners of the image in the kvis window.  It is a good compromise
between slight undersaturation for some pixels without wasting grayscale values on noise.  As with IRAF, the maximum
scale value is determined by inspecting the data values in the brightest regions of interest, in this case in NGC~604.  For the \halpha\ dataset the peak data value in NGC~604 is about $3400$.  As in the IRAF example, the structure in the
\halpha\ dataset is best shown with the logarithmic function.   A log cycle value of $2.4$ is chosen as a balance between
the structure in the faint and bright portions of the image.  Note that the histogram is less useful for determining the maximum scale value
because in most datasets there will be high pixel values that are to be ignored, e.g., hot pixels, cosmic rays and saturated stars.  
The minimum and maximum scale values for the other datasets are similarly determined.  


Once an ideal scaling system has been determined for a dataset, an image can
be exported as a SUN rasterfile, PPM data image, or a PostScript file under the export menu in the kvis window.  The
GIMP is capable of importing any of these formats, although  Photoshop, as of version 7.0, can only import EPS files directly.  SUN rasterfiles and PPM data images must first be converted into an alternative, non-lossy format, such as TIFF, that Photoshop can open.  Several file-conversion programs are
available on many platforms, including {\it convert} for UNIX and GraphicConverter for Macintosh.  As
in the IRAF example above, SUN rasterfiles are exported.

\subsection{Color Image-processing in Layers with The GIMP}

\subsubsection{Layering and Alignment}

\begin{figure}[hbtp]
\begin{center}
\includegraphics{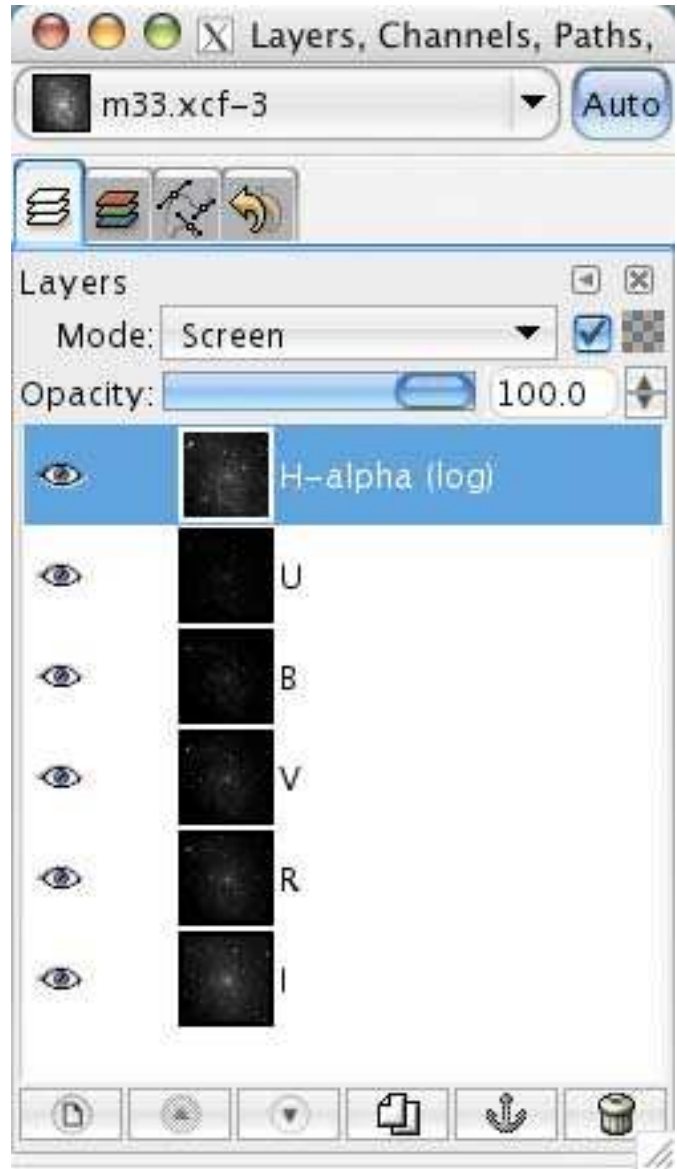}

\caption{The {\it Layers, Channels, Paths and Undo} palette in The GIMP after each of the images has been copied in as a layer.  Note
that all of the layers are visible, as indicated by the eye next to each layer.  The \halpha\ layer is currently selected.}

\label{fig-16}            
\end{center}
\end{figure}

In this example The GIMP will be used to assemble the color image.  
Each SUN or PPM rasterfile is opened with The GIMP as a separate image.  The images are then copied into a single master
file wherein
each image is a separate layer.  For each image, choose the {\it Select / All} menu option, copy it and then paste it into the
master file.  Each image will appear as a separate layer in the {\it Layers, Channels, Paths and Undo} window (Figure~\ref{fig-16}).  For bookkeeping, rename the layers after their respective datasets.  This can be done by double-clicking on the layer, which edits the layer attributes.  Each layer will default to the ``normal" mode and an opacity of 100\%.  To see all the layers the mode for each layer must be changed to ``screen."  

The layers are aligned by selecting each layer and using the {\it Move layers and selections} tool to shift it to a common reference layer.  For this image, alignment is straightforward since all of the data were obtained with the same telescope and camera during the
same epoch.  
The U-band layer was arbitrarily chosen as the reference layer; and the other datasets were aligned to it.
This is done by turning off the visibility of each layer, by clicking on the eye icon to the left of each layer, so that only the
U-band layer and the layer to be shifted are visible.  Select the layer to be shifted and use either the mouse or arrow keys to shift the image until the stars are aligned.  Zoom in to at least 200\% to better align the stars.  Each the layer needs to be shifted to the U-band layer in this manner.    

At this point all of the datasets are combined into a single master file, with each dataset as its own layer set to screen and 100\% opacity.  The layers are also aligned.  The master file should now be saved to disk in the file {\it m33.xcf}.  This master file should be kept unmodified; and a duplicate file
should be used for scaling and colorization.  Making a duplicate is important before proceeding
because The GIMP does not currently use adjustment layers; thus the layers are being modified directly.  Before colorization is possible the image mode needs
to be changed from grayscale to RGB.  This is done by selecting RGB under the {\it Image~/~Mode} submenu.  Use the ``Save as..." option to save a duplicate image to {\it m31color.xcf}.

If desired, rescale the duplicate image to a smaller size for testing multiple schemes before applying the
final scheme to the full-resolution data.  In Photoshop it is straightforward to copy adjustment layers from the low-resolution
test images to the full-resolution final image.  Currently The GIMP does not support adjustment layers, thus it is
critical to keep detailed notes on the steps followed.  These datasets are
about 4~megapixel in size and are relatively easy to manipulate on current computers; thus for this example rescaling the image is not done.

The flexibility of adjustment layers can be mimicked in The GIMP by duplicating a layer before modifying it.  
Compare the modified layer to the original by toggling the visibility of the layers.  It is recommended
to keep an original, unmodified version of each layer, in the image file but invisible, in case it is necessary to
start over on any layer.

\subsubsection{Rescaling the Intensity}

\begin{figure}[hbtp]
\begin{center}
\includegraphics{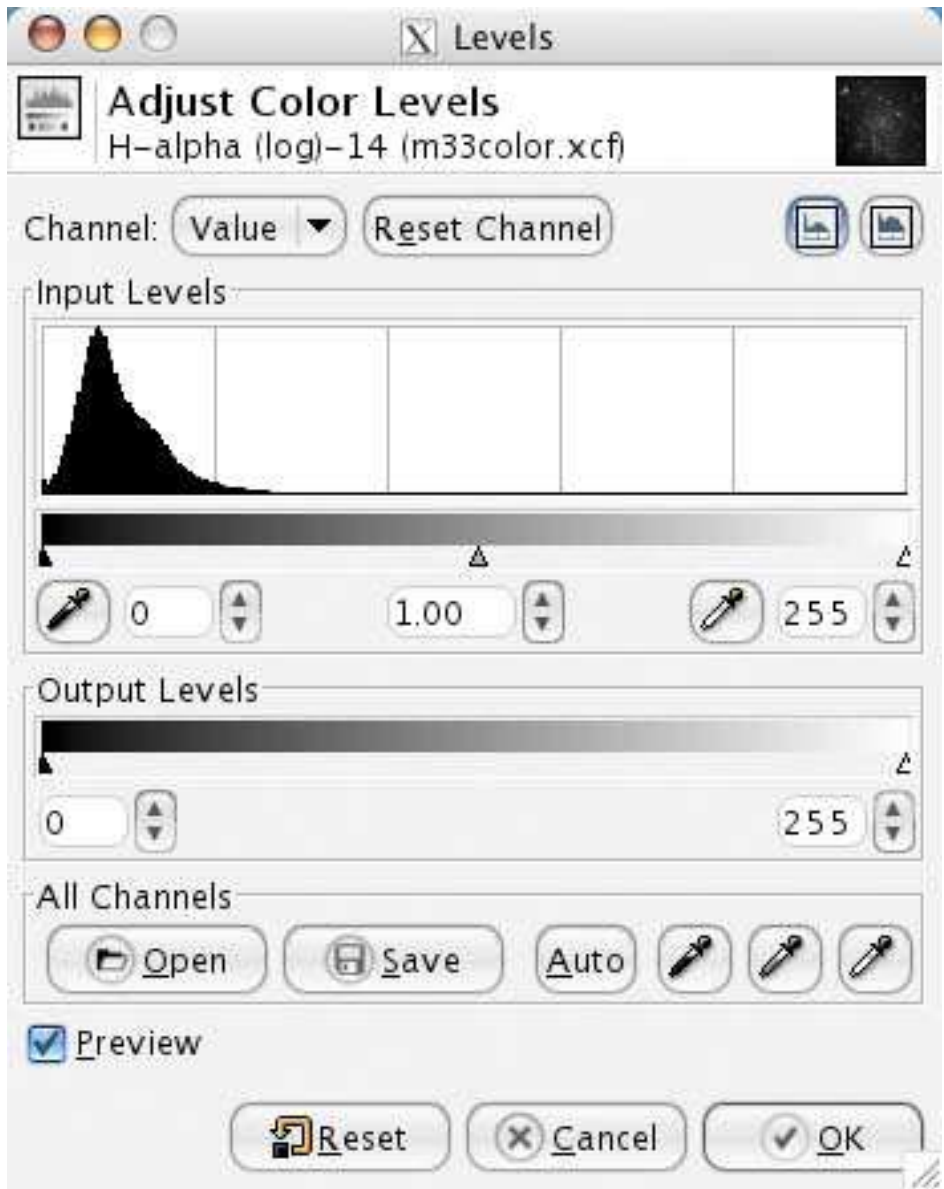}

\caption{An example of the levels dialog box in The GIMP.  In the Input Levels, from left to right, are the settings for the black point, gamma value and white point.}

\label{fig-17}            
\end{center}
\end{figure}

Each layer is rescaled using the {\it Levels...} option under the {\it Layer~/~Colors} submenu as shown in Figure~\ref{fig-17}.
The levels box shows a histogram of the pixel intensity values.  The minimum  and maximum pixel
intensity values can be set here, as well as a gamma correction value which affects how the grayscale values
are mapped to the pixel intensity values.  If the minimum and maximum scale values were correctly determined
in IRAF they should not need to be changed by much here; and in this example these values are retained.
However, the gamma values are adjusted to balance the overall intensity of each layer.  Since the U and B layers
are darker, the gamma values for these layers were greatly increased to better show the faint structure.  Conversely, the
gamma values for the R and I layers are lower because they were brighter.  The result is a good
color balance in the overall image.  The assigned gamma values for each layer are given in Table~\ref{tbl-B}.  

\subsubsection{Colorizing}

\begin{figure}[hbtp]
\begin{center}
\includegraphics{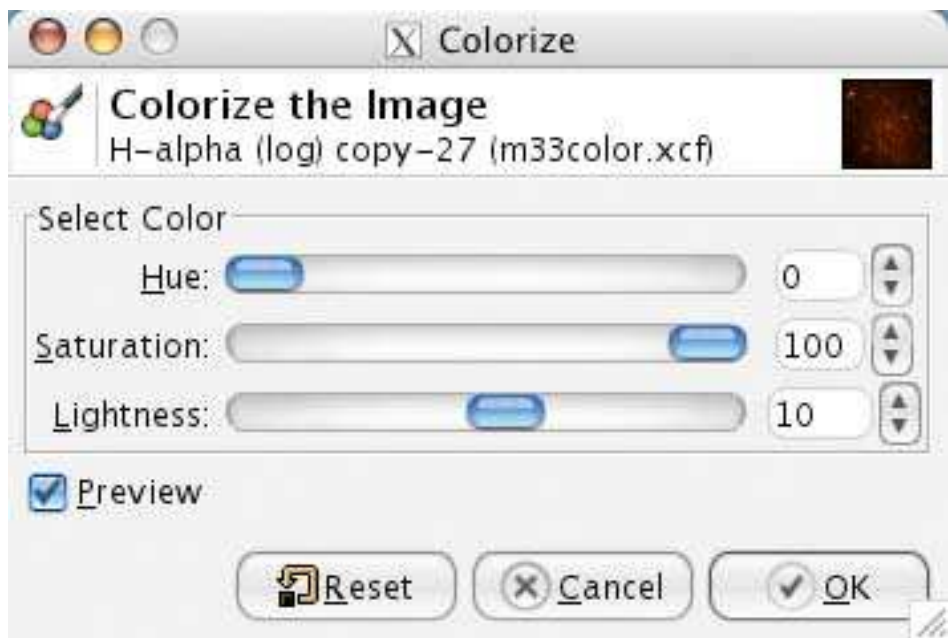}

\caption{An example of the Colorize dialog box in The GIMP.  Here it is set to give the \halpha\ layer a red color (0\arcdeg\ in hue).}

\label{fig-18}            
\end{center}
\end{figure}

The next step after rescaling is to colorize each layer.  This is done using the {\it Colorize...}  option under the {\it Layer ~/~Colors} submenu, as shown in Figure~\ref{fig-18}.  
Table~\ref{tbl-B} lists the colors assigned to each layer and the hue values used to achieve these colors.  Column~1 is the name of the layer; column~2 is the gamma value for each layer; column~3 is the color
of each layer; and column~4 is the hue setting for each layer.  For each colorization the saturation was set to 100\% and the lightness set to $+10$.  

The chosen color scheme is close to a natural color scheme.  The B and V filtered images are assigned blue and green colors respectively,
as they would appear to the human eye when viewed against a bright, white light.  However, the R and I filtered images are given yellow and orange colors
respectively to differentiate them from the \halpha\ filtered image, which is assigned a pure red.  Note that technically this
image cannot be a natural color scheme because it includes the U and I datasets, which are outside of the wavelength
range of human vision.  

\subsubsection{Removing Layer-Specific Defects}

The next step is to fix cosmetic defects that are particular to each layer.  Since each of the V, R and I image layers was generated from a single exposure, cosmic ray removal was not performed with IRAF.  Thus, there are numerous cosmic
rays which appear as bright green, yellow and orange specks on the V, R and I layers respectively.  The clone tool is used
on each layer to remove the cosmic rays.  The size of the clone brush can be set in the {\it Brush Selection} window.
The other layers are also cleaned for other cosmetic defects.  Cosmetic defects which affect multiple layers, such as CCD 
charge bleeds from bright stars, are fixed after the image is flattened into a single layer.


\subsubsection{Generating a Flattened Image}

As shown on the layers palette, the image still includes the original and modified layers for 
each filter, although only the modified layers are visible.  
Since tools in The GIMP only modify the selected layer, to apply overall adjustments the
image must first be flattened or ``copy visible" into a new file.  
The image can be flattened by right-clicking on a layer in the {\it Layers, Channels, Paths and Undo} window and selecting {\it Merge visible layers}.  Save a copy of the layered, unflattened file in the native GIMP (.xcf) format as {\it m33flat.xcf}.

\subsubsection{Color Balance, Composition and Cosmetics}

The final steps are a final color balance, cropping, and cosmetic cleanup.  
The image is too low of contrast, so increase the contrast with the {\it Brightness--Contrast...} option under the {\it Layer~/~Colors} submenu.  The goal is to balance the image so that it has regions of pure black and pure white.  In most optical images, the bright stars will serve as regions of pure white, so it is only necessary to establish regions of pure black.  In this image the contrast is set to $+40$ so that the lower-left and upper-right regions of the image are close to pure black.
Next, it is necessary to adjust the color balance with the {\it Color Balance...} option under the {\it Layer~/~Colors} submenu.  When applying a color balance it is important to remember the different color contrasts and how they can affect your image.  In this case, the image is slightly too red, so set the red to $-5$ for both the shadows and the midtones.  
Note that one may choose other values because of personal aesthetic as well as the color response of your
monitor.  Unfortunately The GIMP, as of version 2.0, doesn't color manage; thus images generated by The GIMP may look significantly different when viewed on different computers.

\begin{figure}[hbtp]
\begin{center}
\plotone{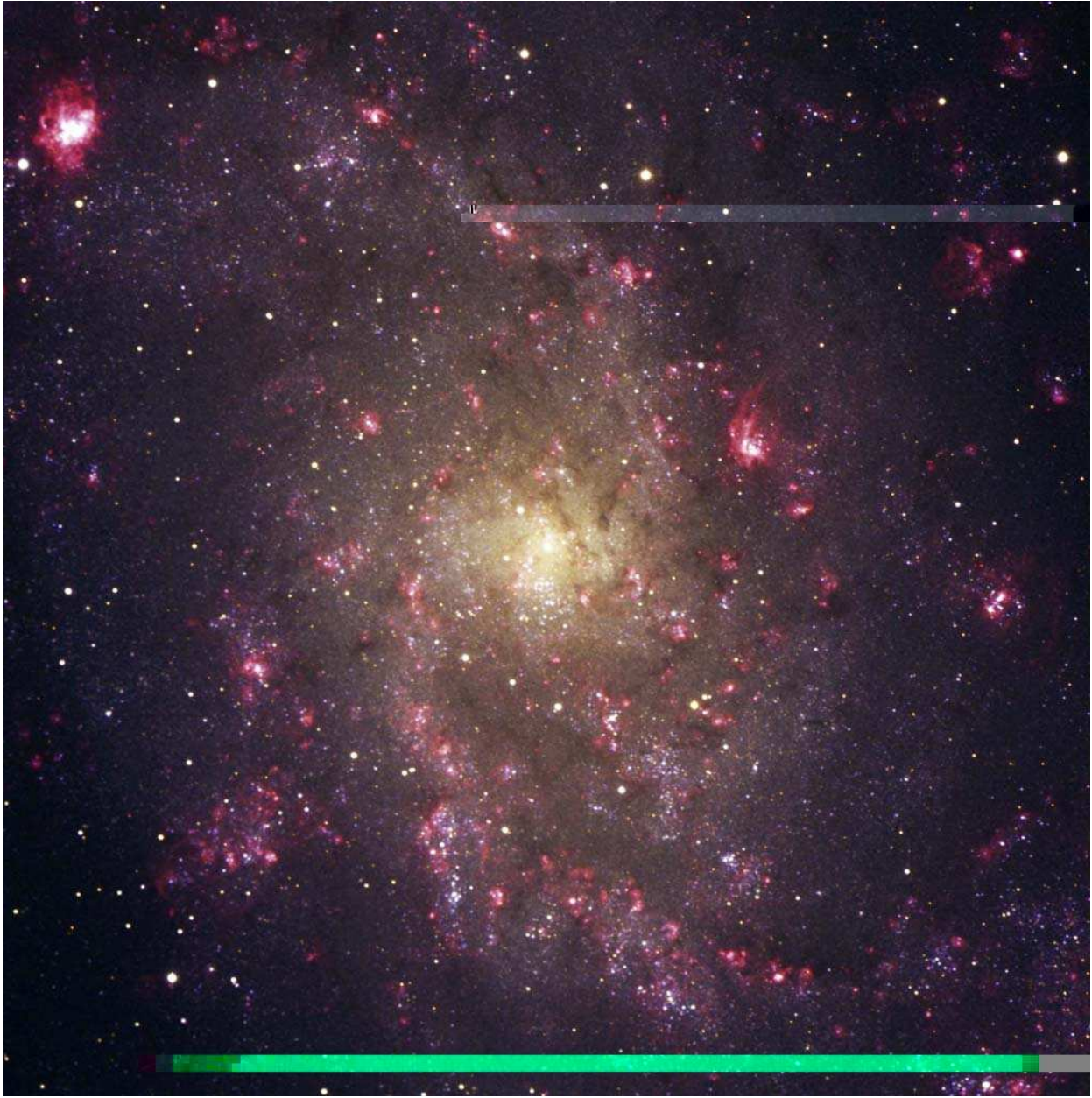}

\caption{The final color image of M33, saved in `m33final.tif.'  After each layer is colorized and rescaled, the image was cosmetically
cleaned, flattened and cropped.  Final color balance and brightness~/~contrast adjustments were also applied.}

\label{fig-19}            
\end{center}
\end{figure}

Aside from the cosmic rays in the V, R and I layers, this image has few cosmetic defects.  There are about a dozen
bright stars which have CCD charge bleeds.  These are repaired as described in \S 2.4.1 and in Appendix~C.  Next, the image is cropped
to remove the left and top edges of the image that are the result of shifting the images.  The noise on the right side of the image is also cropped.  M33 fills the entire image, so it is not necessary, nor desired, to crop the image further.  Finally, the image is rotated.  The data from the S2kb camera are read out in the traditional orientation of north is up and east is to the left.  Other orientations should be considered, but in this case the image is not
rotated because, in the current orientation, the spiral arms create diagonal lines which keep the viewer's eye within the image as discussed in \S 3.3.  The eye is also drawn to NGC~604 in the upper left corner.  At this point the image is considered done, although
additional processing is likely necessary to prepare the image for release on the Internet and for press production.
The final color image is shown in Figure~\ref{fig-19}.  It is saved as {\it m33final.tif}.  It is saved in TIFF format so that software other than The GIMP may open it.

\section{NGC~6369:  An Example with IDL and  Photoshop}

The second tutorial provides an example of producing a color image from FITS
data of NGC~6369, a planetary nebula observed by the WFPC2 camera aboard \hst.
Data were obtained through the F502N~[OIII], F656N~\halpha\ and F658N~[NII] filters.
A slightly different version of this image was released as a {\it Hubble Heritage} image on November 7th, 2002 \citep{her02b}.

In this tutorial, the image will be created using IDL and Photoshop. 
Using IDL, images will be scaled and
projected into an image file that Photoshop can open. Then, these images will be
combined in Photoshop to produce a color composite. An additional
step will be to enhance overall tonal range in the final image by
producing two different intensity scalings and combining them in
Photoshop.  This example was produced in IDL~5.5 and Adobe Photoshop~CS (8.0) 
in the Windows~2000 operating system, but the
functionality is identical with other versions and platforms.  The  necessary functionality
is available in Photoshop versions 6.0 and later as well in the Mac~OS version. Note that the detailed 
appearance of the figures below depends on the OS and versions of software.
It is assumed that the user has properly installed, and has basic familiarity with, these software packages.
The tutorial makes use of procedures that are part of the {\it GSFC} IDL Astronomy Library.
This library may be downloaded from {\tt http://idlastro.gsfc.nasa.gov/}.
Photoshop actions, i.e., macros, can be defined to simplify repetitive sets of
commands such as colorizing. Examples are available on-line at
{\tt http://opostaff.stsci.edu/$\sim$levay/color/actions/}.  Shortcuts are available for some but not all of the 
steps described here.  The original FITS files, as well as the IDL and Photoshop files, used in this tutorial are available for download at: {\tt http://www.stsci.edu/stsci/resources/imaging/papers/rector2005/}.

\subsection{Intensity Scale and Export of Images with IDL}

\subsubsection{Opening and Displaying FITS Files in IDL}

The datasets must first be opened in IDL. The {\it readfits} procedure 
in the IDL Library 
reads FITS data into an IDL variable. In this example, open the
FITS file $f656n1.fit$, an HST/WFPC2 dataset obtained through the F656N~\halpha\ 
filter, with the IDL command $h=readfits('f656n1.fit')$. 
The \halpha\ data now reside in the IDL variable $h$.
The F658N~[NII] and F502N~[OIII] datasets are similarly loaded into the IDL variables $n$ and $o$ respectively.

To determine the appropriate scaling ranges and functions, the data in variables $h$, $n$ and $o$ are displayed
by using the basic IDL procedure $tvscl$, which scales the input data into an
8-bit display range and projects an image into the active image display
window.  Small procedures may be written to provide even simpler commands, such
as writing a TIFF image. Examples are available on-line at 
{\tt http://opostaff.stsci.edu/$\sim$levay/color/idl/}.  More sophisticated display 
facilities may be used permitting zoom and pan
if desired. A subset of the data may also be displayed by specifying a pixel range
in the image, e.g., $h[xmin:xmax,ymin:ymax]$.

\subsubsection{Intensity Scaling to Show Detail in Bright Regions}

The next step is to determine the scaling range or ``stretch," i.e., minimum and maximum scale
values, and the transform function, e.g., linear, square root, and log, that will
render the brightness range ideally.  It is an iterative
process of trying new ranges and functions until an ideal one is found.  
There are numerous tools and techniques to help with this process, such
as histogram plots to visualize the distribution of pixel values in the image.
After several iterations, the dataset $h$ is chosen to be displayed on a linear scale 
with minimum data value $0$ and maximum
data value $450$. That is, pixels of value $\leq 0$ are rendered black, pixels
$\geq 450$ are rendered white and pixels between these values are
rendered gray on a linear scale.  This can be done with the IDL command 
$tvscl, h>0<450$.  Alternatively, the user-defined procedure $tvgw$ can be used with
the command $tvgw, 2, 800, h>0<450$.  
The procedure opens a window of specified size and identifier, scales
the input image to the window size and displays the image.
It is defined as: \\
\indent	$pro \ tvgw, wi, ws, img$ \\
\indent \indent		$window, wi, xs=ws, ys=ws$ \\
\indent \indent		$tvscl, congrid(img,ws,ws,/interp)$ \\
\indent	$return$ \\
\indent	$end$ \\
The $tvgw$ procedure, like other user-defined procedures, can be input directly into
IDL or added to an IDL procedure library in a text file named $tvgw.pro$.  

Displaying the $h$ dataset linearly with a range of $0$ to $450$ renders a pleasing
image of the brighter parts of the nebula without saturating the
brightest emission (Figure~\ref{fig-20}). However, the fainter regions of the nebula 
are barely visible; thus a separate scaling will be necessary to show the detail in these regions.

\subsubsection{Saving the Image as a TIFF}

Now the image will be saved as a TIFF image file.  The TIFF format is preferred because it can be 
opened by Photoshop, it does not suffer image degradation due to lossy compression, 
and it supports 24-bit (8-bits per channel) color depth.  
Applying the same transform function and
data range used to display the image with the $tvscl$ command, use the
$write\_tiff$ command to project the dataset in variable $h$ to an image file in TIFF format. This
command takes two parameters: a string specifying the output file
name and the IDL variable containing the image in a two-dimensional array. 
Note that support for 24-bit color images is possible with $write\_tiff$ by specifying 
three separate two-dimensional arrays for the red, green and blue color channels.
We do not recommend using IDL to create the initial color image because $write\_tiff$ projects
each dataset into separate channels instead of layers.
Also note that the convention for the origin of the pixel coordinates differs
between FITS and TIFF. A reflection on the vertical axis is
required in the output TIFF to match the image orientation of the input
FITS. This reflection is accomplished using the IDL function $rotate$,
with 7 for the value of the direction argument.  Thus, saving the image as a
TIFF can be accomplished with the command
$write\_tiff, Ôhlin1.tifÕ, bytscl(rotate(h>0<450),7)$.  Alternatively, a
user-defined procedure called $graytiff$ can be used.  Itflips the image vertically, scales to 8-bits and writes
a TIFF file with the much-simpler command $graytiff, h>0<450, Ôhlin1.tifÕ$.    The $graytiff$ procedure
is defined as: \\
\indent	$pro \ graytiff, img, tiff$ \\
\indent \indent		$print, tiff$ \\
\indent \indent		$write\_tiff, tiff, bytscl(rotate(img,7))$ \\
\indent	$return$ \\
\indent	$end$ 

\subsubsection{Intensity Scaling to Show Detail in Faint Regions}

As part of this example, two differently-scaled images will first be produced
from the same \halpha\ dataset: one that best shows the detail in the bright regions of the
image and one that best shows the details in the faint regions.  These will then be combined in Photoshop
to take advantage of the detail in both, to produce a pleasing range of tones over a wider dynamic range. 
An image which shows the bright regions well has already been created, so now display the $h$ dataset with
different transform and stretch to better show detail in the faint regions. In this case, the square
root (IDL function $sqrt$) is used to compress the dynamic range in the
brighter emission while stretching the gray levels at the fainter areas
in the image. Again, experiment with various scale values and
functions to get a pleasing result. To avoid the failure of the square
root, log (IDL function $alog10$), and other functions for values $\leq 0$, use a
minimum $>0$ or add a constant bias. This bias value can provide an
additional opportunity to adjust the tonal range. Larger values will
more closely approximate a linear scale, while smaller values will more
strongly enhance the contrast in the darker pixel values and compress
the dynamic range at the brighter end.

\begin{figure}[hbtp]
\plottwo{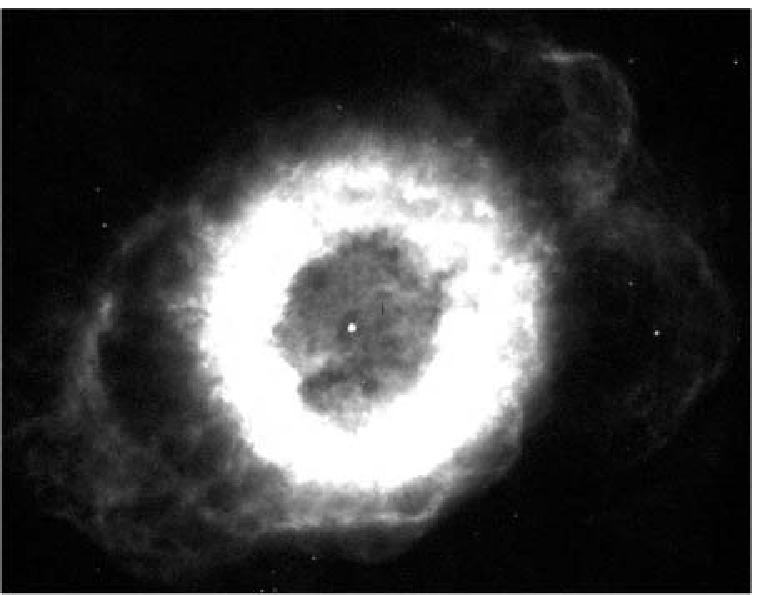}{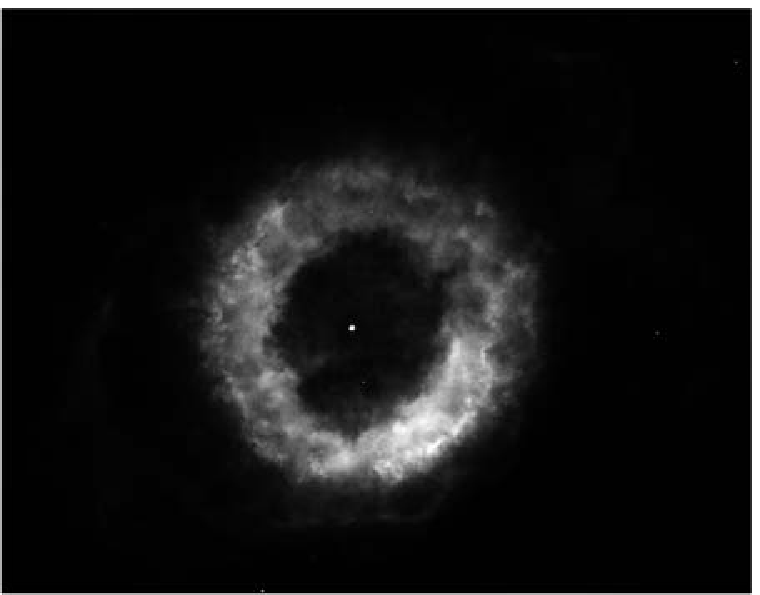}
\caption{The F656N \halpha\ datset of planetary nebula NGC~6369 (HST/WFPC2) intensity-scaled to enhance fainter features (left) and brighter features (right).  The image on the left optimizes the tonal range and contrast in the darker parts of the image, but at the expense of highly oversaturated highlights.  The other optimizes the tonal range and contrast in the brighter regions but at the expense of undersaturated shadows.  \label{fig-20}}
\end{figure}

From the same \halpha\ dataset above, display a second grayscale image that
is scaled for the darker areas with the command $tvscl, sqrt((h+1)>2<150)$, as 
shown in Figure~\ref{fig-20}.  This image deliberately oversaturates the bright
regions to show the faint structure missed in Figure~\ref{fig-20}.  This image
can be saved with the command $write\_tiff, Ôhsqrt1.tifÕ, rotate(sqrt((h+1)>2<150),7)$ or with
the user-defined command $graytiff, sqrt((h+1)>2<150), Ô hsqrt1.tifÕ$.

The scaling values are then determined separately for the $n$ and $o$ datasets.  Two scalings are used for each filter, a linear scaling to show detail in the bright regions and a logarithmic scale to show the detail in the faint regions.  Thus, after scaling each dataset twice a total
of six images should have been generated and saved to disk.  The chosen scaling values and functions determined for each dataset of NGC~6396 are listed in Table~\ref{tbl-C}.  
Column~1 is the filter name; column~2 is the linear range used for the first scaling, in IDL format; and column~3 is the logarithmic range used for the second scaling, also in IDL format.   

Since the $write\_tiff$ procedure can write a 3-color (24-bit) TIFF, it is possible
to produce a full-color TIFF entirely in IDL. However, generating three 
gray images and combining them in color in Photoshop provides much more control and flexibility,
as will be demonstrated below.

\subsection{Control Contrast Using Multiple Intensity Scalings and Masks in Photoshop}

\subsubsection{Loading Images into Layers}

\begin{figure}[hbtp]
\begin{center}
\includegraphics{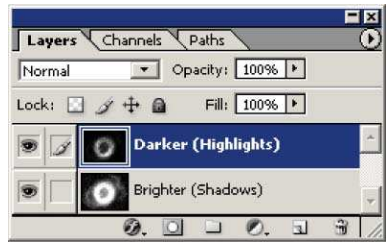}

\caption{The layers palette in Photoshop for a single file that contains the two images from Figure~\ref{fig-20}.  The darker image that shows the details in the bright regions of the nebula is in the layer on top.  The brighter image that shows the detail in the faint regions is in the bottom layer.  Note that the top layer is currently selected, which means that all editing will affect only the top layer.  Also note that visible layers are marked with an eye on the left of the layer.}

\label{fig-21}            
\end{center}
\end{figure}

In Photoshop, the two images generated in IDL for each dataset are combined into a single
grayscale image.  This is accomplished by loading each image into the same file but into 
separate layers.  First, open both of the \halpha\ images, {\it hlin1.tif} and {\it hsqrt1.tif}, in 
Photoshop.  Next, copy the darker image into the file containing the brighter image
so that a single file contains each image as a separate layer.
To do this, click and drag the darker image from its open window to the 
brighter image window.  Alternately, with the darker image window active, duplicate it 
into a new layer in the brighter image window by selecting {\it Duplicate Layer...} from the layers palette menu,
which is accessed by clicking on the button in the upper-right corner of the layer palette.  Make certain that the
destination document is set to the brighter image.
In the layers palette, move the layer containing the darker image above the brighter image. 
Editing the layer names in the layers palette
provides a convenient way to identify the layers.  
This can be done by first selecting a layer and then selecting {\it Layer Properties...} from the layers palette menu.
A shortcut is simply to double click on the layer name in the layers palette.
At this point the layers palette should look similar to Figure~\ref{fig-21}.

\subsubsection{Manually Generating an Image Mask}

Next, a layer mask will be used to determine which parts of each image will be visible.
More precisely, a mask will be created to block regions
of one image and allow the other image to be visible underneath.
A layer mask is a continuous-tone, grayscale pixelmap, also known as
an ``alpha channel," that defines the opacity of its associated image layer.
A white (255 value) pixel in the mask renders the image pixel opaque,
allowing the image pixel to display fully.   A black (0 value) mask pixel value
renders the image pixel completely transparent, allowing the underlying image layers
to be entirely visible. Intermediate-value gray mask pixel values define
intermediate transparencies, resulting in a combination of the
overlapping image layers.

\begin{figure}[hbtp]
\begin{center}
\includegraphics{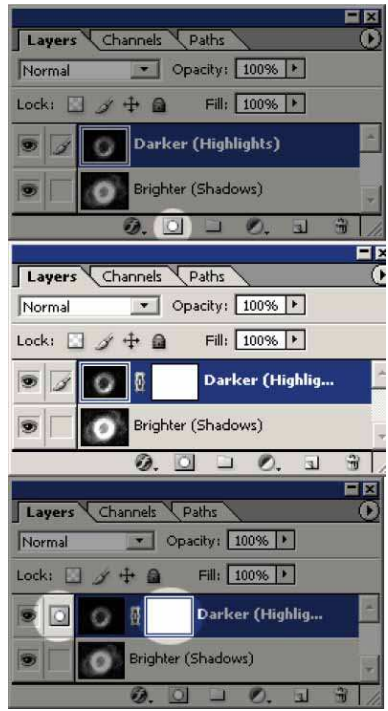}

\caption{A layer mask can be added in Photoshop by clicking on the {\it Add Layer Mask} button in the layer palette, which is highlighted in the top figure.  The middle figure shows that a layer mask has been successfully added to the top layer, as shown as a mask thumbnail (the white square) to the right of the image thumbnail.  However, all edits performed will still be done to the image layer because the image layer is still selected.  To edit the mask it must first be selected by clicking on the mask thumbnail (the white square), as highlighted in the figure on bottom.  Also note that when the mask is being edited the icon to the left of the image thumbnail changes from a brush to a circle, as highlighted in the bottom image.}

\label{fig-22}            
\end{center}
\end{figure}

To add a mask to a layer, first select the layer by clicking on it in the layers palette.  
The layer mask can be added by either clicking on the {\it Add Layer Mask} icon on 
the layers palette or by selecting {\it Add Layer Mask} from the layers menu.  Either
the {\it Reveal All} or {\it Hide All} mask may be selected.  The layers palette should
now look like Figure~\ref{fig-22}.

Since the layer mask is a grayscale pixelmap, it may be manipulated like 
any other image. 
In particular, it may be painted onto with the brush tool by clicking on the mask
icon in the layers palette (Figure~\ref{fig-22}). Painting
black in the mask hides the associated image and allows the
underlying images to show through.  Use a soft-edged brush and paint
with lowered opacity to build up the mask gradually and to avoid sharp
transitions.

\subsubsection{Automatically Generating an Image Mask}

\begin{figure}[hbtp]
\begin{center}
\includegraphics{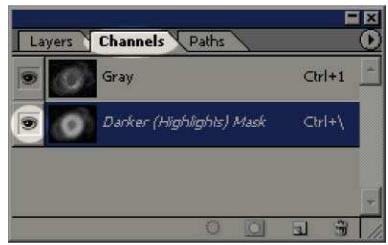}

\caption{The channels palette in Photoshop.  To access it click on the tab to the right of the layers tab (the highlighted tab in the figure).  Make the layer mask visible by clicking on the visibility (eye) icon, as highlighted.}

\label{fig-23}            
\end{center}
\end{figure}

\begin{figure}
\begin{center}
\includegraphics{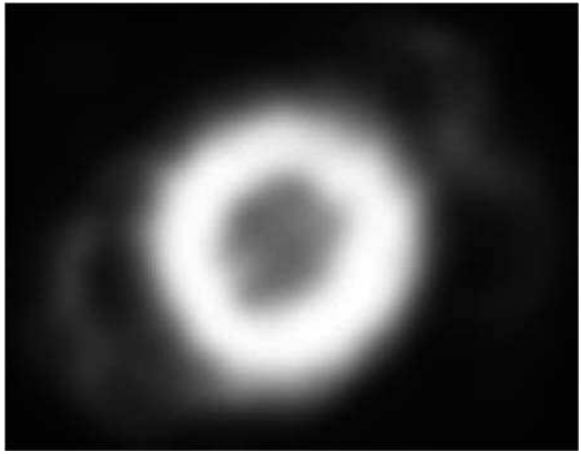}
\caption{The image layer mask for NGC~6369 to combine two differently-scaled images after a Gaussian blur filter has been applied.  \label{fig-24}}
\end{center}
\end{figure}

\begin{figure}
\begin{center}
\includegraphics{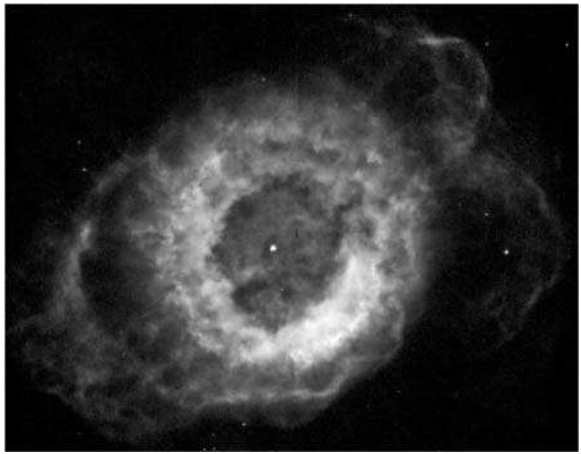}
\caption{The final image of NGC~6369 generated by combining the two images in Figure~\ref{fig-20} with the contrast mask in Figure~\ref{fig-24}.  \label{fig-25}}
\end{center}
\end{figure}

A semi-automated technique can produce a contrast mask
that results in a good transition between the two renditions of the
image.  To generate the mask, add a layer mask to the darker image layer,
which should be the top layer.  Make the bottom, brighter
layer active, select the entire image and copy it to
the clipboard.
Open the channels palette by clicking on the tab to the right of the layers tab, 
and select the layer mask channel as shown in 
Figure~\ref{fig-23}.  Paste
the image from the clipboard into the layer mask.
To blend the images more gradually, apply a Gaussian Blur to the layer
mask (Figure~\ref{fig-24}).  Adjust the radius for the desired effect.  In this example,  choose a radius of
6.5 pixels.  If necessary, paint into the mask
to better blend the two images.
Once the two layers are ideally blended, the image will look similar to Figure~\ref{fig-25}. It is optional to flatten the image before 
proceeding to make the document size smaller.  Due to relatively small image size in this example this is
an unnecessary step.  After combining the two \halpha\ images follow the same steps for
the [NII] and [OIII] images.
\subsection{Assembling the Color Image in Photoshop}

\begin{figure}[hbtp]
\begin{center}
\includegraphics{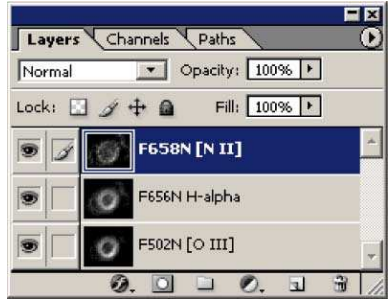}

\caption{The layers palette in Photoshop with all three image layers.  Note that the layers have been renamed for clarity.  Six layers will be present if the highlights and shadows layers for each dataset are not flattened into single layers for each dataset.}

\label{fig-26}            
\end{center}
\end{figure}

If it has not been done so already, place all of the separate images into a single
Photoshop file, with each image as a separate layer.  If each pair
of images was flattened after combining the two scalings for each dataset, there should be
only three layers, similar to Figure~\ref{fig-26}.  Otherwise, all six layers 
should be present.
Also be certain to edit the layer names for clarification.  If necessary, use the move
tool to shift the layers so that the stars are aligned.  

Note that there is a distinction between layers and channels. Channels are the
separate red, green, and blue, i.e., the additive color primaries, pixelmaps containing
the color information for the fully-combined image.  It is analogous to the signals displayed by the
monitor.  Layers each contain a full-color image and therefore contain info that affects each channel.  Multiple layers can reside in a
single Photoshop document.  They can then be combined with a variety of layer blend functions to create
the final, color image.  If a document consists of three gray image
layers with the primary colors applied and rendered in screen layer
blend mode, then the result will be the same as placing the same three gray
images in the three color channels.  However, layers offer many advantages to
directly using the channels, including the ability to assign any color to
a layer.  

\subsubsection{Colorizing Each Layer}
\begin{figure}[hbtp]
\begin{center}
\includegraphics{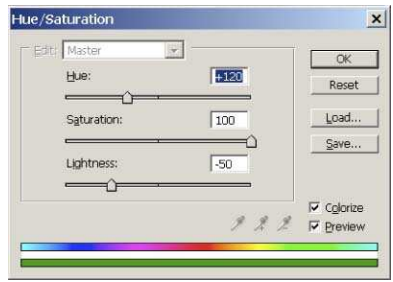}
\caption{The dialog box  in Photoshop  for the {\it Hue/Saturation} adjustment layer for the \halpha\ image layer.  The settings should
be set as shown.}
\label{fig-27}            
\end{center}
\end{figure}

The next step is to colorize each layer.  First, the image must be converted from a grayscale to RGB color
image.  This is done by selecting {\it RGB Color} from the {\it Image~/~Mode} menu.  When prompted, do
not merge the layers.  Now select the \halpha\ layer and then add a Hue/Saturation adjustment layer
by selecting {\it Hue/Saturation...} from the {\it Layer~/~New Adjustment Layer} menu.  
Note that it is generally preferable to use adjustment layers to modify the 
appearance of a image layer because it does not directly modify the image pixels.
Also note that there is a fundamental difference between an image layer and an adjustment layer.  An
image layer contains actual pixel values from an image.  An adjustment layer contains an algorithm  
which modifies the appearance of the image layer(s) below the adjustment layer.
The effect of an adjustment layer is easily removed by either turning off the
visibility of the layer, by clicking on the eye on the left of the adjustment layer,
or by deleting the adjustment layer altogether.  To colorize the layer, 
click on the colorize button and then set the hue to 120\arcdeg\ for green, 
saturation to $100$ and lightness to $-50$, as shown in Figure~\ref{fig-27}.  

\begin{figure}[hbtp]
\begin{center}
\includegraphics{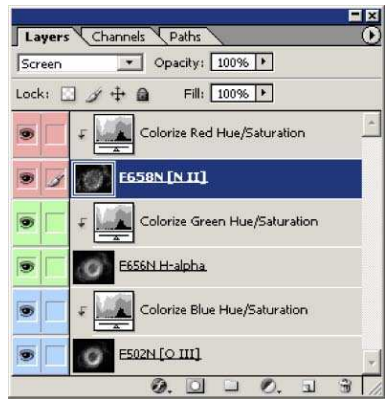}
\caption{The layers palette in Photoshop after each image layer has been colorized with a Hue/Saturation adjustment layer.  Note that the adjustment layers are correctly grouped to their respective image layers and that the layer properties for each layer has been changed so that they are colorized for clarity.}
\label{fig-28}            
\end{center}
\end{figure}

Group
the adjustment layer with the \halpha\ image layer by selecting {\it Group with Previous} from the {\it Layer}
menu.  In doing so the adjustment layer 
affects only that layer.  Otherwise, all of the layers below the \halpha\ layer
will also be affected.  Grouping the Hue/Saturation 
adjustment layer to the \halpha\ layer creates a ``clipping mask."
Photoshop refers to it as a clipping mask when a layer, rather than
an alpha channel, is used to limit or affect the visibility of another layer. A clipping mask 
can be any other type of layer, e.g., an image, text, paths or adjustment layer. Clipping 
masks, i.e., grouped layers, may be enabled or disabled in the layers palette 
by clicking on the boundary between the layers.
It can be useful to annotate the adjustment layer by editing the layer name. for
further clarification, it is also possible to apply a color to the layer in the layers
palette by using the {\it Layer Properties} 
dialog box, which is accessible from the menu 
button in layers palette.  This does not change the appearance of the image, 
but is simply a means for managing the Photoshop layers.  Follow the steps above again
to colorize the [NII] image layer red by setting the hue to 0\arcdeg\ and the [OIII] image
layer blue by setting the hue to 240\arcdeg.  The layers palette should look similar to Figure~\ref{fig-28}.

\subsubsection{Combining Layers to Generate a Color Image}
\begin{figure}[hbtp]
\begin{center}
\includegraphics{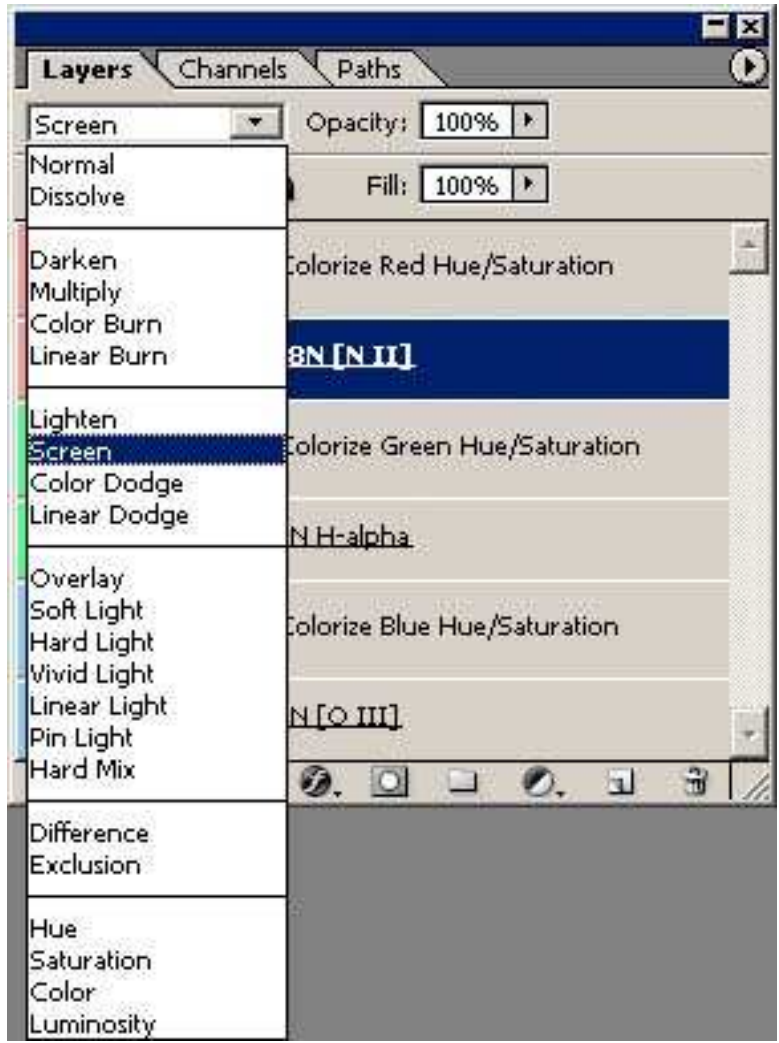}
\caption{The layer blending mode menu on the layers palette in Photoshop.  Change the blending mode for each image layer from normal to screen (as shown).  The blending mode of the adjustment layers should not be changed.}
\label{fig-29}            
\end{center}
\end{figure}

The default setting for the blending mode of an image layer is ``normal," which
obscures all layers below it if the opacity is set to 100\%.
To see the individual, colorized layers combined into a single, full-color image,
set the blending mode of each image layer to ``screen" as illustrated in Figure~\ref{fig-29}. With the
screen algorithm, Photoshop combines each visible layer, as modified by its associated
adjustment layers, to build up the full image.  Photoshop includes an extensive menu of layer blend modes. In this 
example, only the screen mode will be considered.  But the others are 
useful under certain circumstances to achieve various effects.

\begin{figure}[hbtp]
\begin{center}
\plottwo{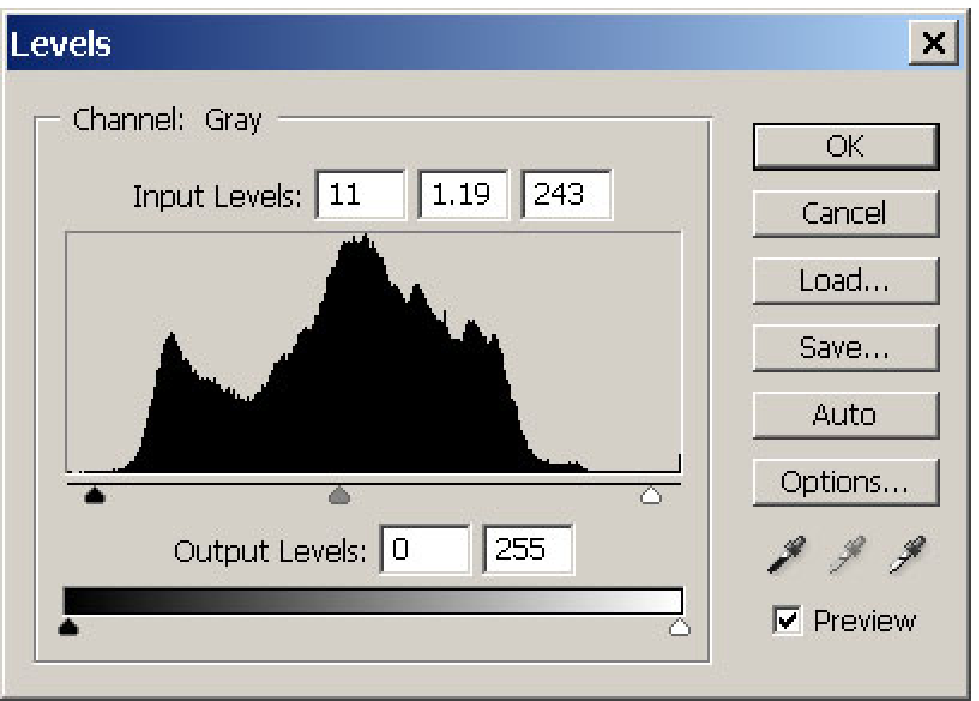}{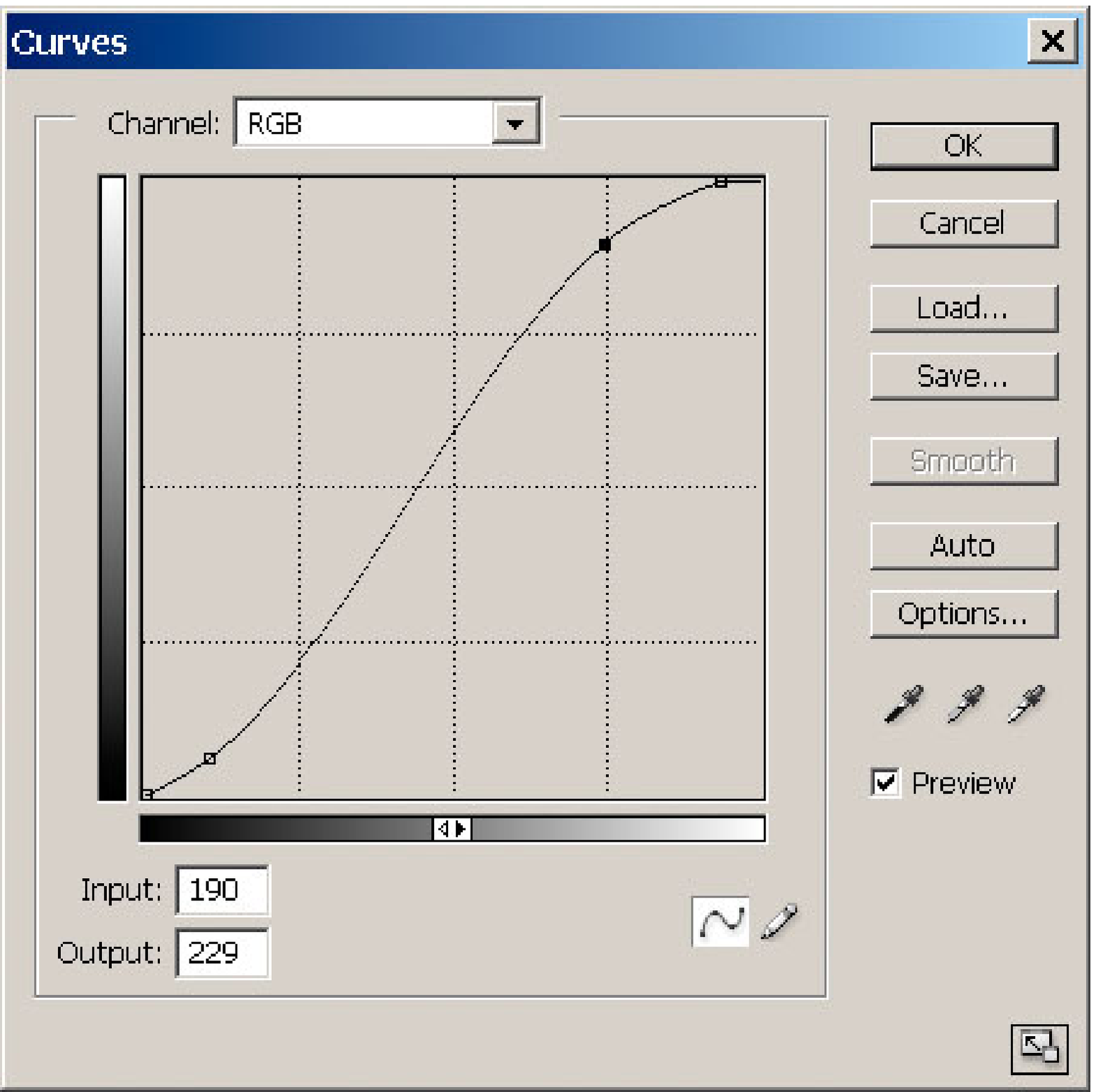}
\caption{An example of levels and curves adjustment layers in Photoshop.  On the left, the levels dialog box shows a 
histogram of the image pixel values.  The black, white and gray sliders below the histogram are used to adjust the black point, white point and gamma values respectively.
On the right, the curves dialog box shows the transfer
function that maps input pixel values (horizontal axis) into displayed
brightness (vertical axis) from black to white.  The shape of the curve can
be modified by adding control points through which a smooth curve is
constructed.  The control points may be moved arbitrarily to adjust the
brightness and local contrast.  The contrast depends on the slope of the line.  
 In this example, the brightness has been
lowered for the darker areas in the image and increased for the brigher
regions, causing the contrast to be higher in the middle levels but lower
for both the darkest and brightest areas in the image.}

\label{fig-30}            
\end{center}
\end{figure}

\begin{figure}[hbtp]
\begin{center}
\includegraphics{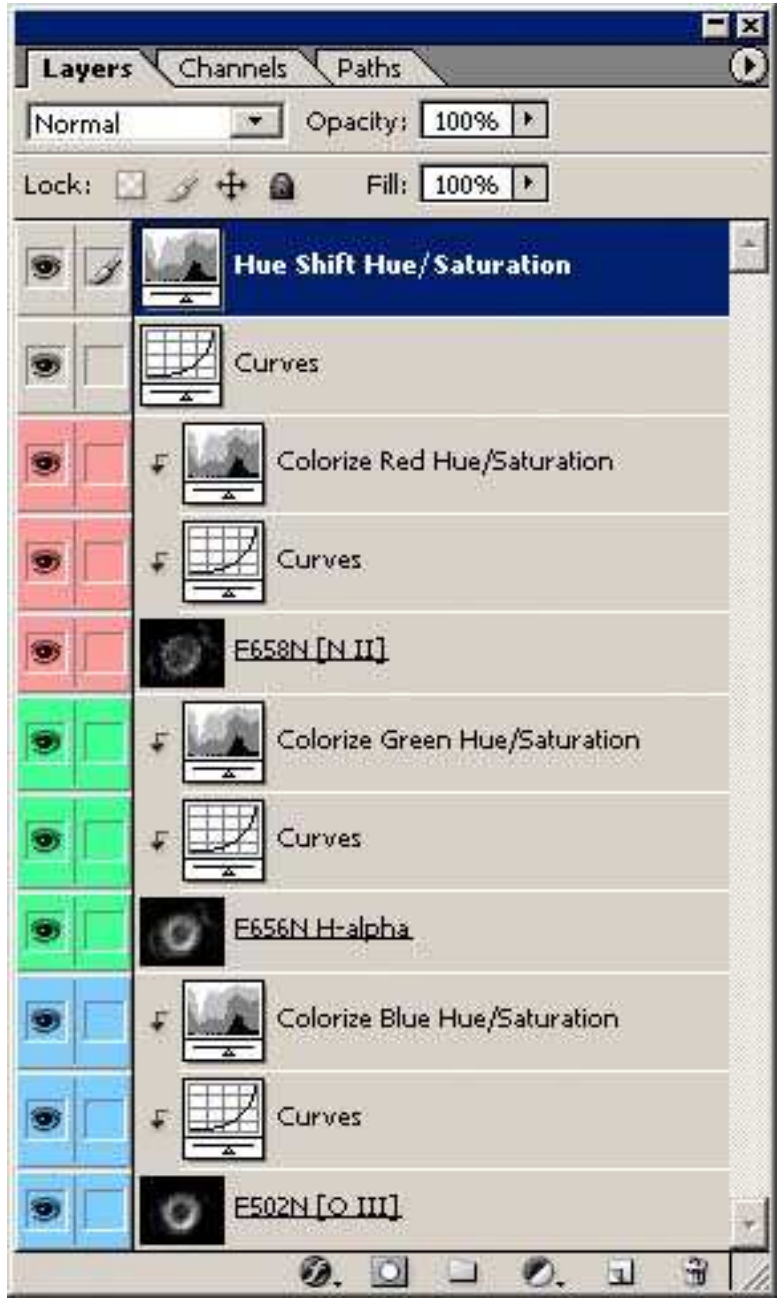}
\caption{The layers palette in Photoshop after each image layer has had a fine-tune intensity scaling with a curves adjustment
layer added.  Each layer also has a text layer as a label.  Since each text layer is underneath the colorize adjustment layer it too is colorized.}
\label{fig-31}            
\end{center}
\end{figure}

At this point a full-color image has been generated, although the results may not yet be pleasing.  To balance the colors,
adjust the brightness and contrast separately for each image layer with a {\it Levels} or {\it Curves} adjustment layer.  
This is done by selecting the grouped {\it Hue/Saturation} adjustment layer for the \halpha\ image
layer.  Now create a new levels or curves adjustment layer and 
modify the parameters for the desired result.  Examples of levels and curves adjustment layers 
are shown in Figure~\ref{fig-30}.
When done adjusting the layer, 
group the adjustment layer so the adjustment applies only the \halpha\ image layer.
Repeat for the [NII] and [OIII] layers.  After each image layer has had a curves
adjustment layer added, the layers palette will look similar to Figure~\ref{fig-31}.

A useful technique is to annotate each image layer with some
description of its source and the color in which it is rendered. A
convenient way to do this is to add a text layer above the relevant
image layer, and then group it to that layer. If the text is rendered
in white, it will appear in the color selected in the {\it Hue/Saturation}
adjustment layer above it.  

With the proliferation of layers in Photoshop, it is often useful to use layer
sets to help manage them. This does not change the appearance of the
image, but can make it easier to keep track of many layers and permit
operating on sets of layers together. For example, it is helpful to group
several adjustment layers or annotations, e.g., text and paths, into a layer set.
The view of the set in the layers palette can be collapsed.  And all the layers
can be linked, moved and resized together. With photoshop~CS v8.0 and later, layer
sets may include subsets.

\subsubsection{Finishing the Image}

Separate adjustments should be added for each image layer to subtly
adjust the appearance of the image.  However, once the image is close
to the desired result, each layer should be cleaned of cosmetic defects.
Due to its compact size and the paucity of bright stars in the field of view,
the image of NGC~6369 has remarkably few cosmetic defects.  The data
for the image is from the WF3 chip only, so there are no chip seams.
Only the cleaning of a few residual cosmic rays was necessary.
To offer more insight on this important step, Appendix~C gives examples 
of how to fix many of the different defects which appear in optical and near-IR data.

Once the cosmetic cleaning is done, the color, brightness and contrast of the image
as a whole is adjusted by adding adjustment layers above all of
the others.  The color balance and brightness/contrast adjustment layers
are very useful to fine tune the image.  When applying these adjustment layers to
the entire image, do not group the adjustment layer or it will only affect the
image layer below it.  The final color and intensity scaling is done after
the repair of cosmetic defects because large and bright defects can significantly
affect the appearance of the image.  Once the image is finished it should be
saved as a separate file that is noted as the final version.  This file should be
saved in Photoshop format.   After this file has been saved, the image needs to
be flattened by choosing {\it Flatten Image} from the layers palette menu.  The
image should then be saved as a TIFF-format file.  Be certain not to overwrite
the final Photoshop-format file that was saved prior to flattening.  It is important to 
keep the layered, Photoshop-format file because it may needed again 
should further changes be required at a later date.  When saving the TIFF-format
file choose to embed the color profile.  This will make it easier for the image to be displayed
properly on other computers and to be printed correctly.

\begin{figure}[hbtp]
\begin{center}
\plotone{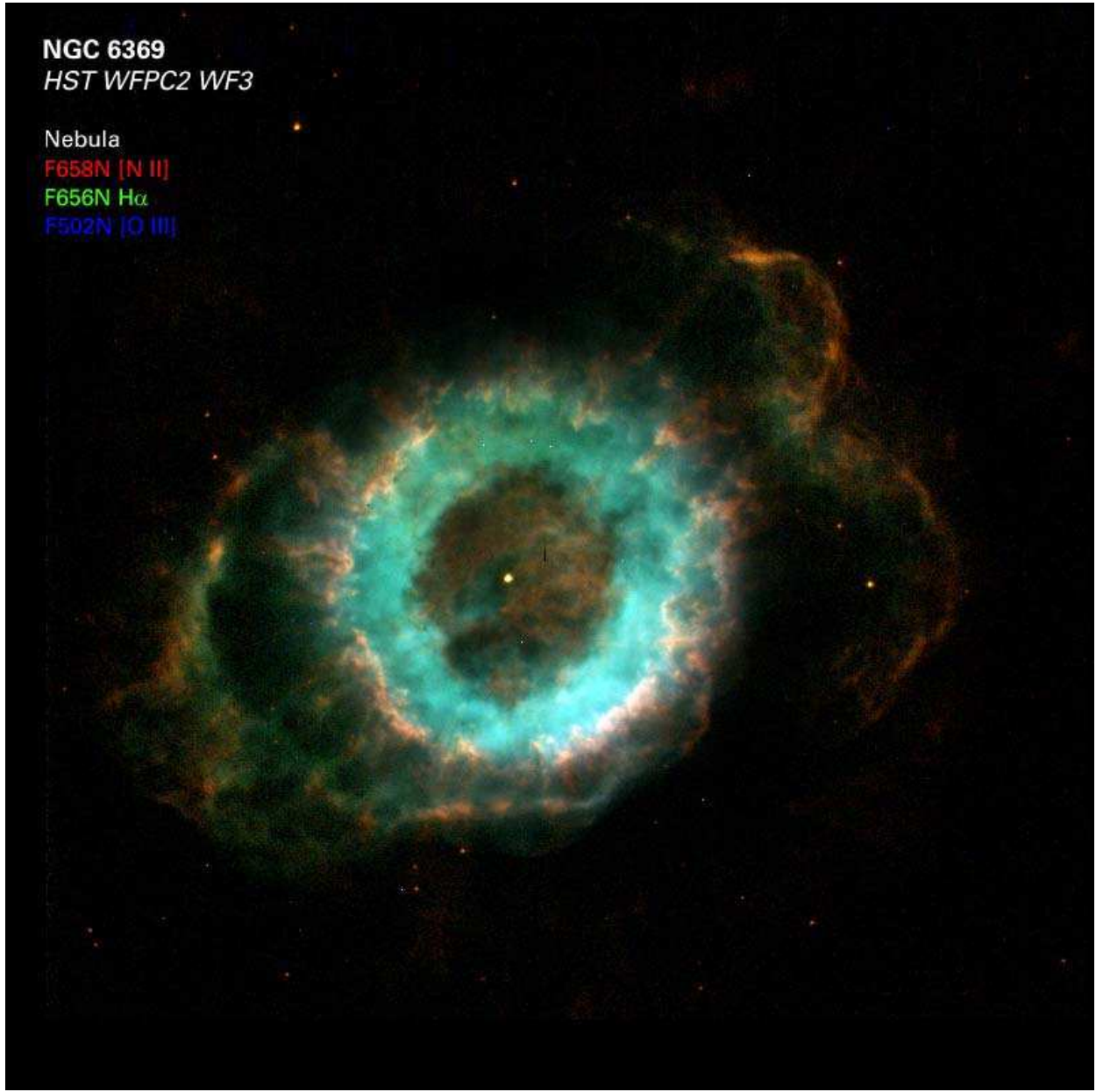}
\caption{The final color composite of NGC~6369.  Note that the color assigned to each layer is shown by the color of the text.}
\label{fig-32}            
\end{center}
\end{figure}

The final image is shown in Figure~\ref{fig-32}.  Note that this image is different
from the November 2002 Heritage release.  The image for the Heritage release also
included data through the F436W~B, F555W~V and F814W~I broadband filters to
better show the stars.  The image was also rotated 180\arcdeg\ to achieve a better
composition.

\section{Cosmetic Cleaning of Astronomical Images with Photoshop}

\subsection{Manual Removal of Pixel Artifacts}

Pixel artifacts in data from optical and near-IR arrays are ubiquitous.  A list of common pixel artifacts is given in Table~\ref{tbl-3}.  The removal of most of these artifacts must be done by hand.  It is a careful, time-consuming process, particularly with noisy or large images.  Fortunately, several tools in Photoshop are particularly useful for fixing undesirable pixels. The clone stamp tool, which is available in Photoshop as well as The GIMP, copies pixels from a preselected sample region that is set by holding down the option key and clicking in the region to be sampled.  The clone tool is useful for fixing smaller defects, such as bad pixels.  

The healing brush tool, introduced in Photoshop v7.0, also copies pixels from a preselected sample region.  It goes a step farther than the clone stamp tool by matching the texture, lighting, and shading of the sampled pixels to the region to be replaced.  It often performs a superior fix, particularly when there is a significant gradient in the background, e.g., in the halo of a bright star, so that the lighting of the sampled region is significantly different than the region to be fixed.

The patch tool, also introduced in Photoshop v7.0, blends pixels in the same manner as the healing brush.  The primary difference is that the patch tool can fix pixels in a selected region with pixels from the sampled area.  The patch tool is particularly useful for fixing larger defects, such as bad columns and CCD chip seams.  It is useful to first feather the selected region with the {\it Feather...} command under the {\it Select} menu.  This will smooth the transition between the fixed region and its surroundings.

\begin{figure}[hbtp]
\begin{center}
\plottwo{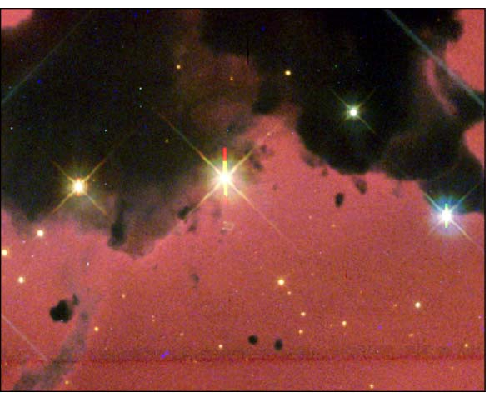}{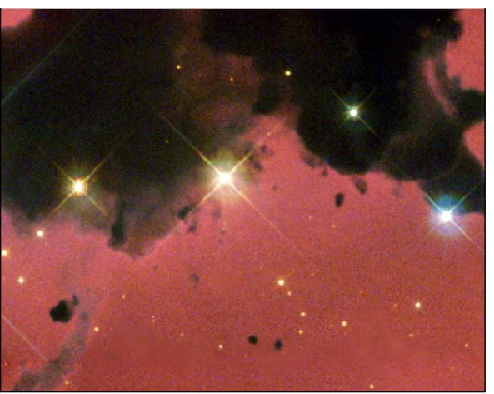}
\caption{An \hst\ image of IC~2944 before it has been cosmetically cleaned (left) and after (right).  Notice that several cosmetic flaws have been fixed, including residual cosmic rays (that appear as small specs), several bad columns and pixels, the chip seam along the bottom, CCD charge bleeds in the bright stars, and a blue line in the upper-right corner.  The diffraction spikes are not removed because they serve as a strong visual cue that this is an astronomical image.  Notice how much the cosmetic defects distract when comparing the uncleaned to the cleaned image.  }
\label{fig-33}            
\end{center}
\end{figure}

These tools allow for varying brush size, hardness and opacity.   In general use a soft brush no larger than necessary.  To achieve smoother-looking results with all of these tools it is recommended that one apply many short strokes rather than fewer long, continuous strokes.  It is also important to continually redefine the sampling source point so that textures in the image are not unduly repeated.  The eye is an excellent detector of patterns, and such repetitions are often noticeable even when very subtle.  Figure~\ref{fig-33} shows an example of an image that has been cleaned of pixel artifacts with the clone and healing brush tools.  It demonstrates the importance of cosmetically cleaning an image.

\subsection{Reducing Noise in an Image}

Unfortunately many datasets suffer from poor signal to noise. In Photoshop, there are several filters available which are effective at reducing noise and removing residual cosmic rays when a treatment by hand, such as with the clone stamp tool, is not possible or excessively time consuming.  Under the {\it Filter/Blur} menu, the {\it Blur} and {\it Gaussian Blur...} filters can be used to apply an overall smoothing to noisy selections, layers or color filter images.  The Gaussian blur filter allows control over the radius of blurring.  Increasing the radius will further reduce noise but at the cost of resolution.  These blurring filters are useful for layers which suffer from an overall poor signal to noise ratio.  To minimize the loss of resolution, apply the filter to only the noisiest layers. 

Under the {\it Filter/Noise} menu, the {\it Despeckle}, {\it Dust \& Scratches...}, and {\it Median...} filters should be used in areas where the signal to noise ratio is high but artifacts are still present. All of these filters smooth the image and therefore affect the overall resolution. To minimize the negative impact of the filter, use the lowest level of the filter, i.e., smallest radius and threshold, that produces the intended result.  Too large of a radius can also cause the centers of bright stars
to be altered.  It is also possible to apply the filters to a selected area only, in the manner described below.


\subsection{Fixing Sky Background Mismatches}

\begin{figure}[hbtp]
\begin{center}
\plottwo{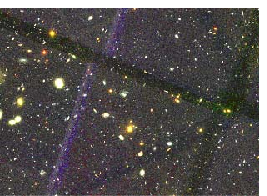}{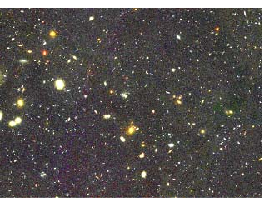}

\caption{A public-release image of the {\it Hubble} Ultra Deep Field before the chip seams have been fixed (left) and after (right).  The overlapping patterns in the dirty image is due to the gaps in the {\it ACS} camera mosaic and the rotation of \hst\ between observations.}

\label{fig-34}            
\end{center}
\end{figure}

\begin{figure}[hbtp]
\begin{center}
\includegraphics{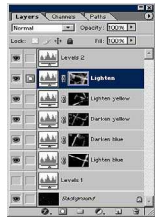}

\caption{The layers palette in Photoshop for the cleaned image in Figure~\ref{fig-34}.  Note that the masks for the adjustment layers have been painted in so that only the chip gaps are lightened.}

\label{fig-35}            
\end{center}
\end{figure}

Mosaics of images from multiple detectors and/or multiple pointings may result in a mismatch in the signal of the sky background, especially if poor sky subtraction was applied during the data reduction.  Carefully select the region to be treated with either a color or shape selection. Shape selections can be done with the marquee and lasso tools.  Color selections can be made with the magic wand tool or with
{\it Color Range...} under the {\it Select} menu.  Feathering the selection region, with the {\it Feather...} command under the {\it Select} menu, is recommended to avoid obvious edges. 
Apply a levels, curves or hue/saturation adjustment layer to match the color and intensity scaling of the adjacent areas.  When working with a complex selected region, it is often easier to see the effect of the adjustment layer by first hiding the selection boundary for a less cluttered view. View of the selection boundary is toggled by selecting {\it Extras} from the {\it View} menu.  If the mismatch is particular to an individual layer, work on that layer before flattening the image.  Multiple smaller adjustments can be used to refine the overall appearance, e.g., remove a colorcast in one region, brighten another region, etc., with separate adjustments.  Figure~\ref{fig-34} shows how gaps and chip seams were removed from a public release image for the {\it Hubble} Ultra Deep Field.  Figure~\ref{fig-35} shows the associated layer palette for the cleaned image.

\subsection{Fixing Mosaic Seams and Gaps}

\begin{figure}[hbtp]
\begin{center}
\plottwo{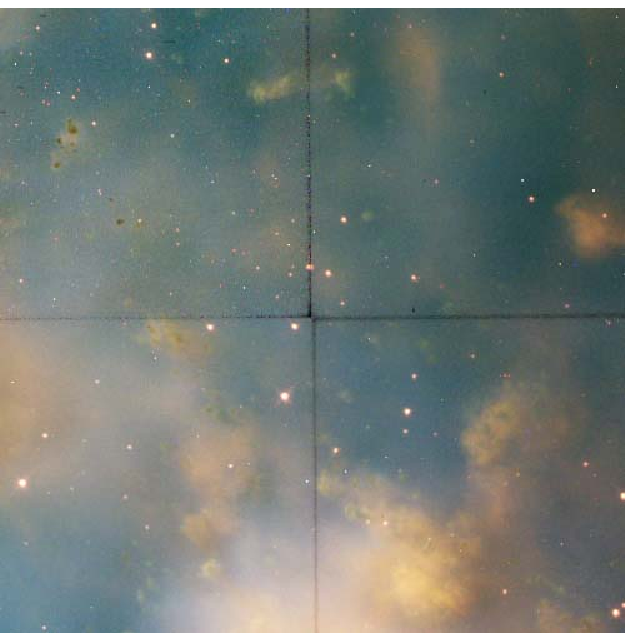}{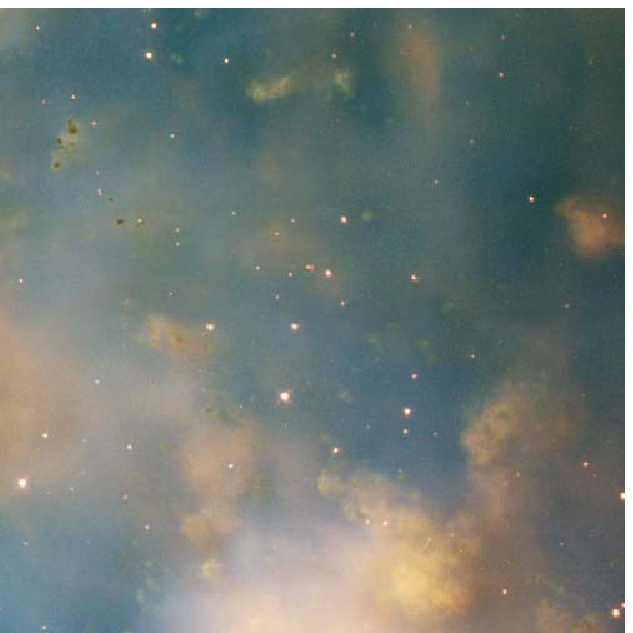}

\caption{An \hst\ image of a portion of M27 before it has been cosmetically cleaned (left) and after (right).  Notice that several cosmetic flaws have been fixed, not just the chip seams.}

\label{fig-36}            
\end{center}
\end{figure}

Many large format CCDs are a composite of individual chips that are mosaiced or stitched together to form a larger field of view. Some telescope detectors have a gap of pixels, e.g., \hst\ Advanced Camera for Surveys, between chips. The resultant artifact will be a seam or gap on the image at the boundary of the CCD chips. The clone stamp and healing brush tools are recommended to repair chip seams and gaps. For a chip seam where data exists on both sides of the seam, the clone stamp can be used in the normal, darken or lighten modes with varying levels of opacity to bridge the gap between the seams.  Experiment with blending a 50\% opacity clone from one side of a seam with a similar clone from the other side for a closer match.  Figure~\ref{fig-36} shows an image before and after chip seams, as well as other cosmetic defects, have been fixed.

\subsection{Removing CCD Charge Bleeds}

\begin{figure}[hbtp]
\begin{center}
\includegraphics{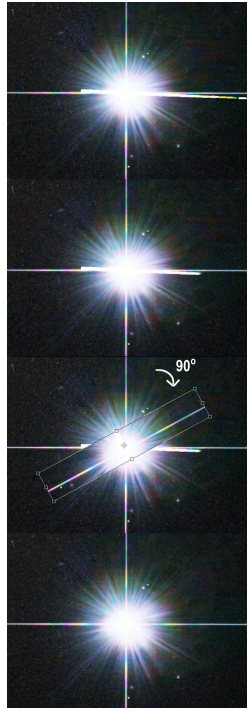}

\caption{A step-by-step illustration of repairing a CCD charge bleed from a bright star.  The top panel shows the
bright star with a CCD bleed that is just a few degrees clockwise from horizontal.  First the clone stamp or healing brush tool is used to truncate the charge bleed (second panel).  Next, the marquee tool is used to select a narrow region along the
vertical diffraction spike and centered on the star.  The edges are then feathered.  Copy the selection and paste it into a new layer.  The layer (not the image) is then rotated 90\arcdeg, as shown in the third panel.  The rotated layer is aligned with the star and then merged down to the original layer.  The clone stamp and healing brush tools are then used to blend any apparent seams (bottom panel).}

\label{fig-37}            
\end{center}
\end{figure}

A common problem for optical images is CCD charge bleeds, i.e., blooming, from bright stars in the field. In contrast to normally symmetric and subtler diffraction spikes from the secondary mirror supports, the single-dimension bleeds are a distraction that should be removed. Bleeds are relatively simple to remove, e.g., with the clone stamp and healing brush tools,  when they appear in a direction diagonal to the diffraction spikes. Bleeds that overlap diffraction spikes can be more difficult to remove because the diffraction spike and the halo around the bright star must be restored.  One can take advantage of the symmetry of the halo and diffraction spikes and copy a small section of the image containing the untainted diffraction spike pair into a new layer. Rotate the copy 90\arcdeg, aligning it with the center of the star and the bleed-affected diffraction spike pair. Erase or mask parts of the copy not needed to replace the bleed.  And then blend the edges using clone stamp and healing brush tools.  Figure~\ref{fig-37} shows a step-by-step example of fixing a charge bleed using this method.

\section{Photometric Calibration}

There are cases where, in the image-making process, one wants to retain the relative flux densities between the filters that are being used to create a color image; e.g., to show in a planetary nebula the relative amounts of \halpha\ emission in red and [OIII] emission in green.  Alternatively, one may be creating a mosaic from a number of chips. The photometric calibration of each chip will ensure a more accurate consistency in intensity across the image field. This is especially important at the low intensity levels if the intention is to show faint structures. 

When photometric calibration data are available, one can convert the counts per pixel in each filter into erg s$^{-1}$ cm$^{-2}$ \AA$^{-1}$ pixel$^{-1}$ values for each pixel. This is done at the data processing stage using the arithmetic tasks in a specified astronomy software program, e.g. IRAF.  This conversion will assign a flux density to each pixel for the bandwidth of that filter. This is consistent with each filter subsequently being assigned to a monochromatic color, i.e. to one wavelength or frequency. 

It is beyond the scope of this paper to describe the photometric calibration process because it is unique for every telescope; e.g., the standard technique for optical ground-based photometry is well described in \citet{lan92}.
However, the calibration of the \hst\ WFPC2 will be used as an example of how to convert from instrumental units, e.g., ``counts" sec$^{-1}$ pixel$^{-1}$, to physical units of monochromatic flux density, e.g., erg s$^{-1}$ cm$^{-2}$ \AA$^{-1}$ pixel$^{-1}$. 
The conversion is done by the following equation, which is based upon one given in the WFPC2 handbook \citep{koe03}:

$$ {\rm Flux \ density} = \frac{DN * PHOTFLAM * 10^{18}}{EXPTIME}  $$

In this equation, DN is the instrument's number of counts in each pixel, EXPTIME is the exposure time in seconds, and PHOTFLAM is the calibration constant, in units of erg DN$^{-1}$ cm$^{-2}$ \AA$^{-1}$. 
Each of the four WFPC2 chips has a different PHOTFLAM value.  Normally the exposure time and PHOTFLAM parameters in header of the FITS file for each chip are used. 
Since the PHOTFLAM parameter is of order $10^{-18}$, for ease of data manipulation and viewing, the equation above has been modified such that the units are $\sim 10^{-18}$ erg s$^{-1}$ cm$^{-2}$ \AA$^{-1}$.  Since only relative photometry is of concern here, the absolute values here need not be retained.
The calibration step is applied before a single dataset is projected from the WFPC2 mosaic so that the calibration is constant across the dataset.  Naturally each filter and each exposure time must also be calibrated separately.  After calibration, each dataset will consist of pixel data values that are independent of exposure time, wavelength and the performance characteristics of the telescope and instrument.

Once the datasets are properly flux density calibrated, they are ready to be projected into images as described in \S 2.1.
To retain the relative value of the flux density between filters, first ensure that the same minimum and maximum data values are set in each dataset.  Then, when projecting each dataset into an image, use the identical intensity scaling function for each dataset; i.e., use the same minimum and maximum scale values as well as identical gamma and scaling functions.
Once imported into an IP program, any rescalings of the image must also be applied equally to all of the layers if relative photometric accuracy is to be retained.  

Images generated in the above manner are photometrically calibrated in the sense that each layer is relatively calibrated to the others.  However, it still does not address the issue of the complex response that the human eye has to light and color.  No image-generation system is capable of conveying quantitatively the relative brightness of objects.  Any image will convey that one star is brighter than another, regardless of the intensity scaling, but to convey that a star is, e.g., two magnitudes brighter, is not possible because the eye is very poor at quantifying the intensity of light.  Similarly, it is poor at quantifying color beyond simple relative measures, e.g., ``bluer" and ``much bluer."  Thus, in general photometric calibration of an image may not be worth the effort in most cases for the simple result of an aesthetically pleasing image; and there are significant disadvantages incurred, as discussed in \S 2.1.

\section{Preparing an RGB Image for Prepress in Photoshop}

If an image is to be sent to press at an outside printing facility, it is necessary to establish the color settings in Photoshop before the image is opened.  If the commercial printing company to be used for reproduction of the image has provided an ICC profile it is important to import them into your Photoshop color settings at this point. 

To establish the color settings choose {\it Color Settings...} under the {\it Photoshop} menu. Once the {\it Color Settings} window opens make sure that the advanced mode is selected. If importing a custom ICC profile from a printer click the {\it Load...} button. In the load window select the profile to be loaded and click the {\it Load}button.  The custom profile will now be available in the CMYK drop down menu within the working spaces section of the {\it Color Settings} window.

\begin{figure}[hbtp]
\begin{center}
\includegraphics{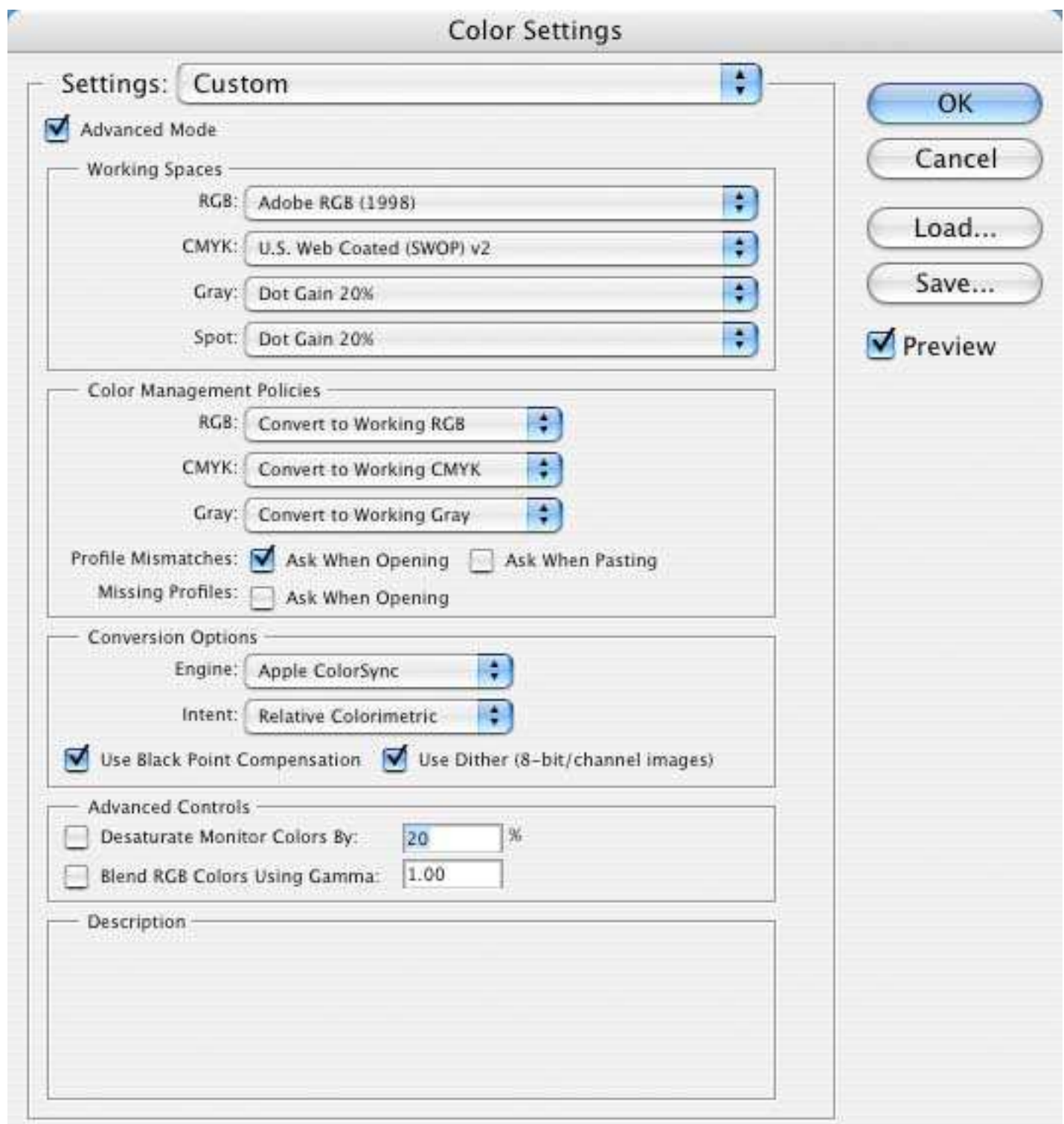}

\caption{The color settings window in Photoshop.  The settings should look similar as to shown here.}

\label{fig-38}            
\end{center}
\end{figure}

The working space for RGB should be set to ``Adobe RGB (1998)."  If a custom ICC profile for the printer was loaded, it should be chosen as the working space for CMYK.  If no custom profile is available, choose an appropriate profile as suggested by the printer.  ``U.S. Sheetfed Coated v2" and ``U.S. Web Coated Standard Web Offset Printing (SWOP) v2" are probably the best standard profiles for their respective types of printing.  The working spaces for Gray and Spot should be set to the dot gain values recommended by the printer.  The color management policies for RGB, CMYK and Gray should all be set to convert to the working space.  Under the conversion options the engine should be set to ``Apple ColorSync" and the intent to ``relative colorimetric."   It should also be set to use black point compensation and to use dither  for 8-bit/channel images. The color settings should look similar to that shown in Figure~\ref{fig-38}.  Save the color settings as the project name in the settings directory for future use.

Once the color settings are set, open the final, flattened image to be converted.  To convert the image select {\it Convert to Profile...} from the  the {\it Image~/~Mode} submenu. In the {\it Convert to Profile} window choose settings that reflect those chosen in the {\it Color Settings} window, i.e.,. set the destination space profile to the working CMYK, e.g., ``Working CMYK Ð U.S. Sheetfed Coated v2."  Also set the conversion options in the same manner, e.g., choose the Apple ColorSync engine and relative colorimetric intent.  Also use black point compensation and dither.  Save the CMYK file to a different filename. 

To see the effects of the conversion, open the original RGB image and place it side by side to the CMYK image.
If the color is inconsistent between the RGB and CMYK images, use the {\it Selective Color...} option under the {\it Image~/~Adjustments} submenu.   With the method set to relative, choose colors from the dropdown menu that need to be adjusted.  Adjust the color balance until the CMYK image matches the RGB image as closely as possible.  Note that because of the different gamuts exact replication isn't likely.

If the image has 16-bits per channel, it needs to be converted to ``8~Bits/Channel" under the {\it Image~/~Mode} submenu. The CMYK image should be saved to either TIFF or EPS format, with the color profile embedded.  The image is now ready to be sent to the press.




\clearpage


\clearpage

\clearpage

\clearpage

\clearpage

\clearpage

\begin{table} 
\begin{center} 

\label{tbl-1} 
\begin{small}
\begin{tabular}{lp{4.0in}p{1.5in}}
Contrast & Description & Example \\
\hline
Hue & The juxtaposition of saturated colors, including black and white.
      These hues can be randomly selected. & Antares \& Rho Ophiuchi \citep{mal79} \\ 

Light-Dark & The juxtaposition of brightnesses.  For example, a 
              dark green beside a light blue. & IC~2944 \citep{her02a} \\ 

Cold-Warm & The juxtaposition of warm colors (those with contain yellow or
              red) with cool colors (those which contain blue). 
             See \S~3.2 for more on these colors. & Trifid Nebula \citep{mal90} \\

Complementary & The juxtaposition of colors directly opposite on the
                 color wheel. See \S~3.2 for more about split complements
                 and compensating tones. & Carina Nebula \citep{her00} \\

Simultaneous & The juxtaposition of colors that are not complements
		in order to generate, e.g., physiological color mixing or the
		appearance of motion along common borders between colors.  & Europa \citep{gal98} \\

Saturation & A pure color is surrounded by dull, less pure colors.  
             The decrease in saturation to produce a dull color can be produced by 
              adding white, or black, or grey, or complementary colors. & SN~1987A \citep{bur94} \\

Extension & Differently sized areas of color are juxtaposed to acheive 
            balance between strong and weak colors. \citet{itt70} lists the appropriate balances for each color. & Tarantula Nebula \citep{bra04} \\

\end{tabular}
\end{small} 

\caption{Color contrasts. These seven color contrasts 
can be used in conjunction with the color
wheel based on Newton's experiments to choose color schemes.  For each contrast an
example image is given as a reference.  Note that many images have more than one contrast.}

\end{center} 
\end{table} 


\begin{table} 
\begin{center} 

\label{tbl-3} 
\begin{small}
\begin{tabular}{lp{4.25in}}
Defect & Description \\
\hline
Bad columns or pixels & Permanent defective areas on the CCD chip; they can be removed via masks and/or dithering. \\ 

CCD chip seams and gaps & Regions of the detector which lack pixels, either because of seams between mosaiced chips or physically damaged pixels; they vary by detector and may be more obvious in some images. \\

Coronograph / occulting masks & Small-aperture openings where a bright star may be
placed so that objects near the star can be imaged with less interference from the bright star. \\

Alignment gaps and ``notches" & A result of the stacking of offset and/or mosaiced images; such regions should be eliminated via cropping. \\

``Hot" pixels & Pixels with a high dark count that are usually saturated; they are periodic noise
features induced by electronics during readout and may be doubled in dithered data rather than be removed. \\

CCD charge bleed (``blooming") & Resultant filling of adjacent pixels due to electron overflow from nearby saturated pixels from a bright source. \\

Diffraction spikes & Lines which extend from the PSFs of bright stars; they are the result of
diffraction of light off of the support structure for the secondary mirror. \\

Cosmic rays & Bright pixels caused by an energetic particle striking the CCD substrate; since they vary in location in each data image, they are usually removed in the dithering process. \\

``Dust donuts" & Large and small rings that appear in the data due to improper flat fielding.  It is usually the result of dust particles on the optics that have moved between flat-field and object observations. \\

Meteor, asteroid and satellite trails & A streak across an image caused by a satellite or space debris crossing the field during an integration; they are less effectively removed in the dithering process. \\

Optical ghosts & Background light and faint images produced by a very bright star; they are the result of
multiple reflections on internal optics and/or crosstalk between amplifiers during readout. \\

``Dragon's breath" & Streaks and ghosts that appear due to internal reflections from a bright star just outside of the field of view. \\

Fringing and noise patterns & Small scale, wave-like patterns can be caused by the interference of sky emission lines within the camera optics.  Fringing patterns can, and should, be removed during data reduction.  Similar patterns can be caused by variable noise during CCD readout.  These cannot be removed during data reduction. \\

\end{tabular}
\end{small} 

\caption{Common artifacts and defects present in CCD data}

\end{center} 
\end{table} 

 \begin{deluxetable}{lllll}
 \tablecaption{The IRAF/KARMA Scaling Values for each Filter for M33 \label{tbl-A}}
 \tablewidth{0pt}
 \tablehead{
 \colhead{Filename} & \colhead{Filter} & \colhead{Minimum ($z1$)} & \colhead{Maximum ($z2$)} & \colhead{Function} 
}
 \startdata
{\it m33\_u.fits} & U &  35 & 700  & linear \\
{\it m33\_b.fits} & B & 236 & 2000  & linear \\
{\it m33\_v.fits} & V & 380 & 4000  & linear \\
{\it m33\_r.fits} & R & 762 & 5000 & linear  \\
{\it m33\_i.fits} & I & 1519 & 4000  & linear \\
{\it m33\_ha.fits} & \halpha & 36 & 3400 & logarithmic \\
 \enddata
 \end{deluxetable}

 \begin{deluxetable}{lllr}
 \tablecaption{The GIMP Scaling Values and Colors for each Filter for M33 \label{tbl-B}}
 \tablewidth{0pt}
 \tablehead{
 \colhead{Filter} & \colhead{Color} & \colhead{$\gamma$} & \colhead{Hue} 
}
 \startdata
U & Violet & 2.76 & 270  \\
B &  Blue & 2.40 & 230 \\
V &  Green & 2.02 & 120 \\
R &  Yellow & 1.74 & 55 \\
I & Orange & 1.00 & 35 \\
\halpha & Red & 1.61 & 0 \\
\enddata
\end{deluxetable}

\begin{deluxetable}{lll}
 \tablecaption{The IDL Scaling Systems for each Filter for NGC~6369 \label{tbl-C}}
 \tablewidth{0pt}
 \tablehead{
 \colhead{Filter} & \colhead{Linear Scaling} & \colhead{Logarithmic Scaling}
}
 \startdata

\halpha\ & $h>(-0.002)<0.20$ & $alog10((h+0.02)>0.0095<0.18)$ \\
${\rm [NII]}$ & $n>(-0.002)<0.09$ & $alog10((n+0.002)>0.0015<0.085)$ \\
${\rm [OIII]}$ & $o>(-0.002)<0.08$ & $alog10((o+0.001)>0.0012<0.075)$ \\
\enddata
\end{deluxetable}

\clearpage 
%
%
%




\clearpage

%




\end{document}